\documentclass[sn-mathphys-num]{sn-jnl}

\usepackage{graphicx}%
\usepackage{amsmath,amssymb,amsfonts}%
\usepackage{amsthm}%
\usepackage{xcolor}%
\usepackage{textcomp}%
\usepackage{natbib}
\usepackage{lmodern}

\usepackage[normalem]{ulem}

\begin{document}

\title[Article Title]{Machine learning based prediction of dynamical clustering in granular gases}

\author*[1,2,4]{\fnm{Sai Preetham}\sur{Sata}}\email{sai.sata@ovgu.de}

\author[2,3,4]{\fnm{Ralf} \sur{Stannarius}}

\author[1]{\fnm{Benjamin}\sur{Noack}}\email{benjamin.noack@ovgu.de}
 
\author[3,4,1]{\fnm{Dmitry} \sur{Puzyrev}}\email{dmitry.puzyrev@med.ovgu.de}

\affil[1]{\orgdiv{Autonomous Multisensor Systems Group, Institute for Intelligent Cooperating Systems}, \orgname{Otto von Guericke University Magdeburg}, \orgaddress{\street{
Universit\"atsplatz 2}, \postcode{39106}  \city{Magdeburg}, \country{Germany}}}

\affil[2]{\orgdiv{Department of Engineering}, \orgname{Brandenburg University of Applied Sciences}, \orgaddress{\street{Magdeburger Str. 50}, \postcode{14770} \city{Brandenburg an der Havel},  \country{Germany}}}

\affil[3] {\orgdiv{Department of Microgravity and Translational Regenerative Medicine (MTRM), Medical Faculty}, \orgname{Otto von Guericke University Magdeburg}, \orgaddress{\street{Universit\"atsplatz 2}, \postcode{39106} \city{Magdeburg}, \country{Germany}}}
 
\affil[4]{\orgdiv{Research Group MARS}, 
\orgname{Otto von Guericke University Magdeburg}, \orgaddress{\street{
Universit\"atsplatz 2}}, \postcode{39106}  \city{Magdeburg}, \country{Germany}}

\abstract{

When dense granular gases are continuously excited under microgravity conditions, spatial inhomogeneities of the particle number density can emerge. A significant share of particles may collect in strongly overpopulated regions, called clusters. This dynamical clustering, or gas-cluster transition, is caused by a complex interplay and balance between the energy influx and dissipation in particle collisions. Particle number density, container geometry, and excitation strength influence this transition. We perform Discrete Element Method (DEM) simulations for ensembles of frictional spheres in a cuboid container and apply the Kolmogorov–Smirnov test and a caging criterion to the local packing fraction profiles  to detect clusters. Machine learning can be used to study the gas-cluster transition, and can be a promising alternative to identify the state of the system for a given set of system parameters without time-consuming complex DEM simulations. We test various machine learning models and identify the best models to predict dynamical clustering of frictional spheres in a specific experimental geometry.}

\keywords{Granular gases, Dynamical clustering, Statistical analysis, Machine Learning}

\maketitle

\section{Introduction}

Granular matter, i.e ensembles of macroscopic particles, exhibits fascinating phenomena when it is subjected to excitation in a confined volume. Generally, granular matter can exist in three different states depending on the volume fraction of the particles, viz. gaseous, liquid, and solid states \cite{Opsomer2012}, and it can undergo transitions from one state to another. When the transition from a homogeneous granular gas to a clustered state occurs, the local number density of particles increases substantially in some region and the dynamics in the cluster resembles rather a liquid than a gas.

The differences to a classical gas arise particularly due to the inelastic character of the collisions of the constituents. Continuous excitation is necessary for the system to maintain a gaseous state \cite{Aumaitre2018} and to avoid the so-called granular cooling by continuous loss of kinetic energy. The experimental realization of three-dimensional (3D) granular gases in stationary heated or in cooling states requires confinement of the particles in a container (usually box-shaped). Vibrating walls \cite{Aumaitre2018} or alternative excitation methods \cite{Adachi2019} provide energy input. To avoid parasitic phenomena such as  accumulation of particles at the bottom of the cell caused by gravity, 3D granular gases are commonly studied under microgravity conditions, namely in a drop tower, in sounding rockets or in parabolic flights \cite{Harth2015,Noirhomme2018}.

As observed earlier \cite{Goldhirsch1993,Falcon2006}, local inhomogeneities can arise and dense patterns, called dynamical clusters, can appear when one increases the particle number density in a granular gas system. This was previously studied with spherical \cite{Opsomer2012,Noirhomme2018,Wu2020} as well as cylindrical particles \cite{PuzyrevDmitry2021}. The transition from the gaseous to the cluster regime is influenced by a combination of system parameters. 

Although a simple visual inspection often provides reliable evidence of clustering, a quantitative criterion of this transition is desirable. It should be applicable to simulated data sets and, ideally, to data obtained in experiments. We consider two established approaches, viz. the Kolmogorov–Smirnov test (KS-Test) \cite{Opsomer2012} and a caging criterion, both based on the analysis of local packing fractions. A certain weakness of these approaches is that they both disregard dynamical features of the ensemble. In addition, the results of the KS-Test are sensitive to its geometric limits and the significance level chosen, which requires a careful selection of the test parameters \cite{Sata2025_preprint_1}.

As a basis of our analysis, numerical simulations are performed on the basis of available software \cite{Hidalgo2013}. The general design of these simulations is derived from the VIP-Gran experiment setup \cite{Aumaitre2018} of the ESA Topical Team {\sc Space Grains} \cite{SpaceGrains}. We consider frictional spheres here, and an expansion to cylindrical (rod-like) particles is planned. Data-driven predictive models play a significant role in understanding  complex patterns observed in granular matter. While with traditional methods, it is often challenging to model the dynamical phenomena exhibited by granular materials, machine learning offers a powerful alternative by leveraging large datasets and predict this complex behavior. By assembling simulated data for multiple sets of system parameters, our goal is to 
test different Machine Learning (ML)  methods using these data in a statistical analysis.
We focus on two system parameters considered as most important for the gas-cluster transition, viz. the number of particles in the system (defining the average particle number density), the excitation strength \cite{Noirhomme2018}, and a third parameter which was not extensively studied before: the phase shift between the oscillations of the two opposite vibrating container walls. This parameter was found relevant in previous work on rod-like particle ensembles in a similar experimental geometry \cite{PuzyrevDmitry2021}.

The next section introduces the main motivation
of our work, followed by a section describing the experimental setup and the DEM based simulation software, the statistical criteria used to quantify dynamical clustering, and the employed machine learning methods. In the fourth section, the results are presented, and the final section contains the conclusions and gives an outlook.

\section{Motivation}

Clustering states in granular gases are observed during both the free cooling phase and during permanent excitation from external sources, yet the particular mechanisms of clustering differ in both cases. Here, we consider only the latter.

When granular gases are subject to external excitation, the energy input from the external sources counteracts the cooling process, it can keep the total kinetic energy in the system at an approximately stable level. At the same time, the excitation is often not applied to all particles uniformly (except for electromagnetic excitation \cite{Yu2019, Falcon2013, Falcon2017}). In most 3D experiments, mechanical excitation is employed by vibrating container walls \cite{Aumaitre2018}. Statistically, this results in higher particle velocities ('hot' particles) near the vibrating walls and regions with higher particle number densities roughly halfway between opposing excitation walls. This reminds one of the Leidenfrost effect of liquid droplets above a hot plate \cite{Leidenfrost1756}. Eshuis et al.\cite{Eshuis2005} found a similar behavior in granular ensembles under normal gravity. 

In microgravity, there exists a transition from the homogeneous gaseous state into such a cluster state depending upon the excitation parameters.  Men\'endez et al.\cite{Menendez2020} established a state diagram using filling fraction and excitation strength as relevant parameters. Sizes, positions and stability of clusters are determined by the balance between energy input and dissipation. The dense regions tend to dissolve when the excitation strength is lowered. Thus, the clusters are of dynamic nature and the phenomenon may be referred to as dynamical clustering.

To model such dynamical behavior of the clusters, machine learning (ML)-aided data-driven predictive models can be trained with simulation data to predict 
complex dependencies on system parameters. Traditional numerical simulations using DEM are computationally expensive, and hence ML can accelerate the overall analysis by learning from past computations and providing fast approximations which significantly reduce overall computational costs. 

In this numerical DEM study, we systematically created ``synthetic'' data sets that can be analyzed with ML software.
This software shall be able to predict the transition from the homogeneous gas to the cluster state (in short, gas-cluster transition), so that costly experiments and further simulations become needless. We primarily focus on testing various machine learning approaches to identify the dependence of dynamical clustering on these system parameters, in order to identify the best ML models for this purpose.

\section{Methods}

\subsection{Experimental reference system}

\begin{figure}[htb]%
    \centering
    \includegraphics[width=13.5cm]{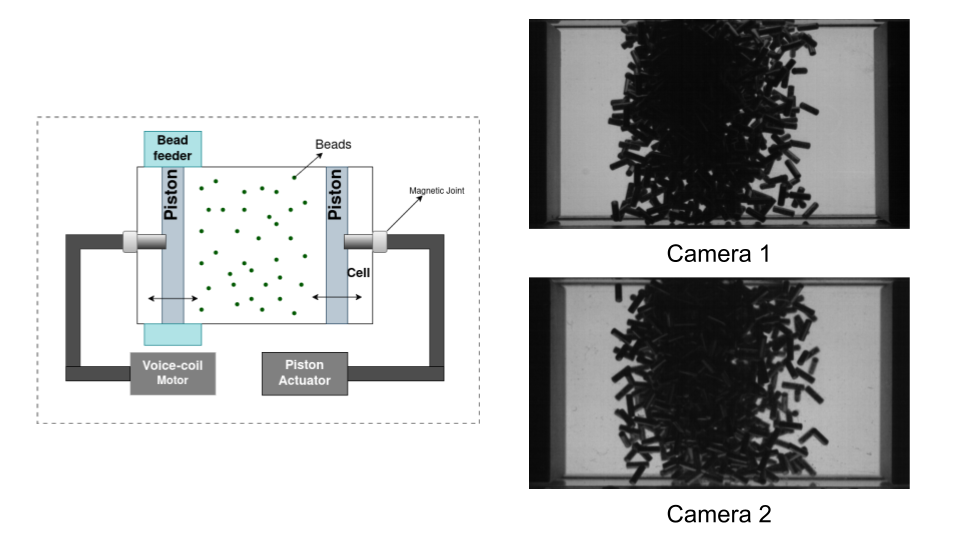}%
    \caption{A cross-sectional view of the of the VIP-Gran experiment with elongated grains is shown at the left. Snapshots from Camera 1 and Camera 2 that were recorded during a Parabolic Flight Campaign (PFC) are shown at the right. The schematic image and and frames of experimental footage were adapted from material kindly provided by the ESA Space Grains  topical team \cite{SpaceGrains}.}%
    \label{fig:VIP_GRAN_setup}%
\end{figure}

Our reference system, the VIP-Gran setup (Fig.~\ref{fig:VIP_GRAN_setup}), consists of a closed cuboid cell containing granular particles that can be excited by two opposing vibrating walls. Two synchronized cameras allow quantitative stereoscopic measurements such as particle tracking (in case of diluted systems) or visual flow and shadow density analysis (in case of dense granular ensembles, see \cite{Noirhomme2018,Puzyrev2021}), to extract packing fraction and mean velocity profiles. In previous experiments, the cameras recorded video footage with a frame rate of 900 frames per second (fps).

The cross section of the container is $30 \times 30$ mm$^2$. The two opposing piston walls are magnetically connected to linear voice-coil motors. The distance between the two opposing pistons can be varied up to a maximum of $L_{x,\max}=60$ mm. The grains can be placed directly into the cells prior to flight or injected into the container using a special bead feeder during flight. The fill fraction is given by $\Phi = (N\pi d^{3})/(6 L_x L_y L_z)$, where $d$ is the diameter of the spherical particles, $N$ is the number of particles in the container, and $L_x, L_y, L_z$ are the dimensions of the container.

The right and left wall positions $\omega_1(t)$ and $\omega_2(t)$ along the excitation axis $x$ are given by  
\begin{equation}
\label{Eq:Wall_coord}
\begin{split}
    \omega_1(t) &= L_x/2 + A\sin(\psi(t)),\\
    \omega_2(t) &= -L_x/2 + A\sin(\psi(t)+\theta),
\end{split}
\end{equation}
where $x=0$ corresponds to the center of the container, $F$ is the excitation frequency, $\psi(t)=2\pi F t$, and $t$ is the time. In individual runs of the experiment, the piston excitation parameters as well as the number of particles $N$ can be systematically varied. The simulation software uses spheres of diameter $d= 1$ mm. 

The same experimental configuration was the basis for several earlier studies. Opsomer et al. \cite{Opsomer2012} considered it as a reference for their numerical model of dynamical clustering. They stated that the gas-cluster transition is identified both by caging at the grain scale (i.e. trapping of particles by adjacent ones) and by a stable increase of the local number density at the system scale. Noirhomme et al. \cite{Noirhomme2018} used this setup experimentally in a low gravity environment during parabolic flights. They obtained a full phase diagram for spherical beads by varying a range of system parameters. The KS-Test proved good agreement of experiment and theory. 

Wu et al. \cite{Wu2020} performed a parametric investigation of the gas-cluster transition in a vibration-driven system, based on numerical simulations with frictional spheres. The authors studied the effects of the global volume fraction, the system size, the friction coefficient, and the restitution coefficients among particles and between particles and walls. The authors fixed the excitation amplitude and found that the coefficient of restitution plays a major role, while the friction coefficient is much less important for the gas-cluster transition. In contrast to the case of spherical beads, the substantial influence of friction was shown for rod-like particles  \cite{PuzyrevDmitry2021}.

Puzyrev et al. \cite{Puzyrev2021} evaluated both simulated and experimental data to extract density and velocity profiles. This encouraged us to model the same experimental setup for spheres to systematically analyze the effects of system parameters, using the same statistical criteria as in previous studies \cite{Opsomer2012, Noirhomme2018}, The simulation data and statistical criteria are employed to train machine learning algorithms. 

The parameters that are kept constant for all simulations in our study are listed in Table \ref{fixedsimulationparameters}. The density corresponds to bronze material; the choice of the elastic modulus is not critical. The ranges of parameters that are varied between simulations are listed in Tab. \ref{variablesimulationparameters}. We kept the frequency of excitation and the resting distance between the vibrating walls constant at $F=15$ Hz and $L_x=40$ mm correspondingly.
\begin{table}[htbp]
\caption{Constant parameters in simulation software}
\label{fixedsimulationparameters}%
\begin{tabular}{@{}llll@{}}
\toprule
Symbol & Parameter Name  & Value \\
\midrule
$\rho$ & Particle density  & 8730 kg/m$^3$  \\
$d$ & Particle diameter & 1 mm  \\
$m$ & Particle mass  & 4.57 mg \\
$Y$ & Young's modulus & 5 MPa \\
$\nu$ & Poisson's ratio & 0.3 \\
$\omega$ & Frequency of the oscillation & 15 Hz  \\
$\mu$ & Coefficient of friction & 0.4 \\
$\varepsilon$ & Coefficient of restitution
& 0.85\\
\botrule
\end{tabular}
\end{table}

\begin{table}[ht]
\caption{Variable parameters in simulation software}
\label{variablesimulationparameters}%
\begin{tabular}{@{}llll@{}}
\toprule
Symbol & Parameter Name & Range \\
\midrule
$N(\Phi)$ & Number of particles (Filling fraction) & [1834, 5643] (2.7\%, 8.2\%) \\
$A$ & Amplitude of the piston & [2 mm, 5 mm]  \\
$\theta$ & Phase shift & [0, $\pi$] \\
\botrule
\end{tabular}
\end{table}

\subsection{Numerical simulations}

\begin{figure}[ht]
    \centering
    \includegraphics[width=1.0\textwidth]{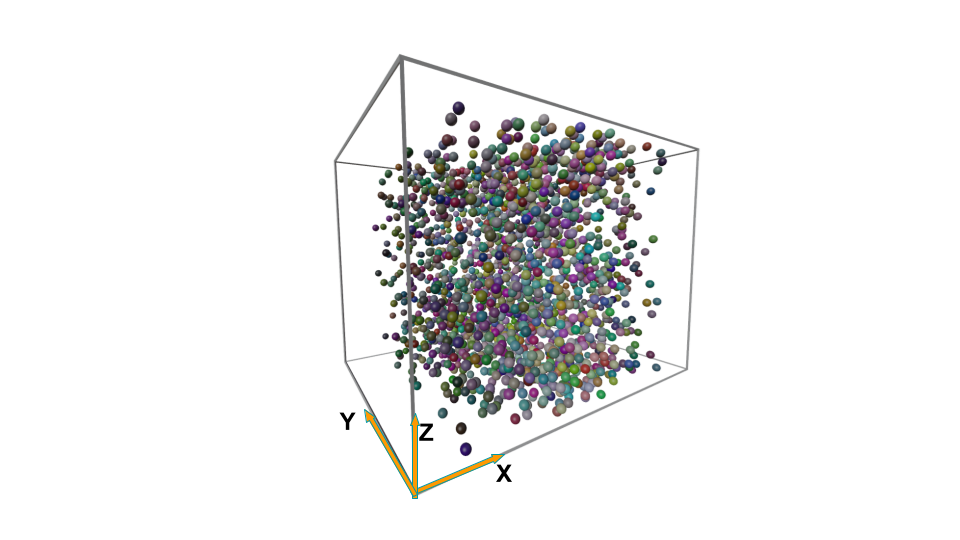}
    \caption{Snapshot of a CPU-GPU hybrid DEM simulation \cite{Hidalgo2013} of a dense granular gas of frictional spheres, rendered using the Blender 3D software. The experiment box contains 2214 spherical particles ($\Phi=3.2 \%$). The piston excitation with movable walls is along $X$. The walls with normals along $Y$ and $Z$ are fixed. The origin of the coordinate system is the center of the container. Colors serve for better visualization, physically, all particles are equivalent.}
    \label{fig:demsimulationtool}
\end{figure}

With the help of the available numerical tools, we simulate the mechanically excited ensemble to obtain particle positions, orientations, and translational and angular velocities using realistic system parameters. In order to model the VIP-Gran experimental setup, a hybrid CPU-GPU based Discrete Element Method (DEM) is used. The moving piston walls are included in the geometry. The DEM software accounts for particle deformations, rotations, and can handle complex geometries. 
A hybrid CPU-GPU based DEM is employed, which takes advantage of the parallelism in GPUs. The GPU-aided methods activate a large number of threads that perform parallel computations simultaneously, and they use fast shared memory to resolve neighboring particle contacts if necessary. This has been successfully demonstrated in previous studies \cite{{PuzyrevDmitry2021}, {Rubio2016}, {Hidalgo2016}, {Rubio2015Disentangling}}.
Each run required around 9-11 hours on a modern semi-professional class GPU (NVIDIA A4000). 
This is one of the major drawbacks of the current DEM simulation software, providing a strong motivation to utilize machine learning approaches to predict the state of the system for a given system parameters without the need for DEM simulation.

For particle-particle and particle-wall interactions, the Hertz-Mindlin contact model was used \cite{Wang2023, pongo-2022} to estimate inter-particle forces $F_{ij}$ between objects $i$ and $j$ in contact. The normal force component is based on Hertzian contact theory \cite{hertz1882contact} and the tangential force component is based on the Mindlin-Deresiewicz theory\cite{Mindlin1949}. The 
restitution coefficient $e_n$ and the dynamic friction coefficient $\mu$ are the essential parameters.

The Young modulus of the grains was set to only 5 MPa for practical purposes, after it was established that this reduction has no critical influence on the statistical results \cite{Puzyrev2021}. This value was large enough to constrain excessive particle overlap and small enough so that collisions are resolved in a sufficient number of simulation time steps $\delta t=  11.1 \cdot 10^{-8}$  \cite{Puzyrev2024}. The shear modulus is set to $10^6$.  The friction coefficient between the particles and the walls was set to $\mu_w=0.4$ \cite{Wu2020}. The collision damping coefficient is 0.0943. 

We adapted the DEM simulation software to model the container of the VIP-Gran experiment described above. The wall vibrations are given by Eq.~\eqref{Eq:Wall_coord}. The choice of phases between the wall vibrations was restricted to in-phase ($\theta=0$) and anti-phase ($\theta=\pi$) piston movement. 

It was observed that after some initial transients, the ensemble dynamics generally synchronizes with the wall excitation (see \cite{Noirhomme2018} for details). For simplicity, the duration of simulation is measured in terms of excitation periods. With $F=15$ Hz and the camera frame rate of 900 fps, it is convenient to define simulation ``frames" similar to the experimental video frames. Then, each full excitation period corresponds to 60 consecutive frames from $\psi(t)=0$ to $\psi(t)=2\pi$. The output of the simulations was saved for every 10000 iterations. Each simulation comprized 120 excitation periods, where the first 60 periods are counted as transient and are discarded in the statistical analysis. Our simulations, along with previous investigations \cite{Noirhomme2018,Wu2019} show that this transient period is more than enough for the system to ``forget" the initial distribution of particle positions and velocities and to reach a stable dynamical state, dependent only on the system properties and excitation parameters.

\subsection{Kolmogorov-Smirnov test (KS-test) criterion}

The Kolmogorov–Smirnov test (KS-Test) can be used to distinguish between gaseous and cluster states. The KS-Test quantifies the distance between the measured cumulative distribution $f(x)$ of the material to the empirical uniform distribution $f(x)$. The supremum distance between the two distributions is
\begin{equation}
    D_{\max} = \sup_{x_1<x<x_2} \left| f(x) - u(x) \right|.
\end{equation}

The KS-threshold is defined as $T=K_\alpha/\sqrt{k}$ \cite{Noirhomme2018}, where $K_\alpha$ is estimated for a particular significance level $\alpha$ and $k$ is the number of classes, which we set to $k = L_x/d$.
Based on previous work \cite{Opsomer2012,Noirhomme2018,Wu2020},  $\alpha=0.05$ was selected. 
When the difference $T-D_{\max}=\delta$ is negative for some $\psi$, the system is in a clustered state at that phase. Otherwise, it is considered to be in the gaseous state. The minimum $\delta_{\min}$ for a given set of system parameters $\{N, A, \theta\}$ is used as a criterion for the {gas-cluster} transition. 
 
\subsection{Caging criterion}

Clusters can also be identified by the caging effect. In the gaseous state, there is a high probability that any selected particle can traverse the entire volume of the cell \cite{Opsomer2012}. For larger fill fractions, this probability is reduced to a large extent. Particles that enter the clustered zone will get caged inside and their kinetic energy is dissipated in inelastic collisions.  
A particle is considered caged when the ratio of the grain volume to the volume of its Voronoi cell (with the particle positions as generators) is greater than a given threshold $T_{\rm caged}= 0.285$ \cite{Opsomer2012}.The number of caged particles $N_{\rm caged}$ is estimated for all $\psi$ and the ratio $N_{\rm caged}(\psi)/N$ is determined. The maximum of this ratio for all phases defines the maximum fraction of caged particles, $\varphi_{\rm caged}$. The system is considered to be in a clustered state if $\varphi_{\rm caged} > 0.05$ \cite{statistical_significance_0_05}, otherwise it is in the gaseous state.

\subsection{Machine learning algorithms}\label{mlalgorithms} 

\paragraph{Regression}

To identify whether the system is in a gaseous or clustered state, $\delta_{min}$ and $\varphi_{caged}$ serve as the relevant criteria. 
Since these parameters are continuous variables, we study machine learning approaches for regression analysis. The five regression algorithms that are evaluated are polynomial regression, random forest regression, support vector regression, XGBoost regression, and neural network regression. The Python-based package \textsc{scikit-learn}  \cite{scikit-learn} was used to implement the ML algorithms in this work. \\

Polynomial regression represents the relationship between independent and dependent variables as an $n^{\rm th}$ degree polynomial. Here, we use second-degree polynomials. 

Random forest regression (RFR) \cite{rf} combines multiple decision trees in parallel for a model with lower variance. In this work, we used RFR with 500 estimators and a maximum depth of 4 respectively. 

Support Vector Regression(SVR) works on the principles of Support Vector Machines (SVM) and transforms the input features into a higher dimension to locate the ideal hyperplane that accurately represents the given data. In this work, SVR with Radial Basis Function (RBF) kernel and regularization ($C$) 2 is used. 

XGBoost regression (XGB) \cite{xgb} utilizes ensembles of decision trees and gradient boosting techniques. XGB with 600 estimators, a maximum tree depth of 5 and a learning rate of 0.01 is used in this work. 

Artificial Neural Networks (ANN) \cite{NN_regression} help in learning non-linear relationships between features and target variables with the help of activation functions. Here, we adopt a Neural Network with three hidden layers that contain 16, 32, and 16 neurons, respectively. The output layer contains one single neuron. A rectified Linear Unit (ReLU) \cite{Agarap2018} is used as the activation function to introduce nonlinearity by outputting the results if positive and zero otherwise. The learning rate, batch size, and number of epochs are set to 0.0001, 8, and 100, respectively.

\paragraph{Classification}

In addition to the prediction of continuous variables $\delta_{\min}$ and $\varphi_{\rm caged}$, we also investigate machine learning approaches to identify the state of the system derived from $\delta_{\min}$ and $\varphi_{\rm caged}$ for given sets of system parameters. We used the classification algorithms from ``scikit-learn'' \cite{scikit-learn} in this study. These algorithms are the Decision tree classifier \cite{breiman1984classification} with maximum depth of 5, the Random forest classifier \cite{breiman2001random}
with 200 estimators and a maximum tree depth of 5, the Support vector classifier
\cite{SupportVectorRegression} with a RBF kernel with $C=2$ and kernel coefficient 0.1, the XGBoost classifier \cite{nvidia_xgboost} with 200 estimators, learning rate 0.01 and maximum depth 5, and a Neural Network based classifier with two hidden layers of 64 and 32 neurons, respectively. Sigmoid layer combined with binary cross entropy was used as the combined loss function \cite{bcewithlogits}.

\subsection{Performance metrics}\label{performancemetrics}

The performance of the regression models is evaluated by calculating how close the predicted values are to the true data in the given sets \cite{nvidia2025}. The data are split into training, validation and test datasets. The models are trained on the training dataset and the performance is evaluated on the validation and test datasets. The validation datasets are used to obtain the best learnable parameters for a given dataset and the model performance is tested on the test sets with the help of these parameters. The important metrics to evaluate the performance of the regression models are Root Mean Square Error (RMSE), Mean Absolute Error (MAE) and Coefficient of Determination ($R^2$).\\

The performance of classification models can be estimated by comparing the number of predicted classes with the true classes. The data sets are split into training, validation and test sets. The training set is used to train the classification algorithms, and the trained algorithm is evaluated on validation and test sets. The popular metrics to evaluate classification algorithms are accuracy, precision, recall, F1-score and confusion matrix \cite{google_metrics, Opitz2024}. We first identify True Positives ($TP$, i.e. correctly predicted cluster states), False Positives ($FP$ gaseous states incorrectly identified as clustered), and in a similar sense True Negatives ($TN$) and False Negatives ($FN$).

The Receiver-operating characteristic (ROC) curve illustrates the performance of the classifier, comparing the true positive rate $TP/(TP+FN)$ with the false positive rate $FP/(FP+TN)$ at various threshold settings. The area under the ROC Curve (AUC) quantifies the performance ranging from 0 to 1. An AUC close to 1 indicates a highly effective model, while a value near 0.5 indicates performances no better than a random chance. An AUC close to 0 would imply the model is consistently incorrect.

\section{Results and Discussion}

In this section, we discuss the performance of the machine learning algorithms and obtain best models for predicting $\delta_{\min}$ and $\phi_{\rm caged}$. We compare the predicted gas-cluster transition with that of the true transition in order to visualize how reliable the predictions are. We identify the best algorithms to predict the state of the system.

\subsection{Prediction of ML algorithms using the KS-test criterion}\label{prediction_ks_test}

We investigate both regression and classification in order to evaluate the performance of machine learning algorithms mentioned in Sec. \ref{mlalgorithms} on the data obtained by KS tests, to select the best models to predict the state of the system from $\delta_{\min}$. Two cases were investigated in order to explore the model's ability to interpolate predicted $\delta_{\min}$ in the given range of the training and validation datasets, and to extrapolate $\delta_{\min}$ beyond that range. 

\subsubsection{Regression}

The independent variables are $N$, $A$, $\theta$ and the dependent (target) variable is $\delta_{\min}$.\\ 

\textit{\textbf{Interpolation}}

 The original dataset contained particle numbers $N$ ranging from 1834 ($\varphi=2.7 \%$) to 5643 ($\varphi=8.2 \%$), amplitudes $A$ from 2 mm to 5 mm with step size of 0.5 mm, and $\theta/ \pi$ from 0 to 1 with step size of 0.25. This parameter combination yields the dataset size of 385 samples, where 25 \% were randomly selected as the test set. Of the remaining 75\%, 65\% were randomly assigned as a training set, and the remaining 10\%  formed the validation set. The ML algorithms of Sec. \ref{mlalgorithms} were then evaluated with the metrics introduced in Sec. \ref{performancemetrics}. Table \ref{testdataset_var_interpolation} shows the results.

 \begin{table}[ht]
\caption{Performance metrics to predict $\delta_{\min}$ from the test dataset}
\label{testdataset_var_interpolation}%
\begin{tabular}{@{}llll@{}}
\toprule
Model & RMSE & MAE & R$^2$\\
\midrule
Polynomial Regression & 0.007  & 0.005  & 0.964  \\
Random Forest Regression & 0.001  & 0.01  & 0.894 \\ 
Support Vector Regression & 0.038  & 0.03 & 0.07 \\ 
XGBoost Regression & 0.0012  & 0.01  & 0.89  \\ 
\textbf{Neural Network Regression} & \textbf{0.007}  & \textbf{0.005}  & \textbf{0.967}  \\  
\botrule
\end{tabular}
\end{table}

The Neural Network (ANN) has the howest RMSE and MAE values and the best R$^2$ of all models. Thus, it was chosen as the best model for this particular scenario. We used a three-layered Neural Network trained for 300 epochs with a batch size of 8. Adam optimizer and Mean Square Error (MSE) loss function were chosen with a constant learning rate of 0.0001 to train the ANN. 

\begin{figure}[htbp]%
\begin{minipage}{0.4\textwidth}     \caption{Comparison of true and ANN predicted values of $\delta_{min}$ in the test dataset. Red line: perfect machine learning model }%
\label{fig:test_loss_curve_neural_network_variable_interpolation}%
\end{minipage}\hfill
\begin{minipage}{0.5\textwidth}    
    \includegraphics[width=\textwidth]{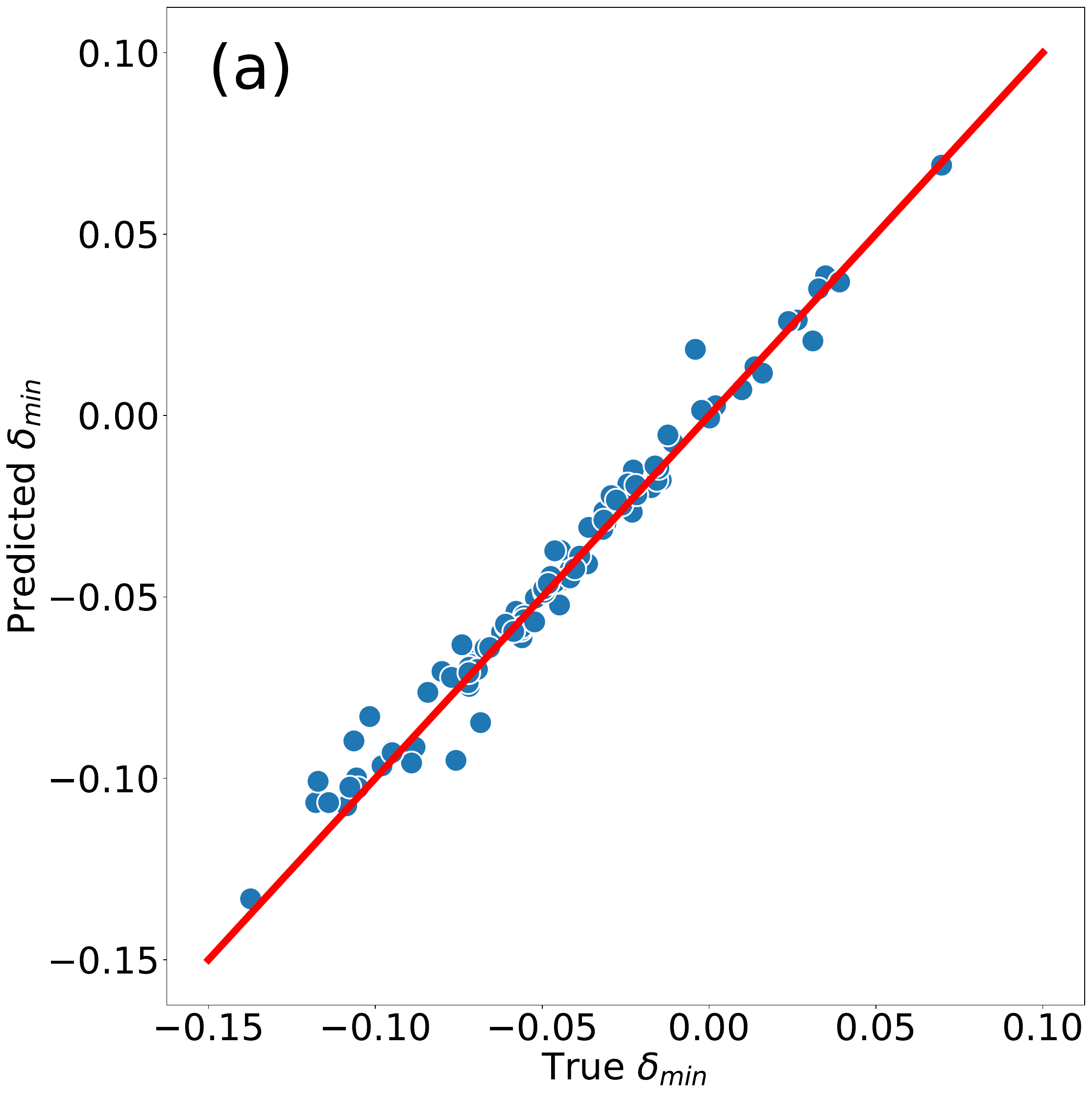}%
\end{minipage}      
\end{figure}

\begin{figure}[htbp]%
    \centering
    \includegraphics[width=6.5cm]{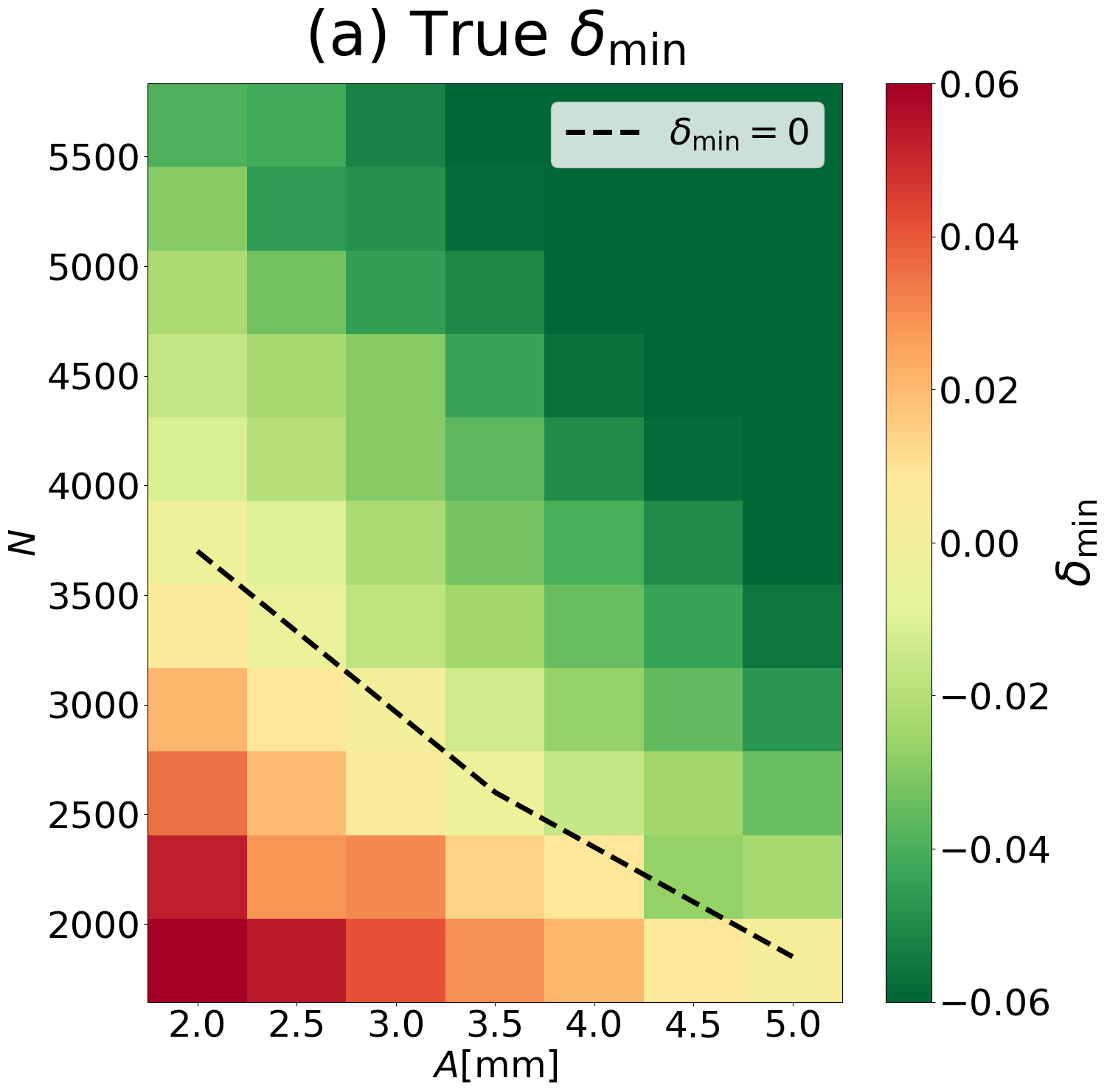}%
    \includegraphics[width=6.5cm]{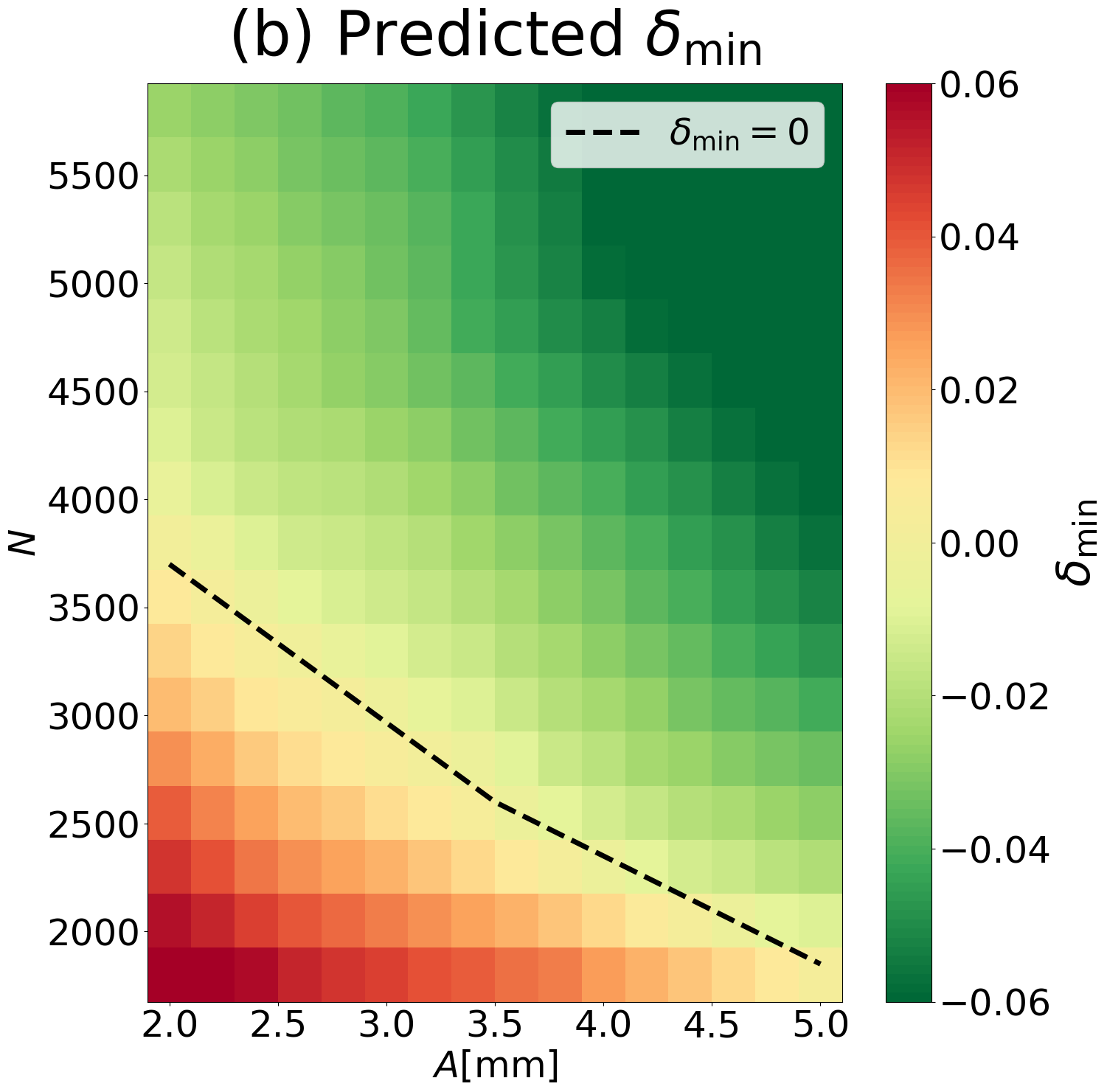}%
    \caption{Comparison of true (a) and interpolated ANN predicted (b) $\delta_{min}$ 
    when the system is antiphase excitation. Red and green reflect gaseous and cluster states, respectively. The  transition occurs near the dashed line (zero of $\delta_{\min}$).}%
    \label{fig:contour_plots_phase_shift_0_1_interpolation_variable}%
\end{figure}

The reliability of the predictions can be validated by comparing the true and predicted $\delta_{min}$ for the test dataset, as shown in Fig. \ref{fig:test_loss_curve_neural_network_variable_interpolation}. The red line represents the performance of a perfect machine learning model,
and the predictions of the ANN are satisfactorily close to that. Figures \ref{fig:contour_plots_phase_shift_0_1_interpolation_variable}(a) and (b) show the true and predicted $\delta_{\min}$ respectively. A transition from the homogeneous to the cluster state for 3700 particles
happens at $A \approx 2$ mm. With $A\approx 5$ mm, the same transition occurs already at a fill fraction of $\varphi=2.7\%$.
The plots show that ANN can predict the gas-cluster transition, with relatively good accuracy, closely resembling the true $\delta_{\min}$. ANN can identify the state of the ensemble reliably without the need to perform numerical DEM simulations.
\\

\textbf{\textit{Extrapolation}}

For extrapolation tests, two cases were investigated. In the first case, the model's ability to extrapolate its predictions for $N$ that are not in the training domain is tested. For this case, the initial data were sampled with $N$ between $1834$ ($\varphi=2.7 \%$) and 4881 ($\varphi=7.1 \%$) particles. Of this dataset, 90 \% were randomly assigned for training and the remaining 10 \% for validation. The trained model was then applied to test data with $N$ between 5262 ($\varphi=7.65 \%$) and 5643 ($\varphi=8.2\%$). The amplitudes $A$ were between $A=2$ mm and $5$ mm with a step size of 0.5 mm, and $\theta/\pi$ was chosen between 0.0 and 1.0 in steps of 0.25. The initial dataset contained 315 samples, the test set 70 samples.

Table \ref{testdataset_var_extrapolation} shows the overall performance of the machine learning models in the extrapolation task. XGBoost and Polynomial regressions have the smallest RMSE, MAE and the largest R$^2$ of all models and were hence chosen as the best candidates. The negative R$^2$ for Support Vector and Neural Network regression models mark failure to fit the given data.  

\begin{table}[ht]
\caption{Performance metrics in the extrapolation to larger $N$}
\label{testdataset_var_extrapolation}%
\begin{tabular}{@{}llll@{}}
\toprule
Model & RMSE & MAE & R$^2$\\
\midrule
\textbf{Polynomial Regression} & \textbf{0.017}  & \textbf{0.012} & \textbf{0.49} \\
Random Forest Regression & 0.018  & 0.016 & 0.39 \\ 
Support Vector Regression & 0.036  & 0.028 & -1.2 \\ 
XGBoost Regression & 0.016  & 0.013  &  0.557 \\ 
Neural Network Regression & 0.063 & 0.051  & -6.0 \\  
\botrule
\end{tabular}
\end{table}
    
\begin{figure}[htbp]%
    \centering
    \includegraphics[width=6cm]{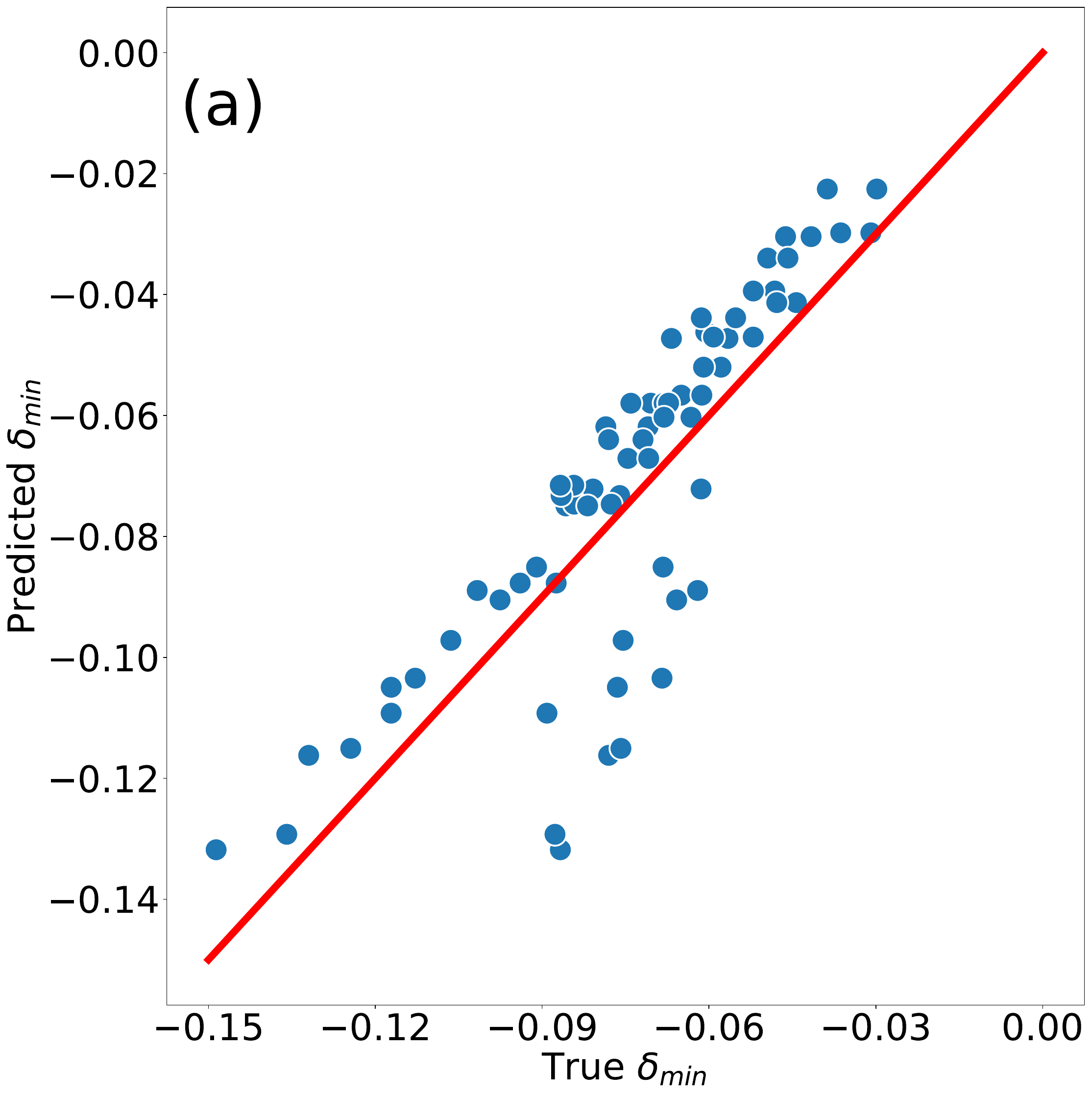}%
    \hspace{0.5cm}
    \includegraphics[width=6cm]{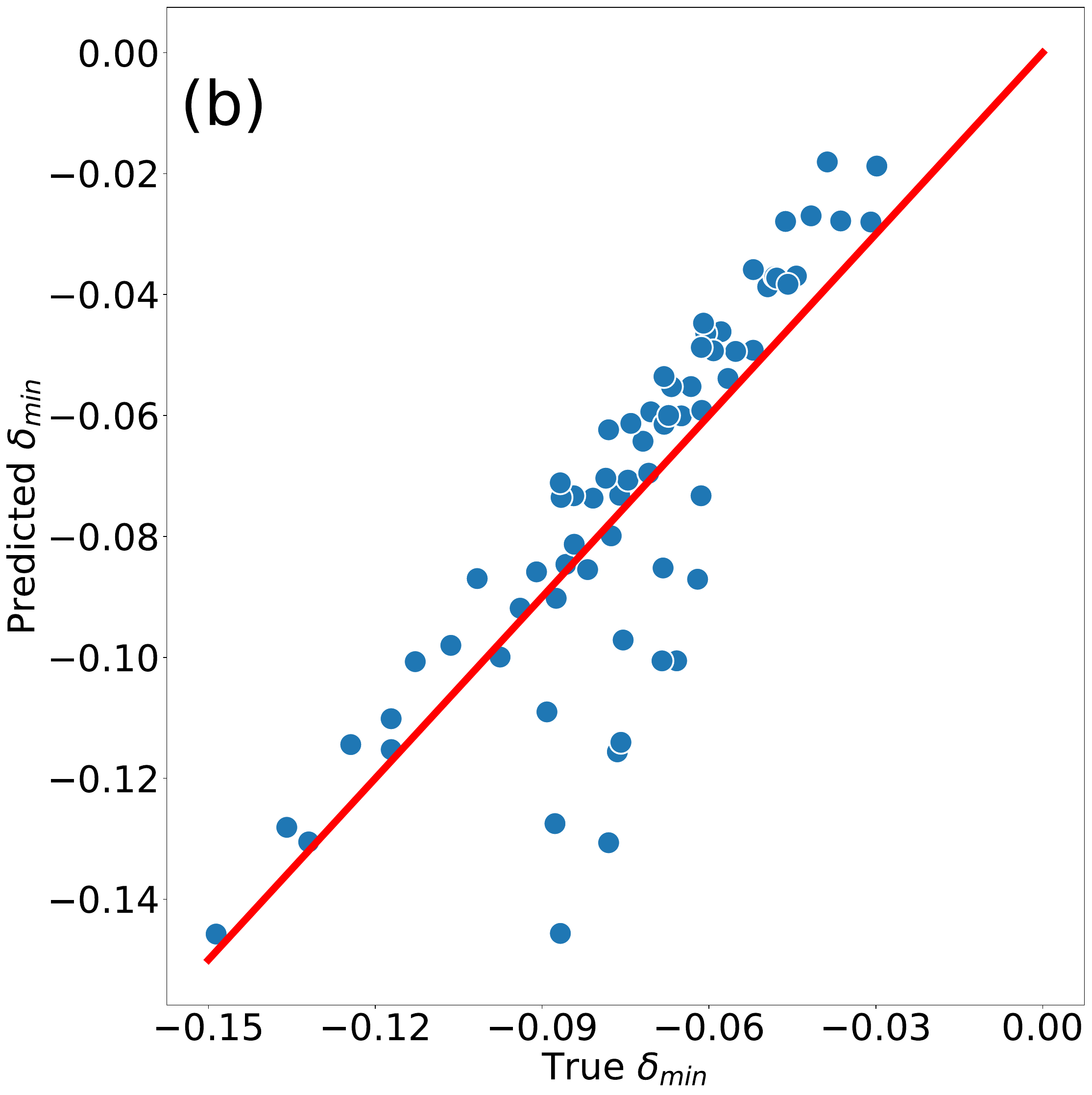}
    \caption{Comparison between true and predicted values of $\delta_{\min}$ for the test dataset obtained by XGBoost regression (a) and Polynomial regression (b) respectively for the extrapolation to larger $N$.}%
    \label{fig:test_xgboost_variable_extrapolation}%
\end{figure}

\begin{figure}[htbp]%
    \centering
    \begin{tabular}{cc}
        \includegraphics[width=7.0cm]{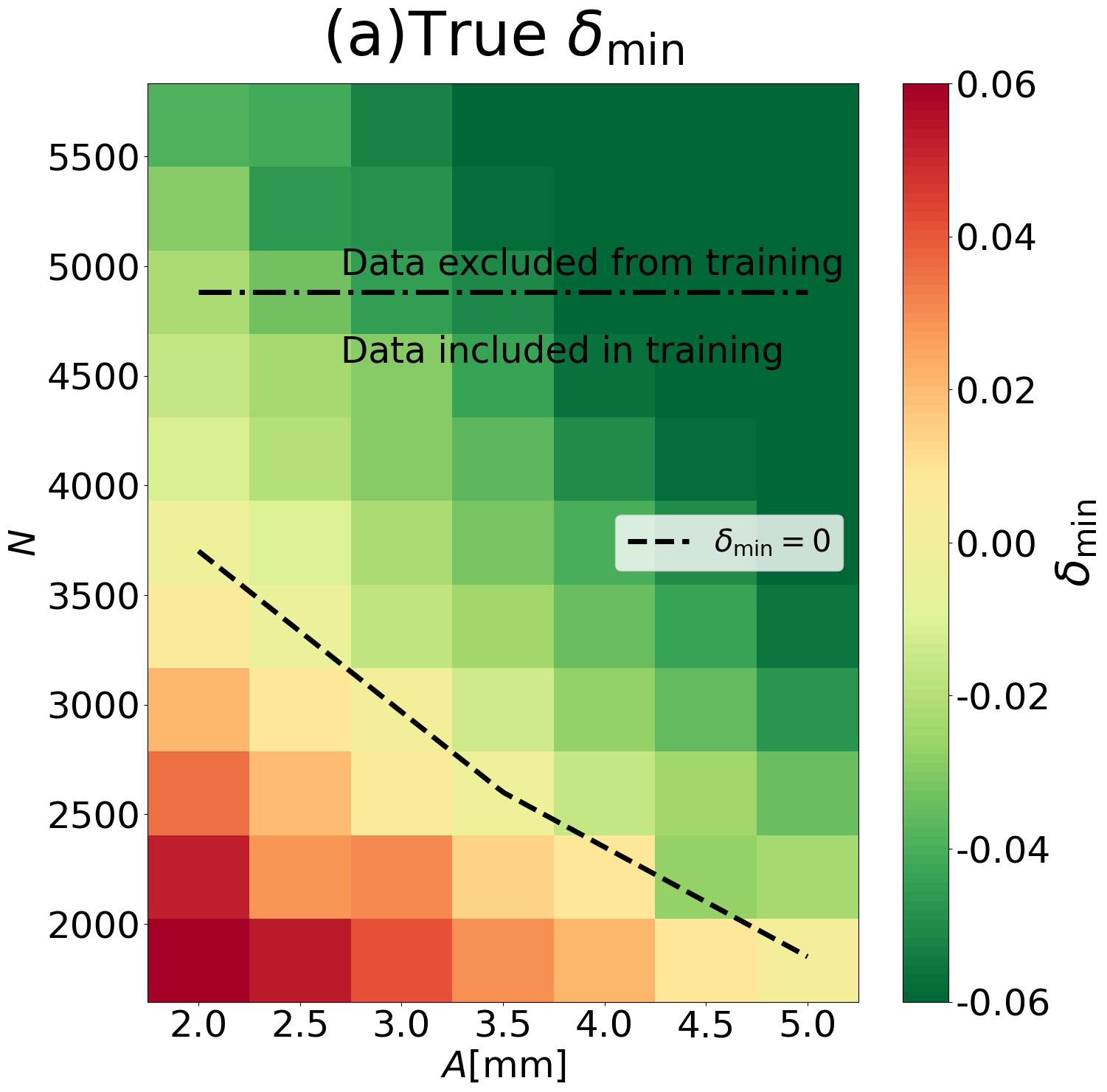} &
        \includegraphics[width=7.0cm]{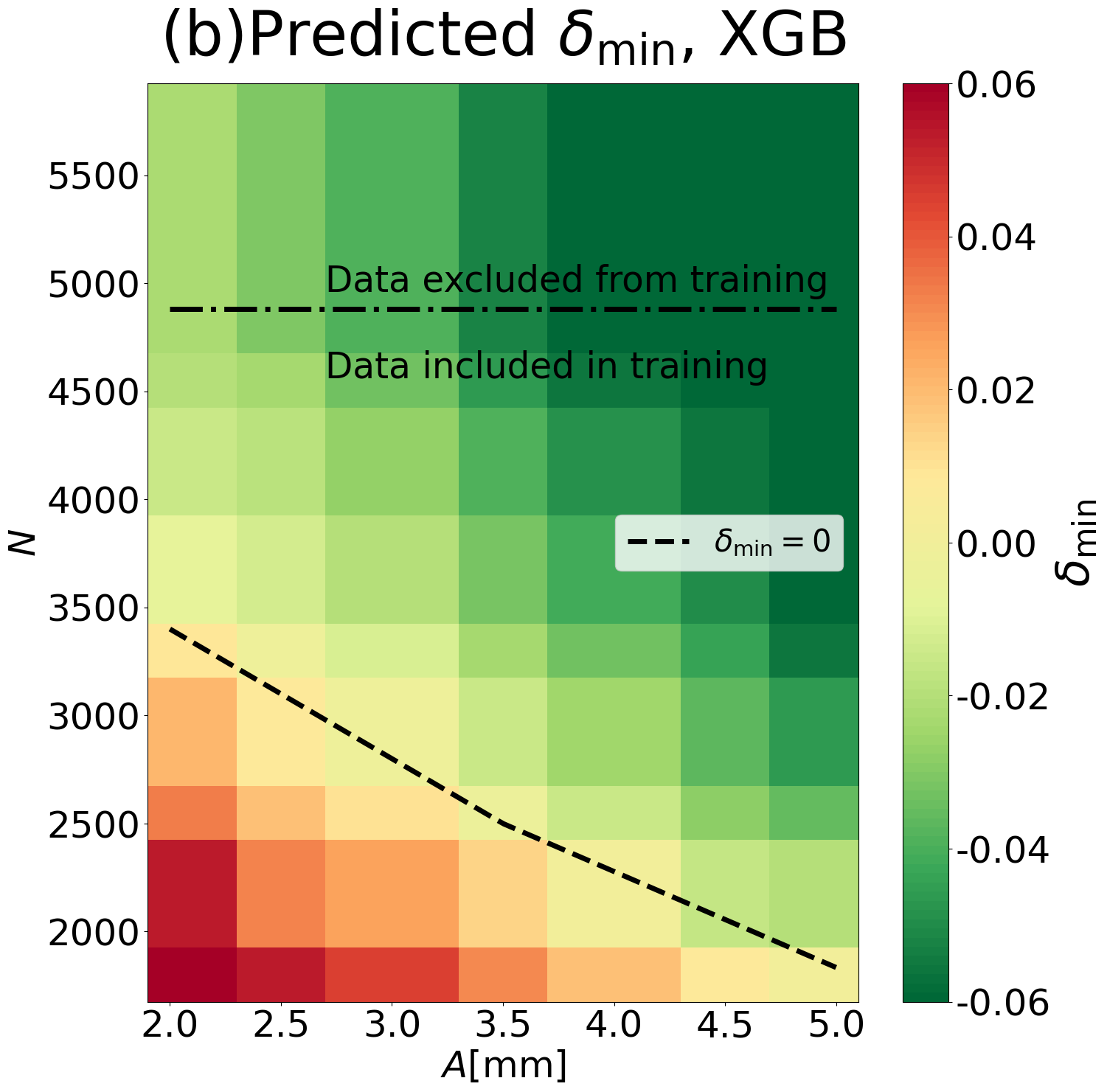} \\
        \includegraphics[width=7.0cm]{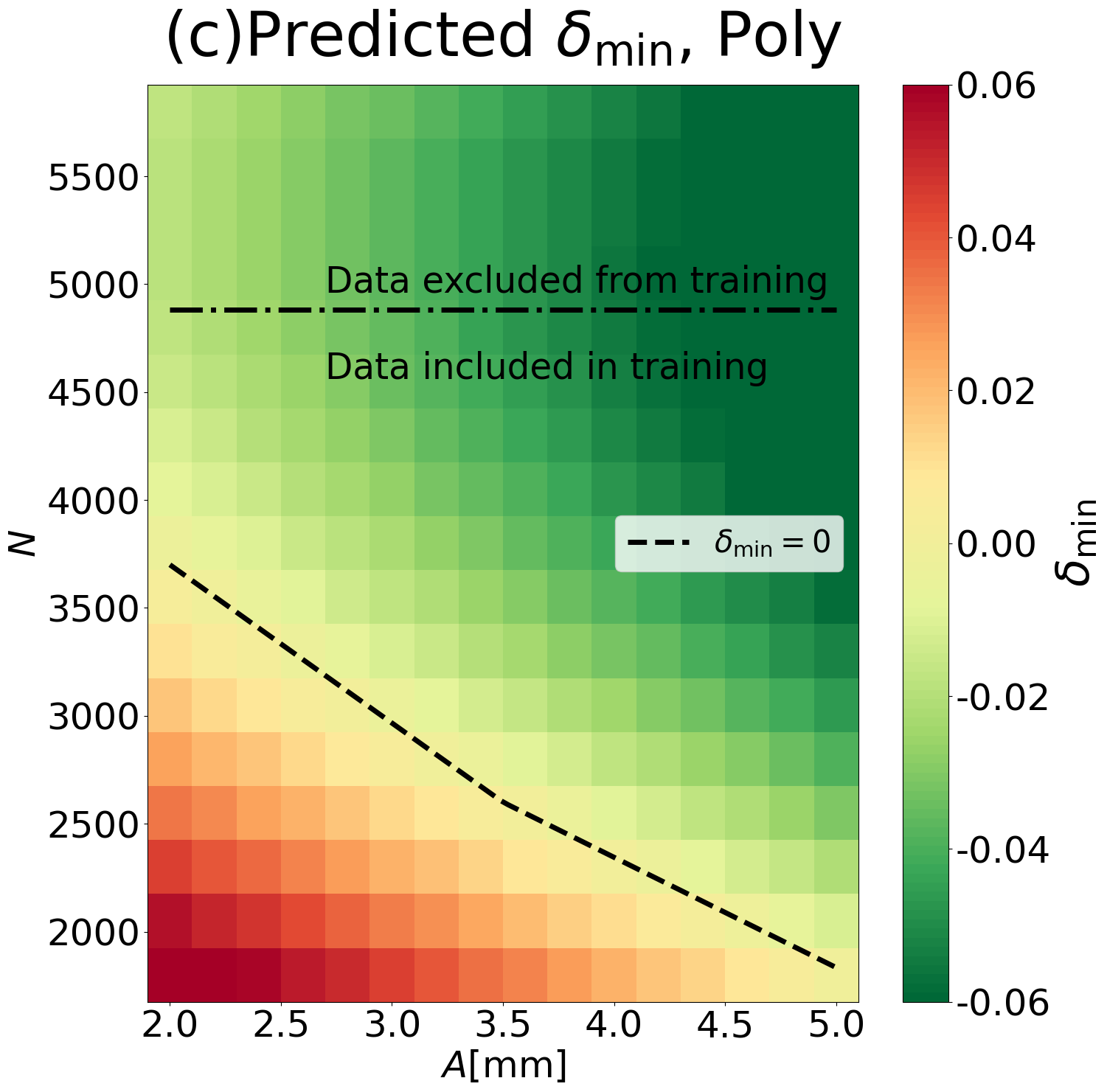} &
        \includegraphics[width=7.0cm]{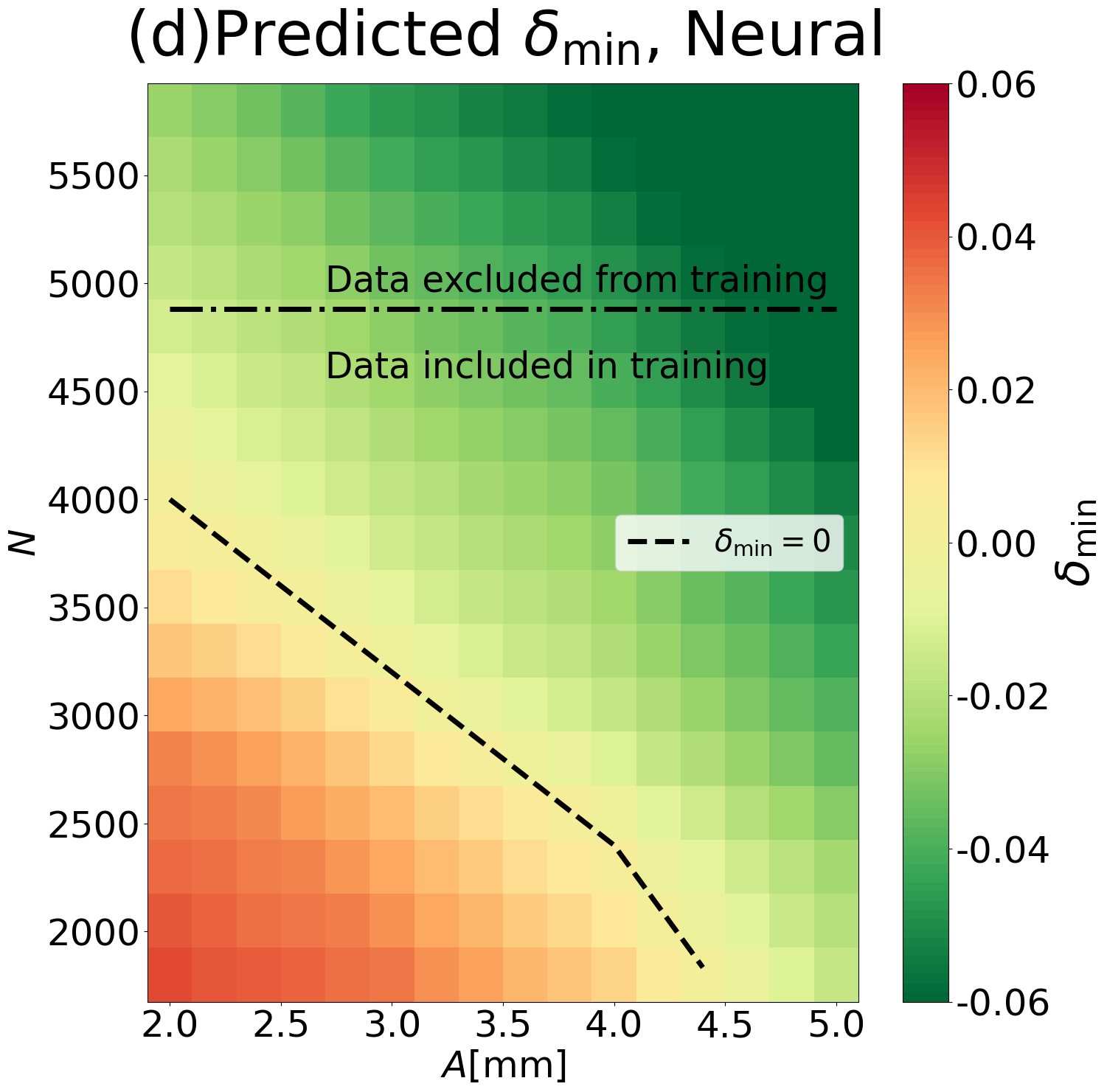}\\
    \end{tabular}
    \caption{Comparison of true $\delta_{\min}$ (a) and data by XGBoost regression (b), Polynomial regression(c), and Neural Network regression (d) in an extrapolation to larger $N$. The gas-cluster transition occurs around the ``zero" dashed line. Data above the dash-dot line were excluded from the training of the machine learning algorithms. The excitation phase-shift was fixed at $\theta=\pi$.}%
    \label{fig:contour_plots_true_predicted_extrapolation}%
\end{figure}

Figure \ref{fig:test_xgboost_variable_extrapolation} shows the comparison between the true and predicted $\delta_{\min}$ of the test dataset obtained with XGBoost regression (a) and Polynomial regression (b). In the extrapolation task, all models show a significantly poorer performance than the Neural network during interpolation (see $R^2$ values in Tabs. \ref{testdataset_var_interpolation} and \ref{testdataset_var_extrapolation}).

Figure \ref{fig:contour_plots_true_predicted_extrapolation} shows contour plots of the true and predicted $\delta_{\min}$ for the test dataset using XGBoost regression, Polynomial regression and Neural Network regression models, respectively. The true data are plotted in (a). 
The predictions obtained with the XGBoost (b) and Polynomial (c) regressions are close to the true transition data (a). 
Even though the predicted $\delta_{\min}$ from the Neural Network regression model (d) roughly resemble the true values, the predicted transition boundary differs from the true data. The Polynomial regression is the best model for $\delta_{\min}$ in the extrapolation to large $N$.

In the second case, the abilities to extrapolate to higher amplitudes $A$ that are not in the training domain are tested. The initial dataset contained between $A=2$ mm and $A=4$ mm is taken with a step size of 0.5 mm. Of this dataset, 90 \% were randomly assigned for training and the remaining 10 \% for validation. The trained model was then applied to test data with $A$ from 4.5 mm to 5 mm. The number of particles $N$ were between $N=1834$ ($\varphi=2.7 \%$) and 5643 ($\varphi=8.2 \%$) and $\theta/\pi$ was chosen between 0.0 and 1.0 in steps of 0.25. The initial dataset contains 275 samples, while the test dataset contains 110 samples. 

Table \ref{testdataset_var_extrapolation_second_case} shows the overall performance of the machine learning models in the extrapolation task for the second case. Similar to the first case, the XGBoost and Polynomial regressions have the smallest RMSE, MAE and the largest R$^2$ of all models and were hence chosen as the best candidates for the second case as well.

\begin{table}[ht]
\caption{Performance metrics on test dataset in the extrapolation task to  larger amplitudes $A$}
\label{testdataset_var_extrapolation_second_case}%
\begin{tabular}{@{}llll@{}}
\toprule
Model & RMSE & MAE & R$^2$\\
\midrule
\textbf{Polynomial Regression} & \textbf{0.011}  & \textbf{0.009} & \textbf{0.88} \\
Random Forest Regression & 0.03  & 0.027 & 0.18 \\ 
Support Vector Regression & 0.06  & 0.052 & -2.1 \\ 
XGBoost Regression & 0.024  & 0.022  &  0.48 \\ 
Neural Network Regression & 0.05 & 0.046  & -1.23 \\  
\botrule
\end{tabular}
\end{table}

\begin{figure}[htbp]%
    \centering
    \includegraphics[width=6cm]{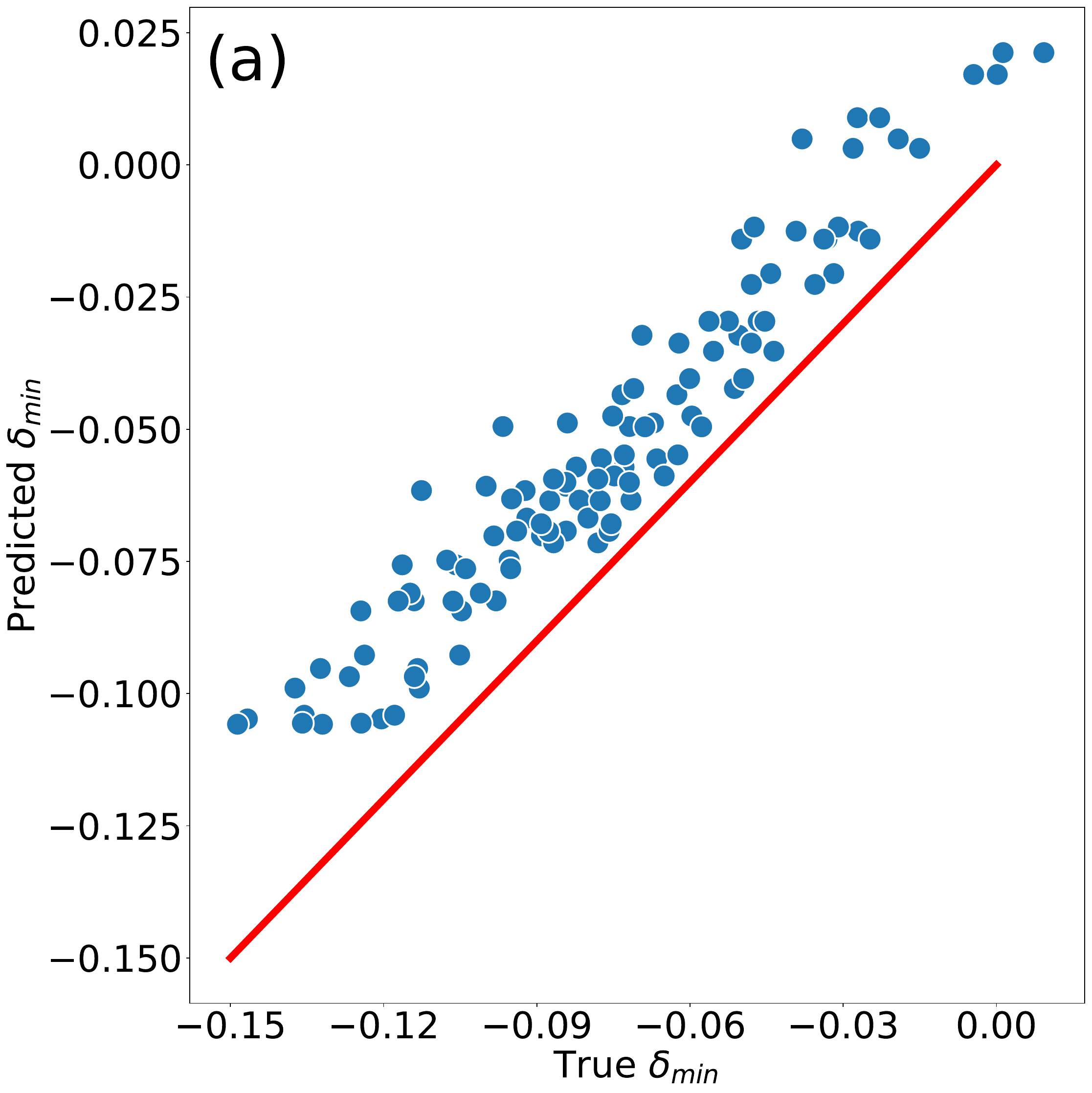}%
    \hspace{0.5cm}
    \includegraphics[width=6cm]{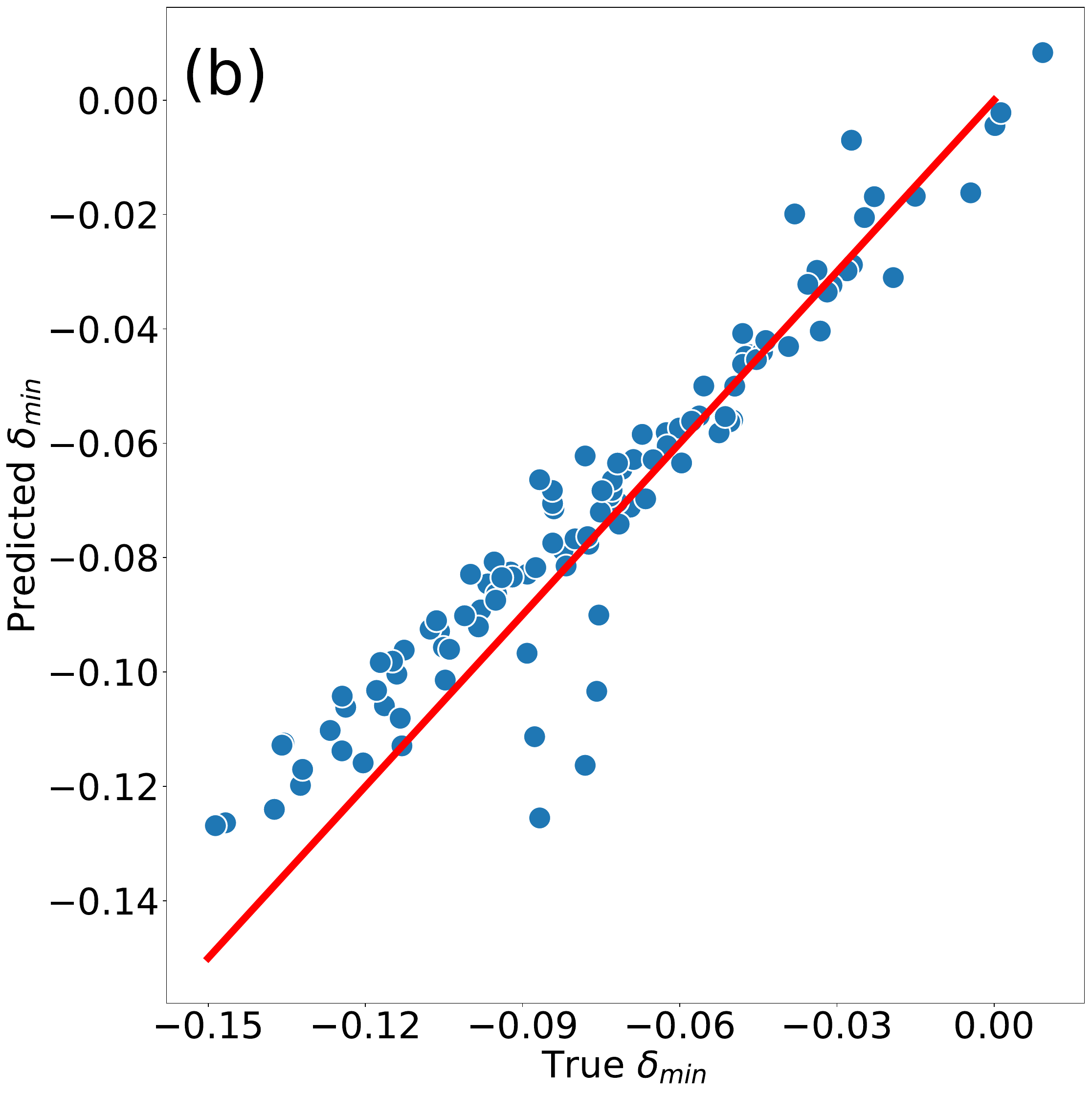}
    \caption{Comparison between true and predicted values of $\delta_{\min}$ for the test dataset obtained by XGBoost regression (a) and Polynomial regression (b) respectively for the extrapolation task to larger amplitudes $A$.}%
    \label{fig:test_xgboost_variable_extrapolation_amplitude}%
\end{figure}

Figure \ref{fig:test_xgboost_variable_extrapolation_amplitude} shows the comparison between the true and predicted $\delta_{\min}$ of the test dataset obtained with XGBoost regression (a) and Polynomial regression (b) in the second case. Similar to previous case, all the models show reduced performance than Neural network during interpolation. However, the models in the second case show enhanced performance in comparison to the extrapolation to larger $N$ (see $R^2$ values in Tabs. \ref{testdataset_var_extrapolation} and \ref{testdataset_var_extrapolation_second_case}).

\begin{figure}[htbp]%
    \centering
    \begin{tabular}{cc}
        \includegraphics[width=7.0cm]{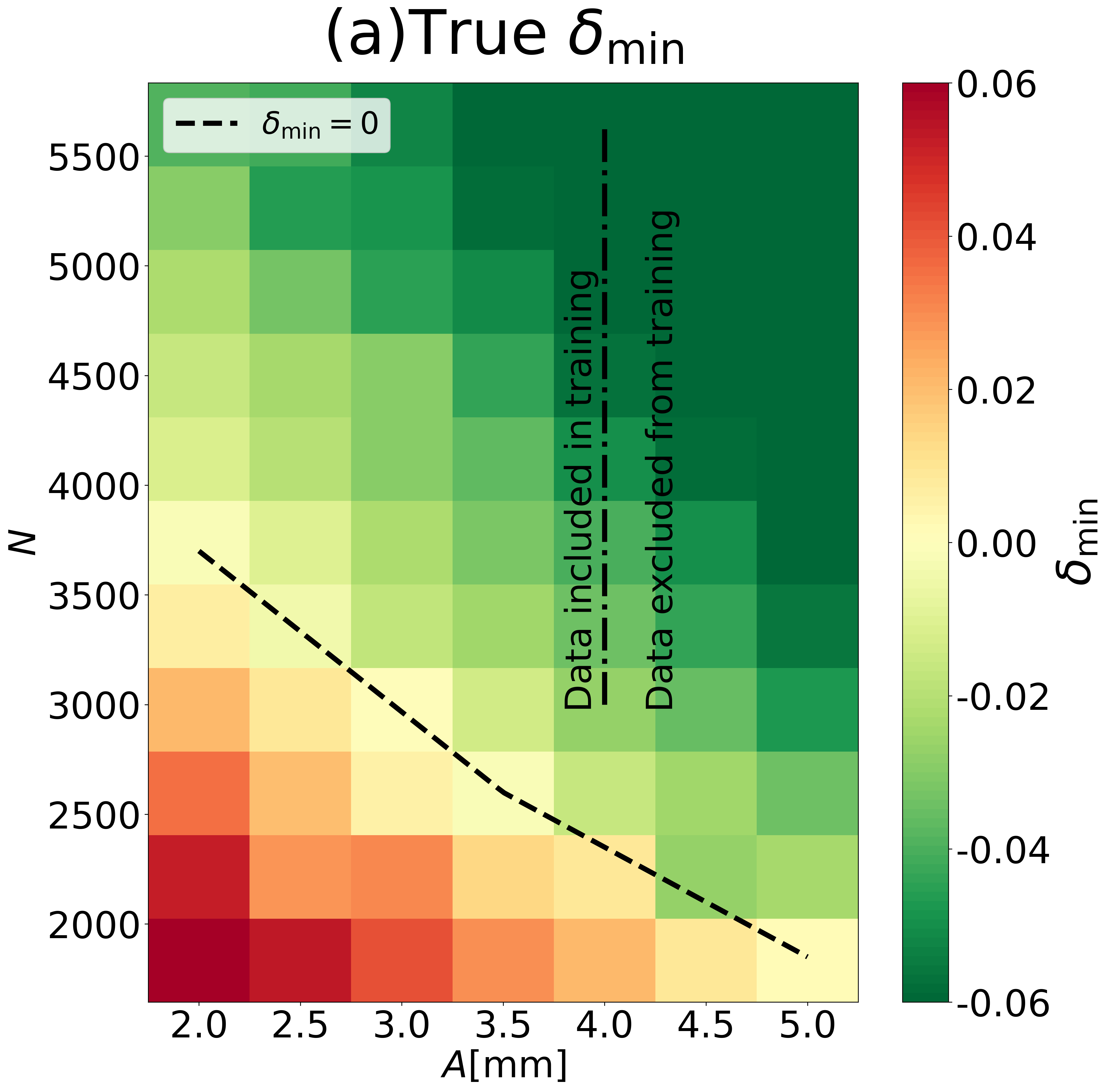} &
        \includegraphics[width=7.0cm]{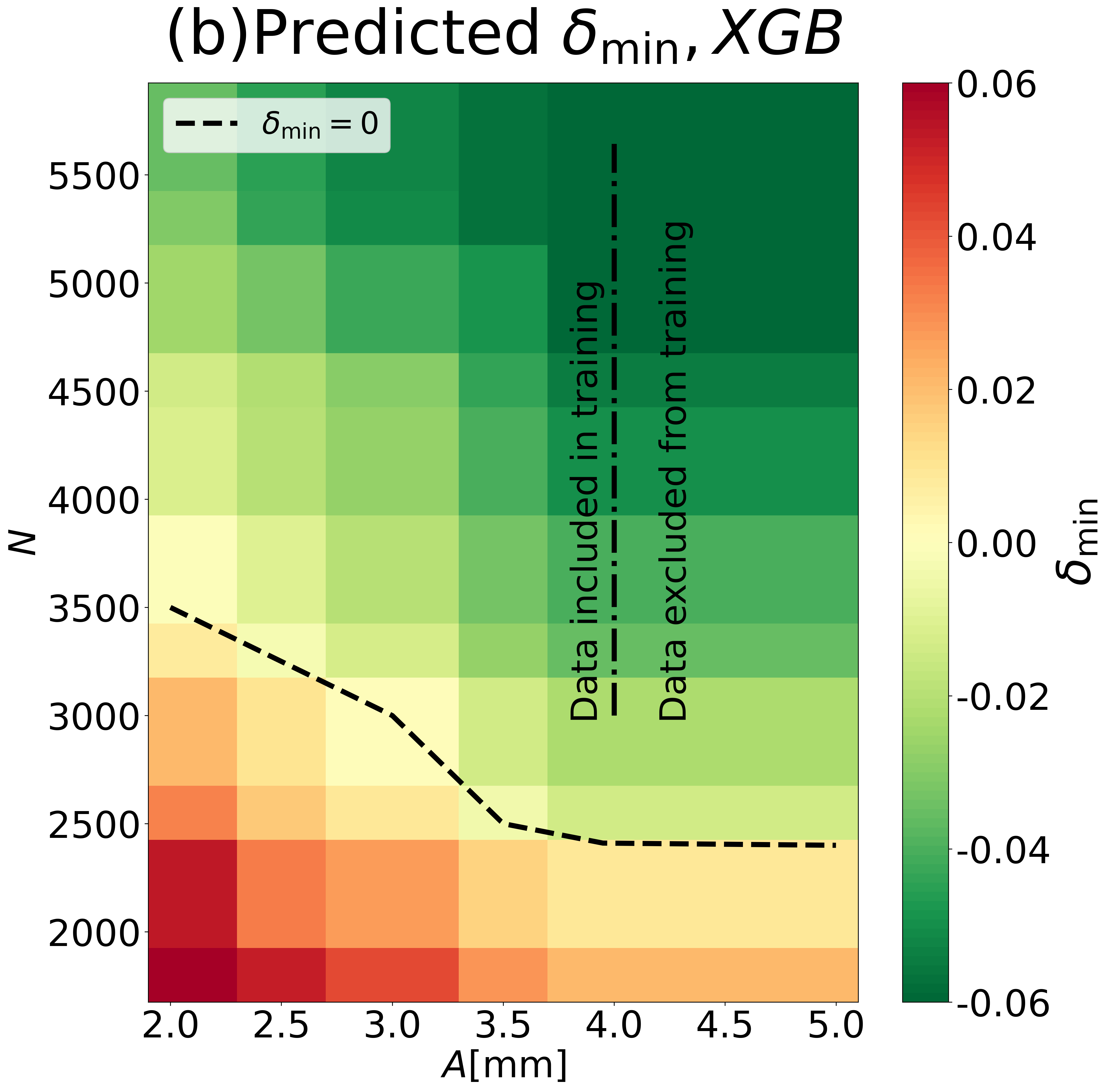} \\
        \includegraphics[width=7.0cm]{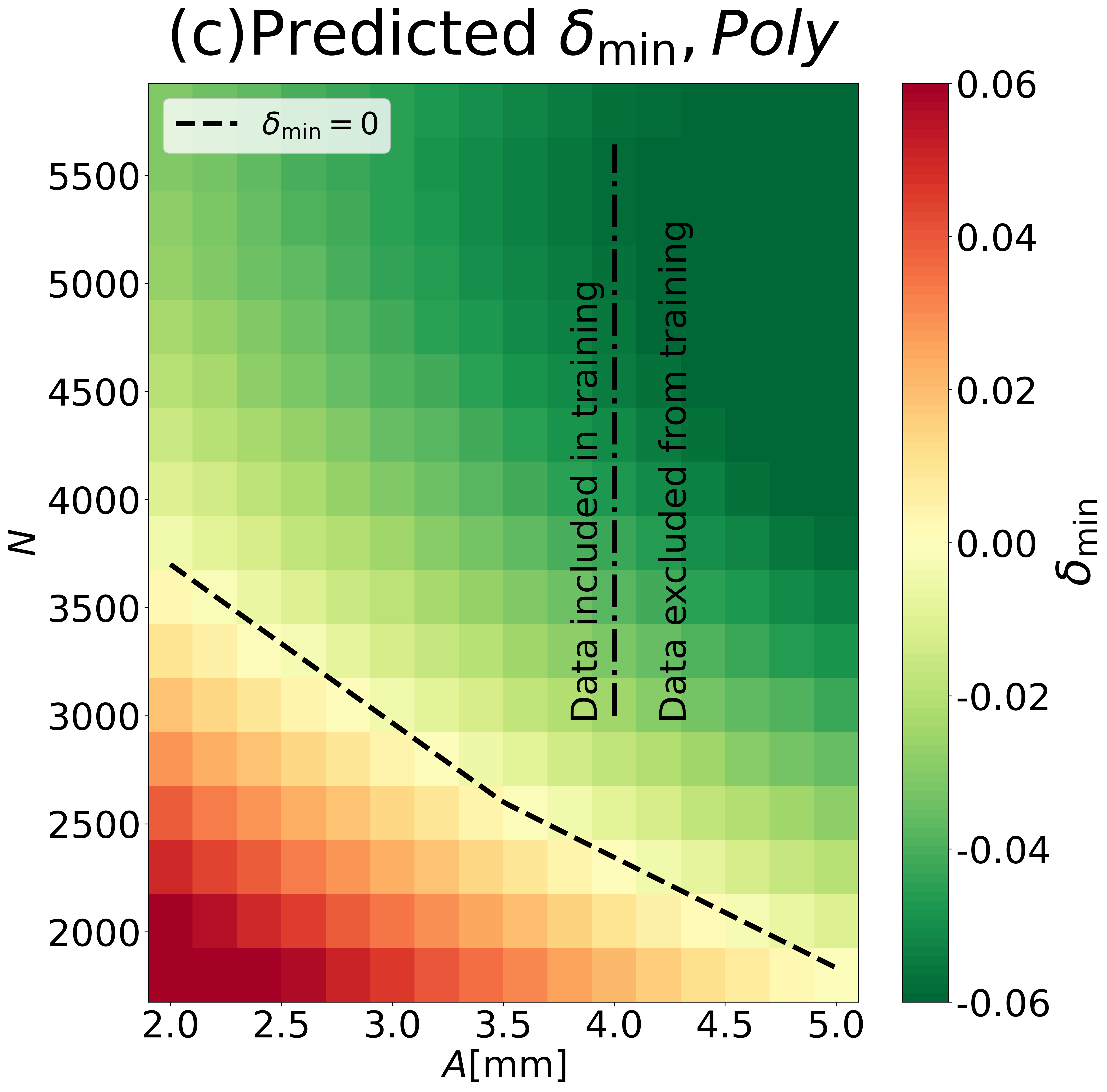} &
        \includegraphics[width=7.0cm]{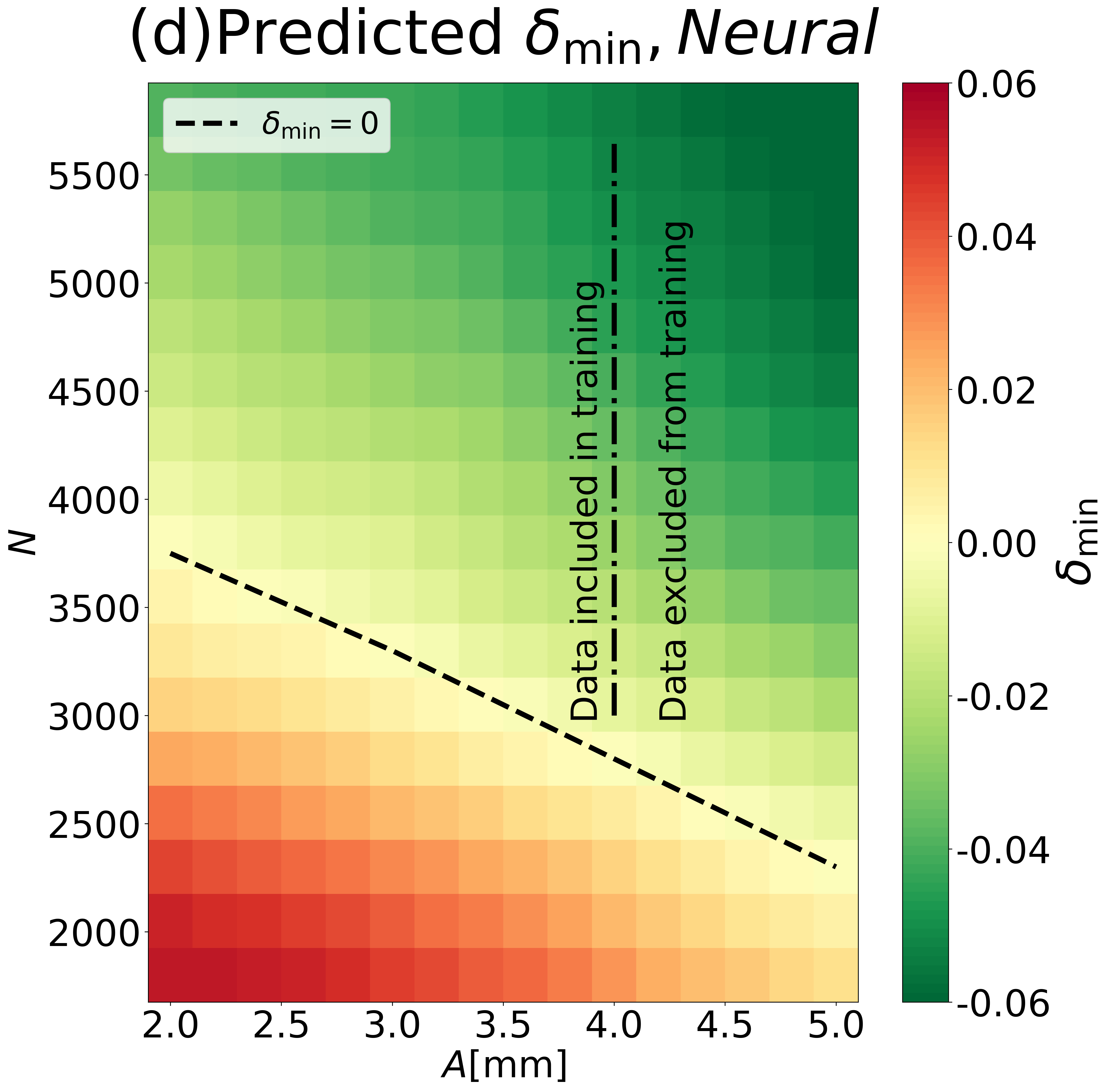}\\
    \end{tabular}
    \caption{Comparison of true $\delta_{\min}$ (a) and data by XGBoost regression (b), Polynomial regression(c), and Neural Network regression (d) in the extrapolation to larger amplitudes $A$. The gas-cluster transition occurs around the ``zero" dashed line. Data right to the dash-dot line were excluded from the training of the machine learning algorithms. The excitation phase-shift was fixed at $\theta=\pi$.}%
    \label{fig:contour_plots_true_predicted_extrapolation_amplitude}%
\end{figure}

Fig. \ref{fig:contour_plots_true_predicted_extrapolation_amplitude} shows the contour plots of the true and predicted $\delta_{\min}$ for the test dataset using XGBoost regression, Polynomial regression and Neural Network regression models, respectively, in the extrapolation to larger $A$. In this case, the predictions obtained from Polynomial (c) and Neural Network (d) regression models resemble the true transition data (a). The predictions from XGBoost regression are somewhat close to the ground truth, but systematically higher than the true values. Because of this, the predicted gas-cluster transition boundary significantly differs from that of the true data. Based on these observations, Polynomial regression is the best model for $\delta_{\min}$ in the extrapolation to larger amplitudes, as well as to larger particle numbers.

\subsubsection{Classification}

The original dataset from the regressions is used for the classification as well. $\delta_{\min}$ greater than or equal to 0 is given a binary label '0' which symbolizes the gaseous state and $\delta_{\min}$ that is less than 0 is given the binary label '1' symbolizing the cluster state. The dataset is split  into the similar proportions as in the regression tasks. The ML algorithms of Sec. \ref{mlalgorithms} are trained and evaluated using the classification metrics.
The Accuracy ($Ac$), Precision ($P$), Recall ($Rc$) and F1-Score ($F1$) of these algorithms are compared in Tab.\ref{testdataset_var_interpolation_classification}. 

\begin{figure}[ht]%
    \centering
    \includegraphics[width=8cm]{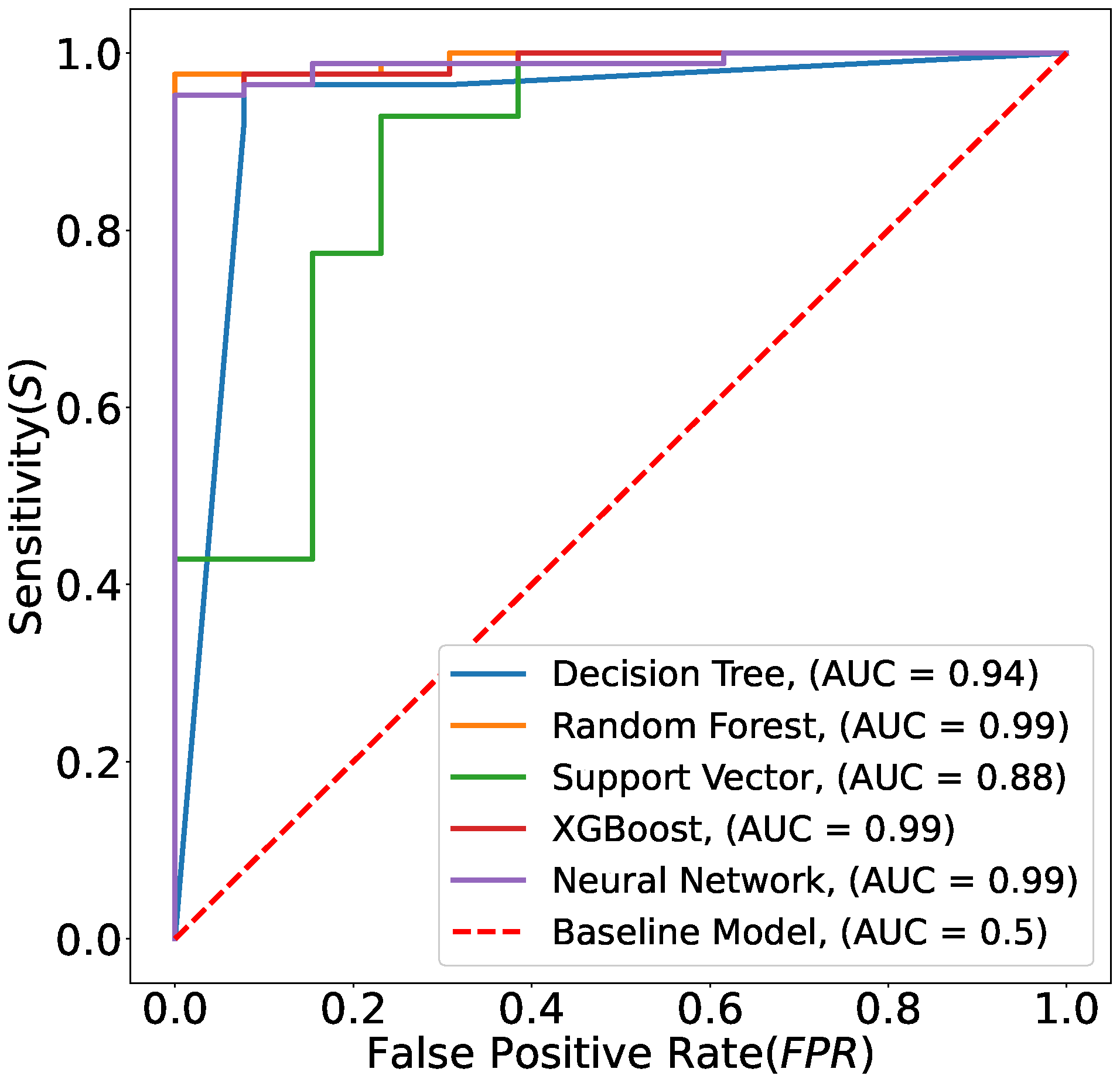}%
    \caption{Comparison of the ML algorithms using the KS test in the interpolation task, based on the area under the ROC curve (AUC). $AUC=0.5$ indicates the baseline model and $AUC=1.0$ defines a perfect model.}%
    
    \label{fig:roc_auc_curve_interpolation}%
\end{figure}

Fig. \ref{fig:roc_auc_curve_interpolation} shows the ROC curves and the area (AUC) under these ROC curves that are estimated for these algorithms and compared with a baseline model shown in orange dashed line. The baseline model is the model that gives the minimum performance, while the perfect model is the one that gives the maximum performance on the classification metrics. From these curves, we derive that the ensemble tree-based algorithms corresponding to the Random Forest Classifier and XGBoost classifiers show the best performances compared to the other models, with $AUC=0.99$. To provide more insight into the performance of these algorithms, Tab. \ref{confusion_matrix_interpolation_ks_test} shows the confusion matrix on the test dataset that was extracted in interpolation task. The Random Forest Classifier results in a minimum number of false predictions ($FN$ and $FP$) compared to XGBoost and Neural Network Classifiers, and is therefore considered the best model to predict the state of the system by interpolations.

 \begin{table}[ht]
\caption{Performance metrics to predict the state by interpolation, using the KS-test criterion}
\label{testdataset_var_interpolation_classification}%
\begin{tabular}{@{}lllll@{}}
\toprule
Model & Accuracy & Precision & Recall & F1-Score\\
\midrule
Decision Tree Classifier & 0.958 & 0.987  & 0.964 & 0.975 \\
Random Forest Classifier & 0.97  & 0.987  & 0.976 & 0.982 \\ 
Support Vector Classifier & 0.917  & 0.941 & 0.964 & 0.952 \\ 
XGBoost Classifier & 0.948  & 0.987  & 0.952 & 0.97  \\ 
Neural Network Classifier & 0.958  & 0.976  & 0.976 & 0.976 \\
\\  
\botrule
\end{tabular}
\end{table}

 \begin{table}[ht]
\caption{Confusion matrix for   interpolation with the KS test criterion} 
\label{confusion_matrix_interpolation_ks_test}%
\begin{tabular}{@{}lllll@{}}
\toprule
Model & $TP$ & $FP$ & $TN$ & $FN$\\
\midrule
Decision Tree Classifier & 81 & 1 & 12 & 3 \\
Random Forest Classifier & 82 & 1 & 12 & 2 \\ 
Support Vector Classifier & 81 & 5 & 8 & 3 \\ 
XGBoost Classifier & 80 & 1 & 12 & 4 \\
Neural Network Classifier & 82 & 2 & 11 & 2 \\
\\  
\botrule
\end{tabular}
\end{table}

\begin{figure}[ht]%
    \centering
    \includegraphics[width=8cm]{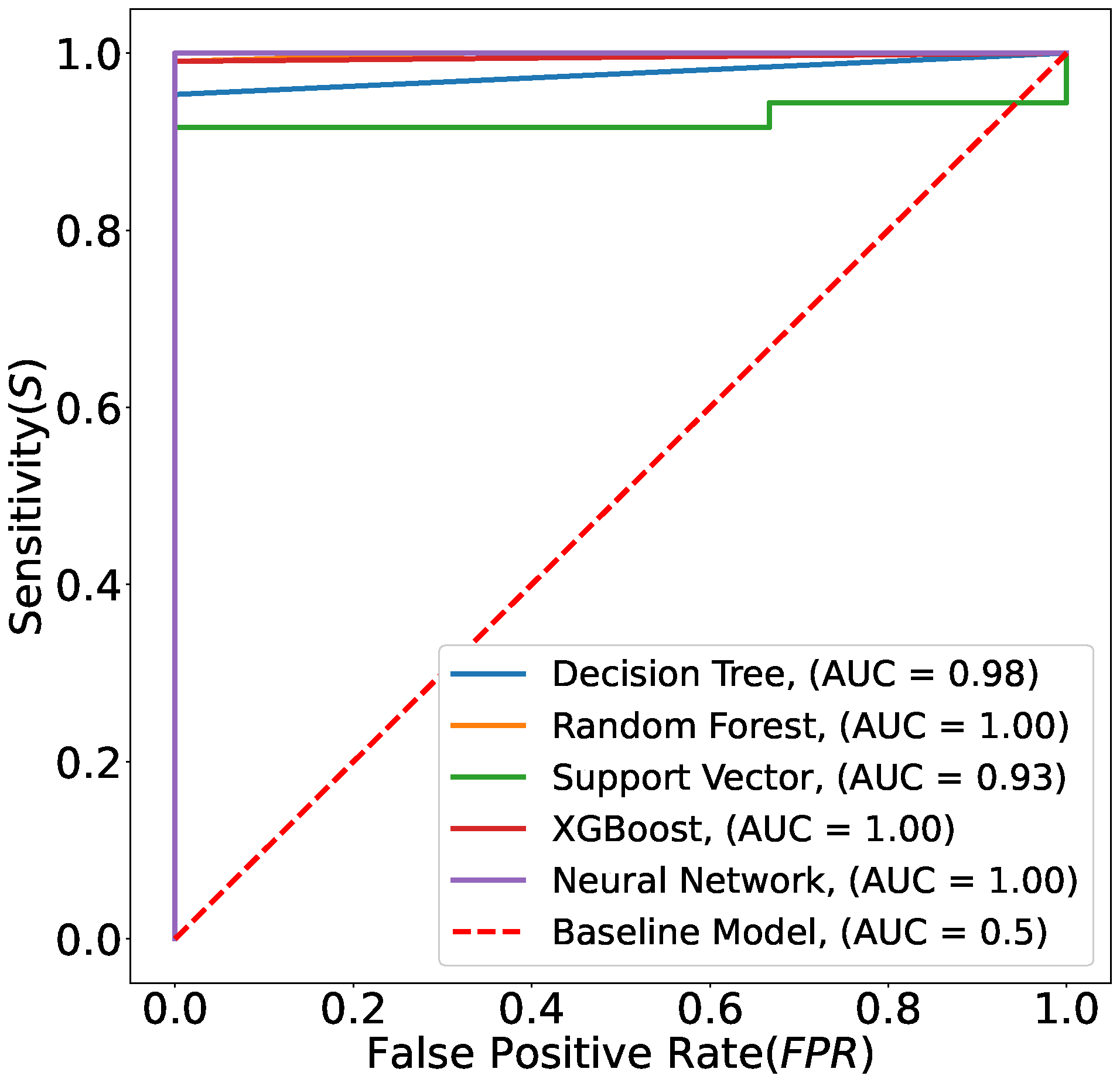}%
    \caption{Comparison of the ML algorithms using the KS test in the extrapolation task, based on the area under the ROC curve (AUC). $AUC=0.5$ indicates the baseline model and $AUC=1.0$ defines a perfect model.}%
    \label{fig:roc_auc_curve_extrapolation}%
\end{figure}

Similar to regression, the performance of the ML algorithms is investigated to predict the state of the system on the test dataset in extrapolation task. Sample data points with amplitudes in between 2 mm and 4 mm were used to train these algorithms, while the ones with 4.5 mm and 5 mm were used in the test dataset. We chose this extrapolation range for the classification tests in particular, because the transition boundary passes this region.
The performance of the tested algorithms is tabulated in  Tab.~\ref{testdataset_var_extrapolation_classification}. The confusion matrix for these algorithms is tabulated in Tab.~\ref{confusion_matrix_extrapolation_ks_test}. Figure \ref{fig:roc_auc_curve_extrapolation} shows the ROC curves and compares the area under these curves for these algorithms. Neural network results in significantly less false predictions ($FP$ and $FN$) compared to other algorithms and is therefore considered the best model to predict the state of the system by extrapolations. 

 \begin{table}[ht]
\caption{Performance metrics to predict the state from the test dataset in the extrapolation task, using the KS test criteria}
\label{testdataset_var_extrapolation_classification}%
\begin{tabular}{@{}lllll@{}}
\toprule
Model & Accuracy & Precision & Recall & F1-Score\\
\midrule
Decision Tree Classifier & 0.92 & 1.0  & 0.91 & 0.95 \\
Random Forest Classifier & 0.93  & 1.0  & 0.93 & 0.96 \\ 
Support Vector Classifier & 0.82  & 1.0 & 0.82 & 0.90 \\ 
XGBoost Classifier & 0.91  & 1.0  & 0.91 & 0.95  \\ 
Neural Network Classifier & 0.99  & 0.99  & 1.0 & 0.99 \\
\botrule
\end{tabular}
\end{table}

 \begin{table}[ht]
\caption{Confusion matrix for the extrapolation task with the KS criterion}
\label{confusion_matrix_extrapolation_ks_test}%
\begin{tabular}{@{}lllll@{}}
\toprule
Model & $TP$ & $FP$ & $TN$ & $FN$\\
\midrule
Decision Tree Classifier & 98 & 0 & 3 & 9 \\
Random Forest Classifier & 100 & 0 & 3 & 7 \\ 
Support Vector Classifier & 88 & 0 & 3 & 19 \\ 
XGBoost Classifier & 98 & 0 & 3 & 9 \\
Neural Network Classifier & 107 & 0 & 2 & 1 \\
\botrule
\end{tabular}
\end{table}

Summarizing the performance on the classification metrics, we find that the Random Forest Classifier and Neural Network Classifier are the most suitable ML algorithms to predict the state of the system on the basis of $\delta_{\min}$.

\subsection{Predictions of ML algorithms using the caging criterion}\label{prediction_caging}

Next, we evaluate the performance of the ML models to predict $\varphi_{\rm caged}$ as the caging criterion. We explore both regression and classification to predict the continuous variable $\varphi_{\rm caged}$ and the corresponding state of the system, respectively. The results for both the interpolation and extrapolation tasks are provided.

\subsubsection{Regression}

As before, $N$, $A$, $\theta$ are the independent variables. In case of the caging criterion, the maximum fraction of caged particles $\varphi_{caged}$ is the dependent variable chosen for the regression tasks.\\ 

\textbf{\textit{Interpolation:}}

The dataset consists of $N$ between 1834 ($\Phi = 2.7\%$) and 5643 ($\Phi=8.2\%$) particles and excitation amplitudes $A$ in between 2 mm and 5 mm, $\theta/\pi$ was chosen between 0.0 and 1.0. This provided a set of 693 data points. The initial dataset was split into training, validation and test datasets. 25 \% of the samples were randomly selected and assigned as test data. From the remaining data, 65\% were assigned randomly to the training set and 10\% to the validation set.  

\begin{table}[ht]
\caption{Performance metrics on test dataset in interpolation method of sampling}
\label{testdataset_var_interpolation_voronoi}%
\begin{tabular}{@{}llll@{}}
\toprule
Model & RMSE & MAE & R$^2$\\
\midrule
Polynomial Regression & 0.019  & 0.012  & 0.961  \\
Random Forest Regression & 0.02  & 0.012  & 0.956 \\ 
Support Vector Regression & 0.06  & 0.054 & 0.64 \\ 
\textbf{XGBoost Regression} & \textbf{0.019}  & \textbf{0.01}  & \textbf{0.964}  \\ 
Neural Network Regression & 0.023  & 0.015  & 0.947  \\ 
\botrule
\end{tabular}
\end{table}

\begin{figure}[htbp]%
    \centering
    \includegraphics[width=6cm]{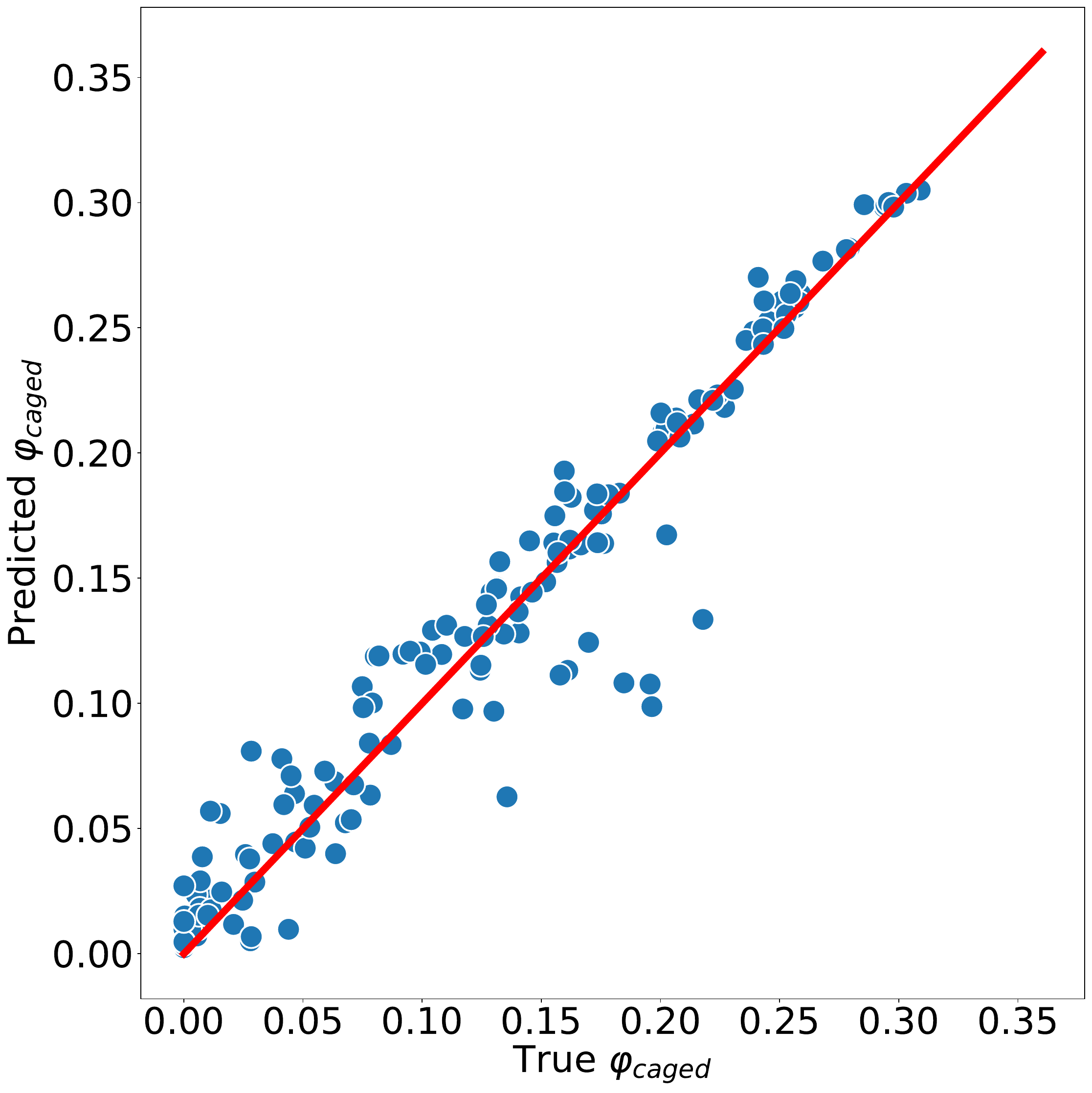}%
    \caption{Comparison of true and predicted $\varphi_{\rm caged}$ using the XGBoost regression model in the interpolation task.}%
    \label{fig:test_voronoi_interpolation}%
\end{figure}

\begin{figure}[htbp]%
    \centering
    \includegraphics[width=6cm]{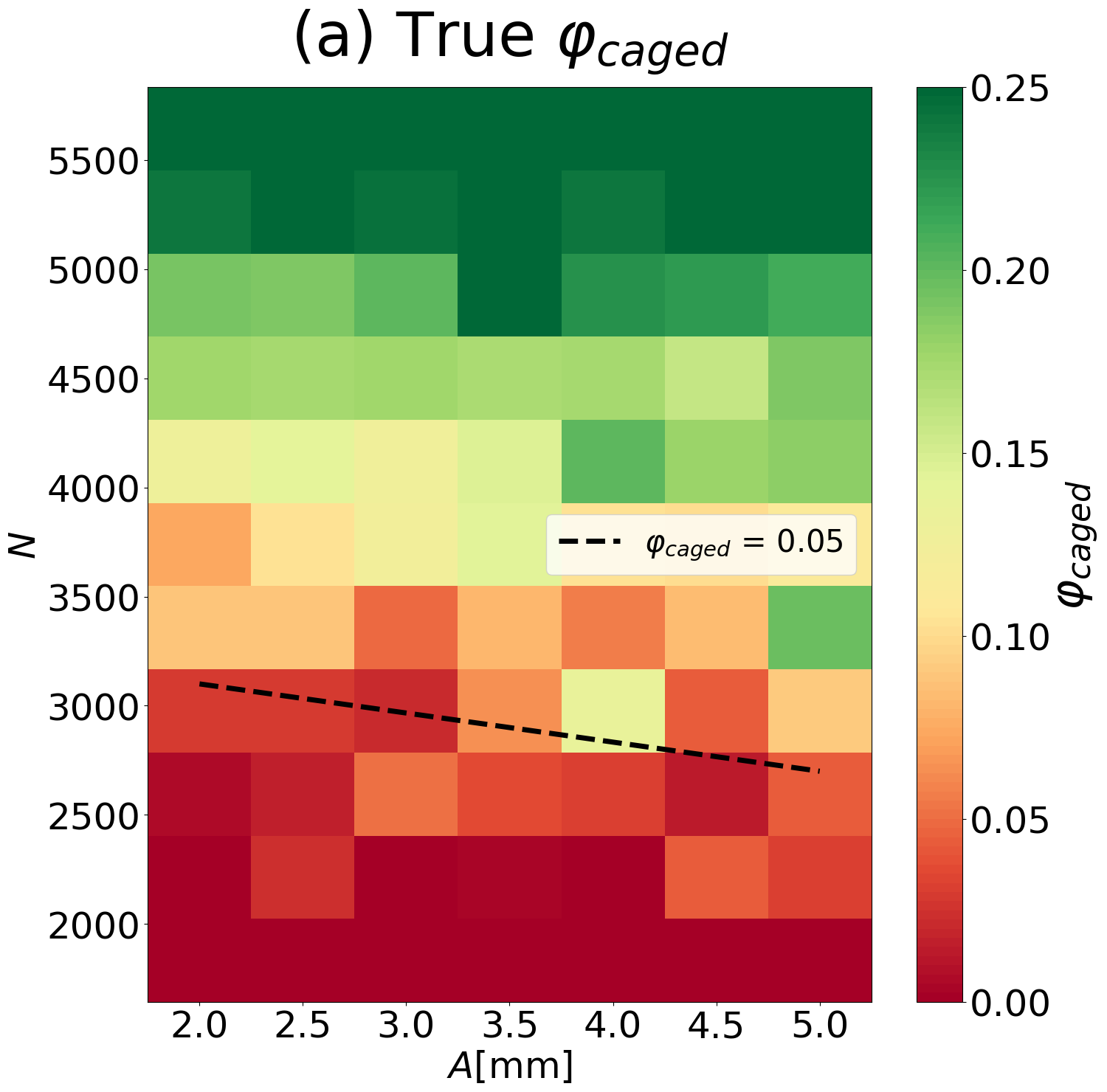} %
    \hspace{0.5cm}
    \includegraphics[width=6cm]{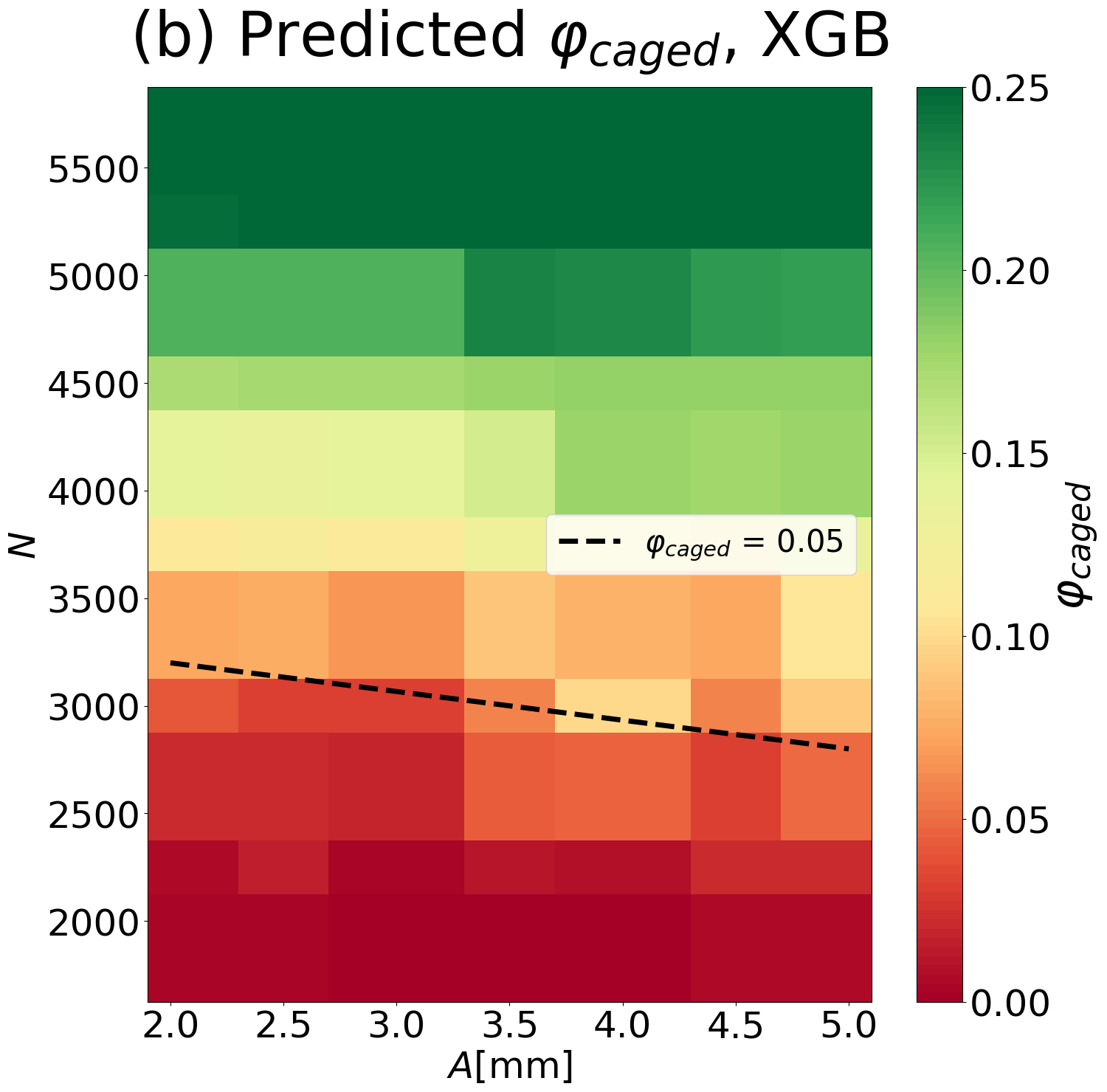}%
    \caption{Comparison of true (a) and predicted (b) $\varphi_{\rm caged}$ with the XGBoost regression model by interpolation. Parameters $N, A$ are varied, $\theta=\pi$ is fixed.}%
    \label{fig:contour_plots_phase_shift_1_interpolation_voronoi_1}%
\end{figure}

The regression performance metrics for our ML models are listed in Tab. \ref{testdataset_var_interpolation_voronoi}. The XGBoost and Polynomial regressions have comparatively higher $R^2$ and lower RMSE and MAE compared to the other models. XGBoost was preferred over Polynomial regression due to its advantages with respect to overfitting, effective handling of complex relationships in the dataset, and others. The comparison of predicted and true $\varphi_{\rm caged}$ for the test datasets is shown in Fig. \ref{fig:test_voronoi_interpolation}. Most of the values predicted by the XGBoost regression are close to the ``true'' data. 
In detail, the performance of the XGBoost regression is seen in Figs. \ref{fig:contour_plots_phase_shift_1_interpolation_voronoi_1}(a) and (b). Figure \ref{fig:contour_plots_phase_shift_1_interpolation_voronoi_1}(a) shows the true $\varphi_{\rm caged}$ for the simulation data. 
The phase-shift was fixed at $\theta=\pi$. 
Fig. \ref{fig:contour_plots_phase_shift_1_interpolation_voronoi_1}(b) shows that the predictions of the XGBoost regression are accurate in the full excitation parameter range by interpolation.

\textbf{\textit{Extrapolation:}}

Similar to the KS-test predictions, two cases were investigated in the extrapolation task for the caging criterion. In the first case, we explore the model's ability to extrapolate to $N$ that were not part of the training data, while in the second case we evaluate its ability to extrapolate to $A$ that are outside of the training domain.

In the first case, the initial dataset is now chosen with $N$ ranging from 1834 ($\varphi=2.7 \%$) to 4881 ($\varphi=7.1 \%$) particles. From this set, 90 \% are randomly selected for training, the remaining 10 \% for validation. After training and validation, the final model is applied to the test data with $N$ ranging from 5262 ($\varphi=7.65 \%$) to 5643 ($\varphi=8.2\%$) particles, $A$ was chosen between 2 mm and 5 mm and $\theta/\pi$ in between 0 and 1. The initial dataset contains 567 samples, and the test set 126 samples.

\begin{figure}[htbp]%
    \centering
    \includegraphics[width=6cm]{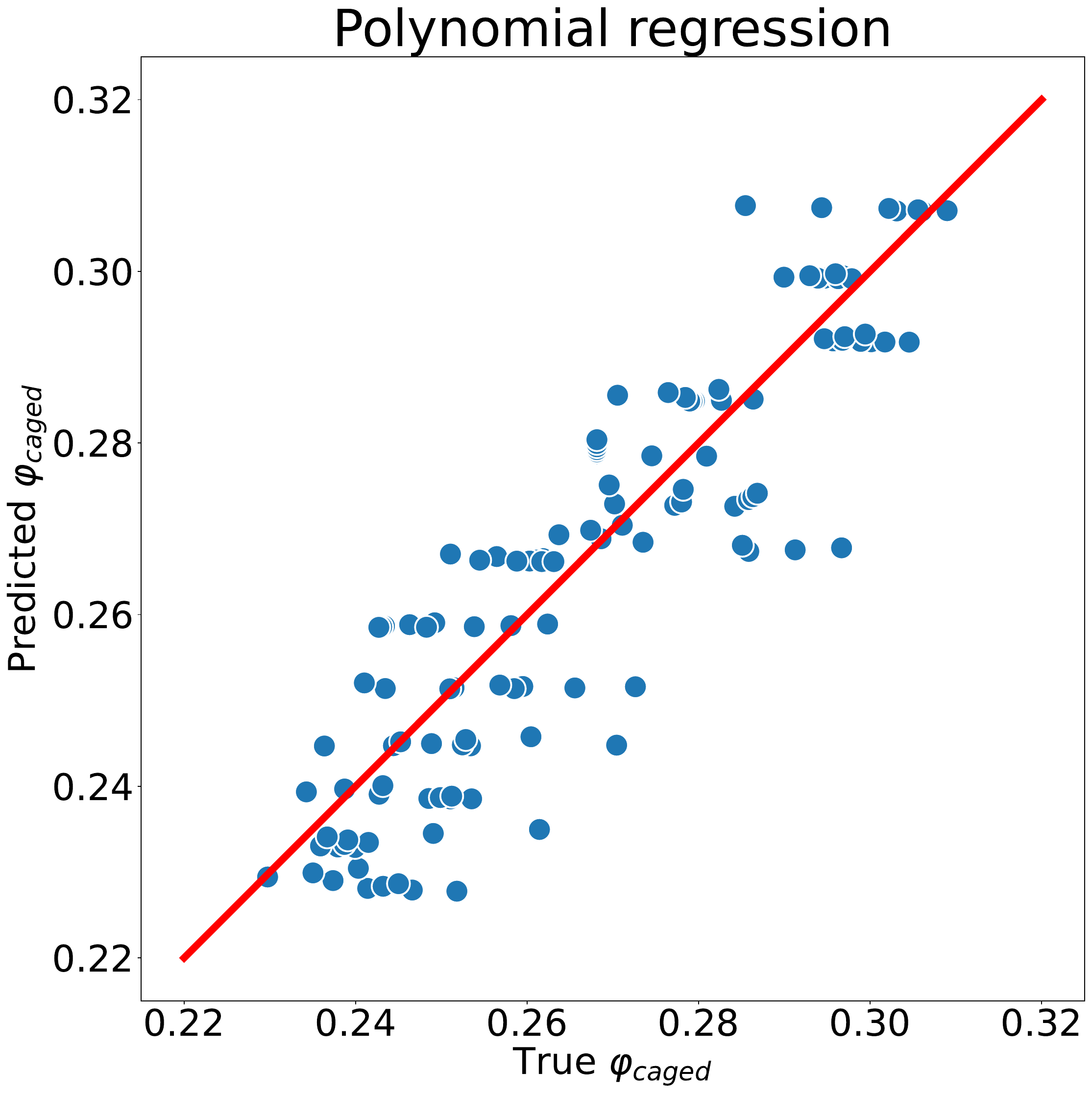}%
    \caption{Comparison of true and predicted $\varphi_{\rm caged}$ using  Polynomial regression in the extrapolation task to larger $N$}.
\label{fig:test_XGB_PR_max_lpf_prediction_extrapolation}%
\end{figure}

The performance metrics are tabulated in Tab. \ref{testdataset_var_extrapolation_voronoi}. Polynomial regression with degree {2} shows the best performance for the given data and is therefore selected. Figure \ref{fig:test_XGB_PR_max_lpf_prediction_extrapolation} 
shows the relation between true and predicted $\varphi_{\rm caged}$ for the Polynomial regression model. It satisfactorily extrapolates $\varphi_{\rm caged}$ to the larger particle numbers.

\begin{table}[htbp]
\caption{Performance metrics on test dataset in the extrapolation to larger $N$. Negative $R^2$ indicate that the models fail to generalize beyond the training data.}
\label{testdataset_var_extrapolation_voronoi}%
\begin{tabular}{@{}llll@{}}
\toprule
Model & RMSE & MAE & R$^2$\\
\midrule
\textbf{Polynomial Regression} & \textbf{0.01}  & \textbf{0.008}  & \textbf{0.77}  \\
Random Forest Regression & 0.058  & 0.056  & -6.77 \\ 
Support Vector Regression & 0.12  & 0.12 & -33.1 \\ 
XGBoost Regression & 0.057  & 0.053  & -6.32  \\ 
Neural Network Regression & 0.18  & 0.16  & -73  \\
\botrule
\end{tabular}
\end{table}

\begin{figure}[htbp]%
    \centering
    \begin{tabular}{cc}
        \includegraphics[width=7.0cm]{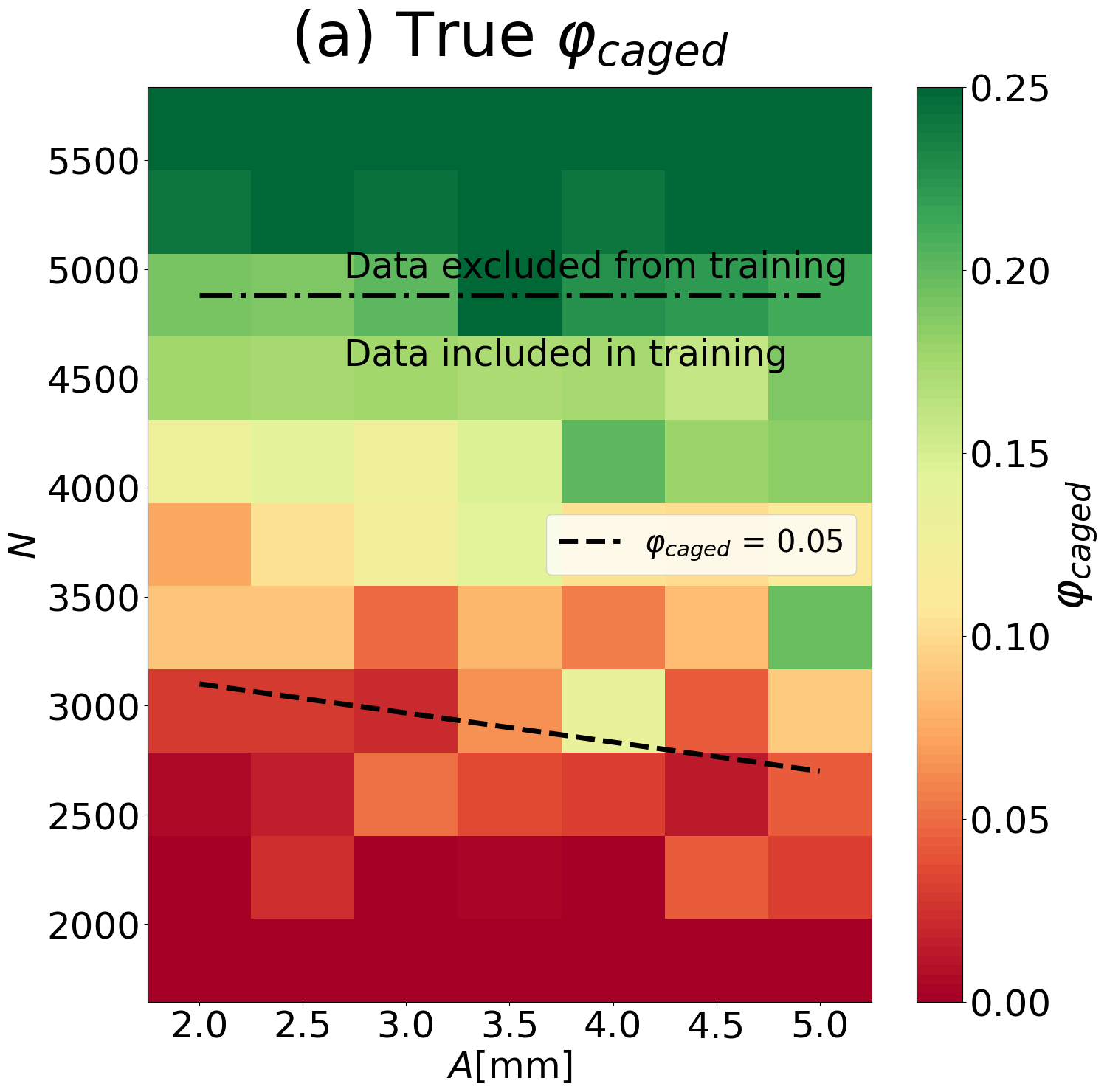} &
        \includegraphics[width=7.0cm]{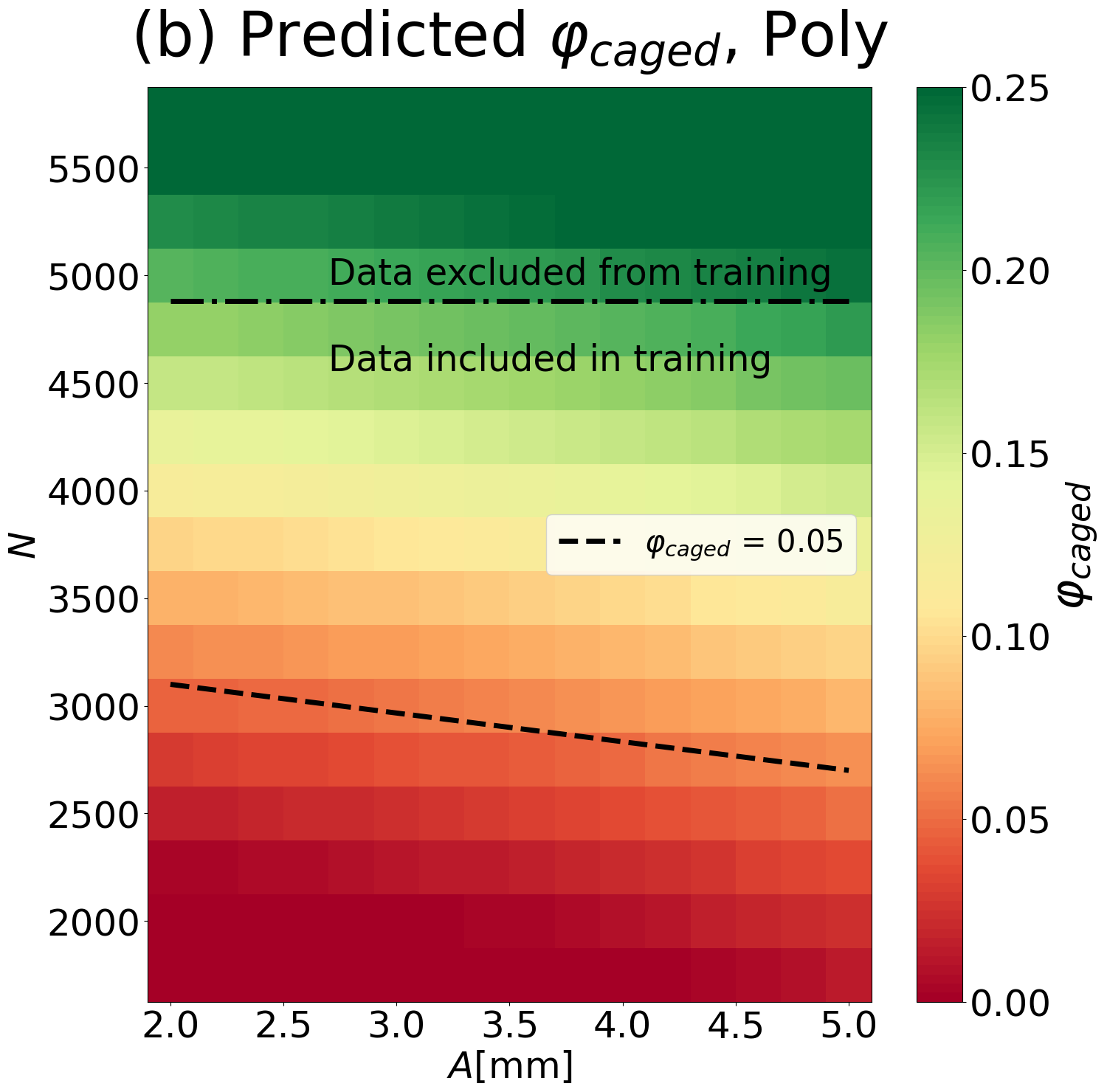} \\
        \includegraphics[width=7.0cm]{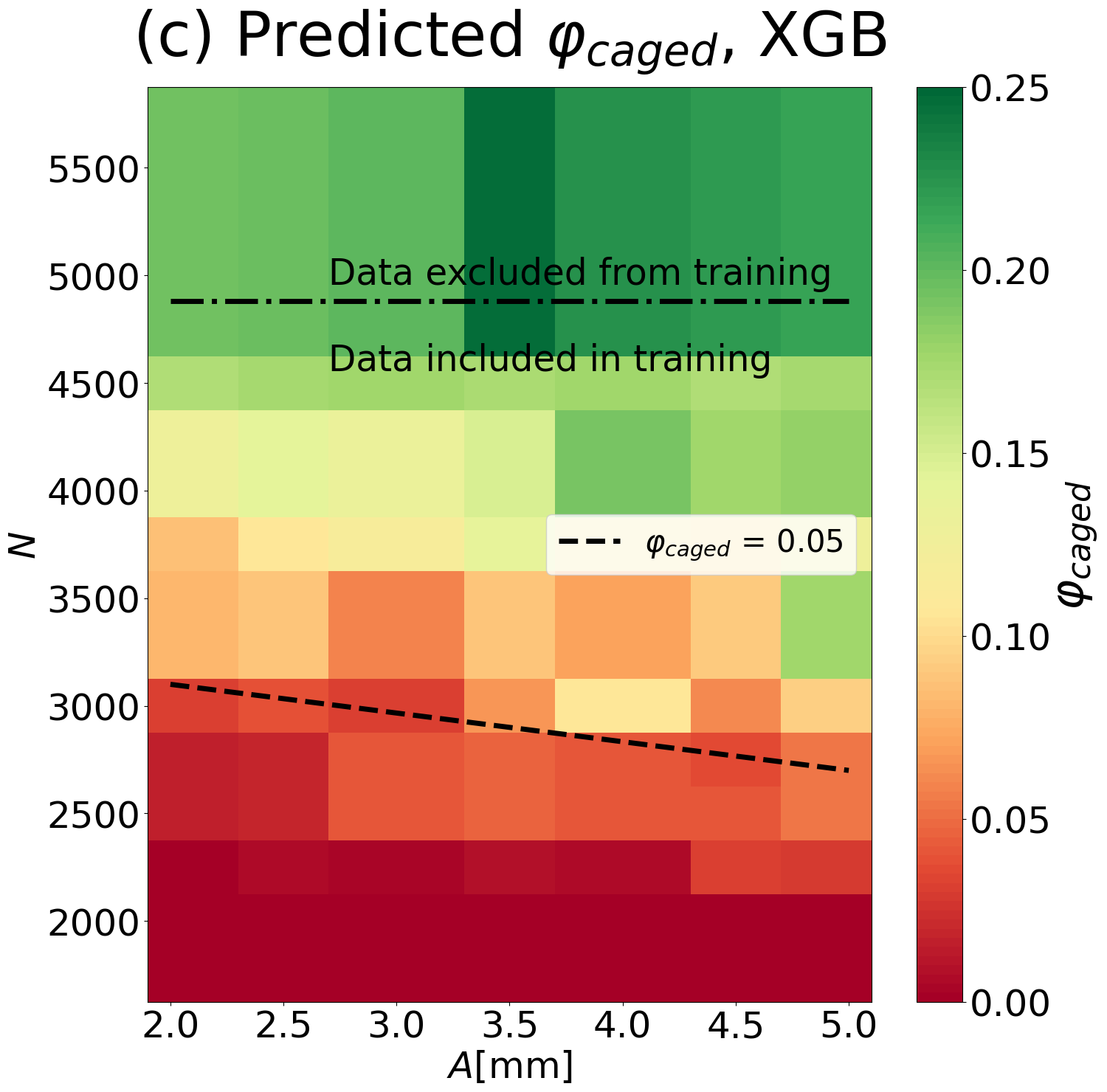} &
        \includegraphics[width=7.0cm]{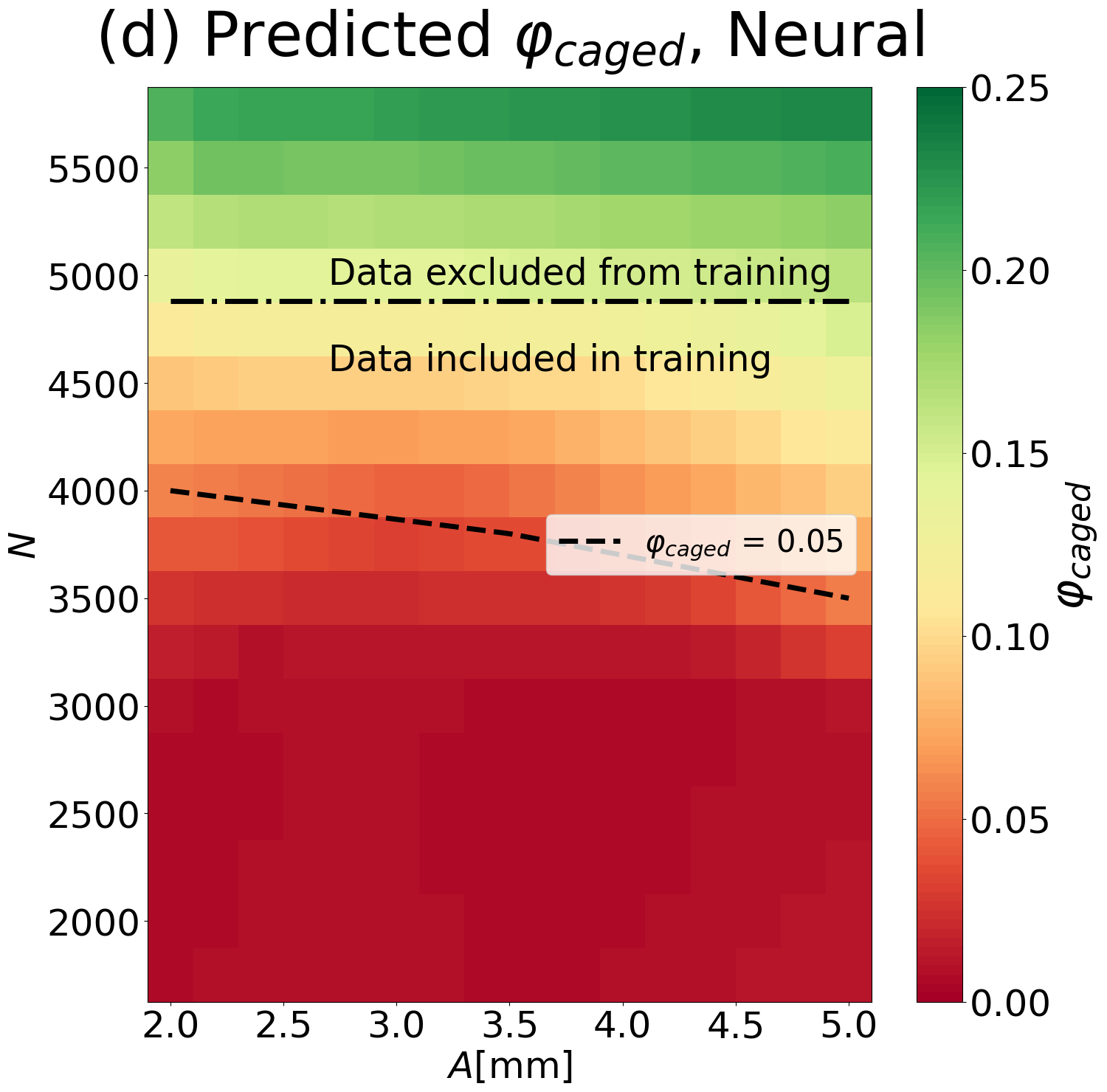}\\
    \end{tabular}
    \caption{True (a) and predicted $\varphi_{caged}$ by Polynomial regression (b), XGBoost regression (c) and Neural Network regression (d) models when $N$ and $A$ are varied in the extrapolation to larger $N$. The gas-cluster transition occurs at the dashed line. The region above the black dash-dot was not used for training. $\theta=\pi$.}
    \label{fig:contour_plots_true_predicted_extrapolation_phi_caged}%
\end{figure}

Figures~ \ref{fig:contour_plots_true_predicted_extrapolation_phi_caged}(a-d) show the true and predicted $\varphi_{\rm caged}$ from Polynomial regression with degree\sout{3} 2, from XGBoost regression and from Neural Network regression models, respectively. The predictions from the Polynomial regression are close to the true data and hence the predicted gas-cluster transition boundary closely resembles the corresponding ``true" curve. The predictions from XGBoost regression and Neural Network regression are not reliable in the extrapolated region.

In the second case, the initial dataset is now chosen with $A$ ranging from 2 mm to 4 mm. From this set, 90 \% are randomly selected for training while remaining 10 \% are chosen for validation. The final model is applied to the test data with $A$ from 4.5 mm to 5 mm. The number of particles $N$ were between $N=1834$ ($\varphi=2.7 \%$) and 5643 ($\varphi=8.2 \%$) and $\theta/\pi$ was chosen between 0.0 and 1.0 in steps of 0.25. The initial dataset contains 495 samples while the test dataset contains 198 samples. 

\begin{figure}[htbp]%
    \centering
    \includegraphics[width=6cm]{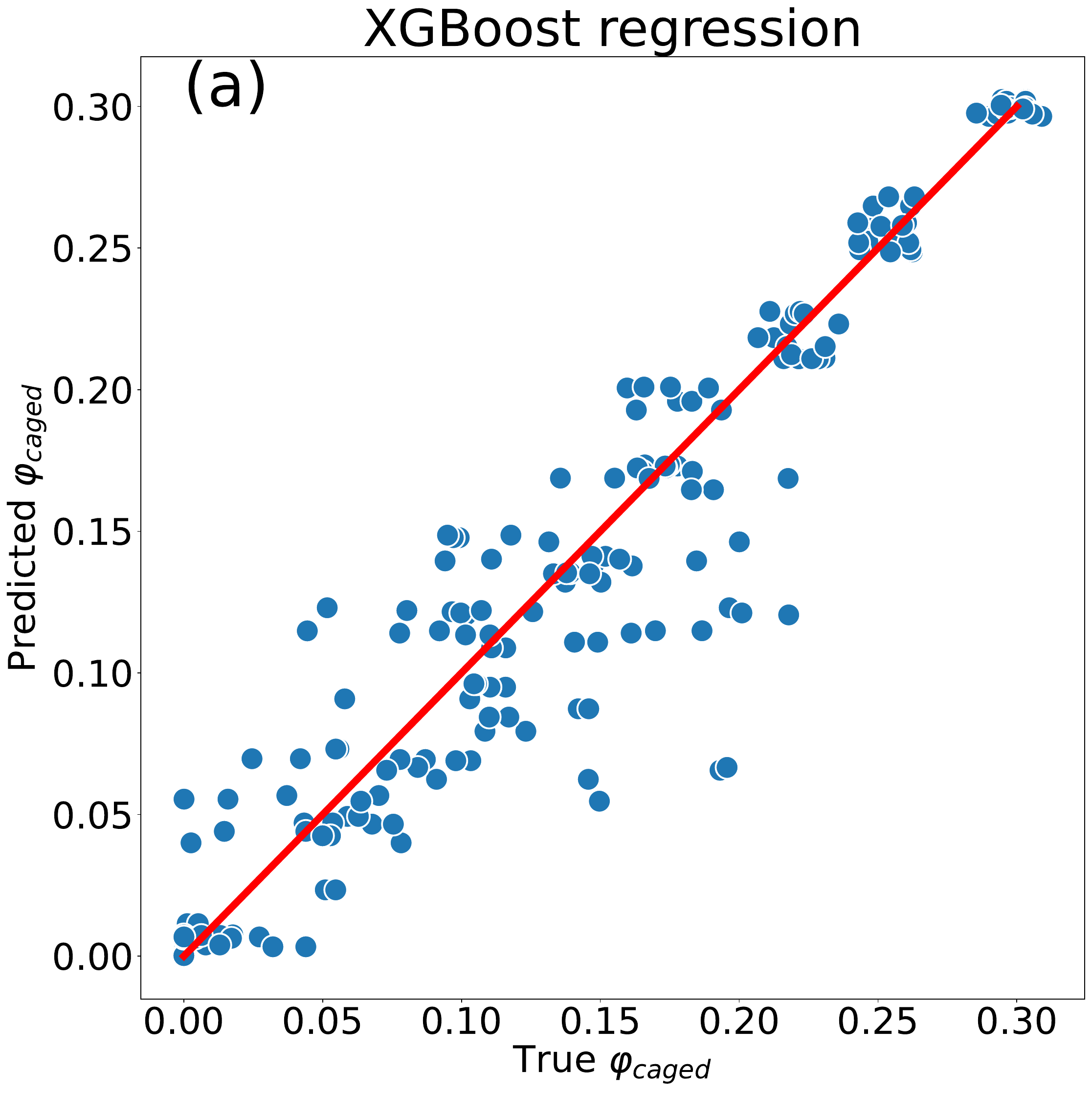} %
    \hspace{0.5cm}
    \includegraphics[width=6cm]{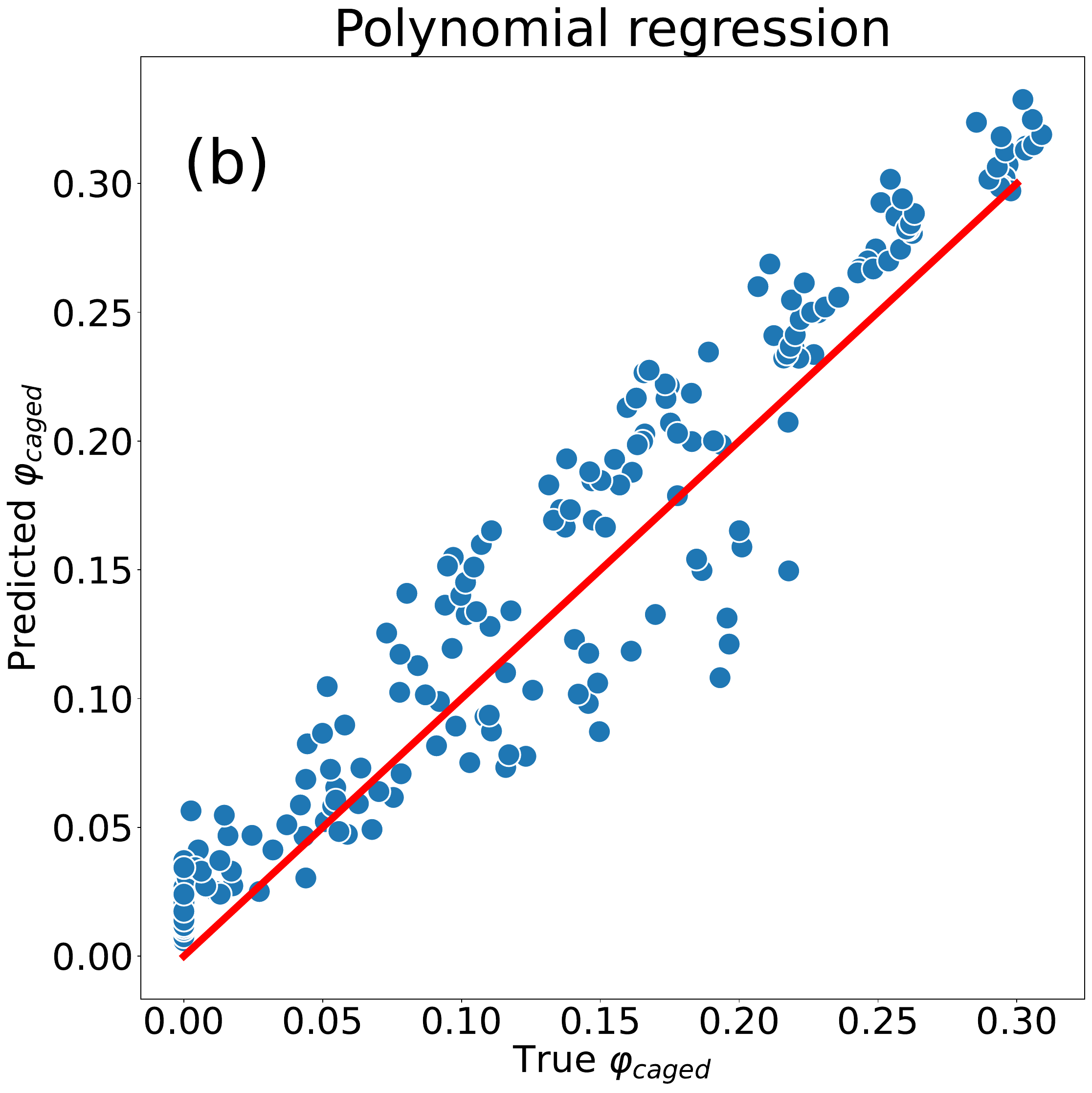}%
    \caption{Comparison of true and predicted $\varphi_{\rm caged}$ using XGBoost regression (a) and Polynomial regression (b) in the extrapolation to larger $A$.
    }%
    \label{fig:test_XGB_PR_max_lpf_prediction_extrapolation_amplitude}%
\end{figure}

Table \ref{testdataset_var_extrapolation_amplitude_voronoi} shows the performance metrics for the extrapolation task to the larger values of $A$. According to the metrics, Polynomial regression shows the best performance in this case, closely followed by XGBoost regression and Neural Network regression.    Fig.~\ref{fig:test_XGB_PR_max_lpf_prediction_extrapolation_amplitude} shows the comparison between true and predicted $\varphi_{\rm caged}$ for XGBoost regression and Polynomial regression models. 

\begin{table}[htbp]
\caption{Performance metrics on test dataset in extrapolation to larger $A$. Negative $R^2$ indicate that the models fail to generalize beyond the training data.}
\label{testdataset_var_extrapolation_amplitude_voronoi}%
\begin{tabular}{@{}llll@{}}
\toprule
Model & RMSE & MAE & R$^2$\\
\midrule
\textbf{Polynomial Regression} & \textbf{0.03}  & \textbf{0.026}  & \textbf{0.89}  \\
Random Forest Regression & 0.034  & 0.022  & 0.86 \\ 
Support Vector Regression & 0.058  & 0.051 & 0.61 \\ 
XGBoost Regression & 0.028  & 0.018  & 0.91  \\ 
Neural Network Regression & 0.033  & 0.022  & 0.87  \\
\botrule
\end{tabular}
\end{table}

\begin{figure}[htbp]%
    \centering
    \begin{tabular}{cc}
        \includegraphics[width=7.0cm]{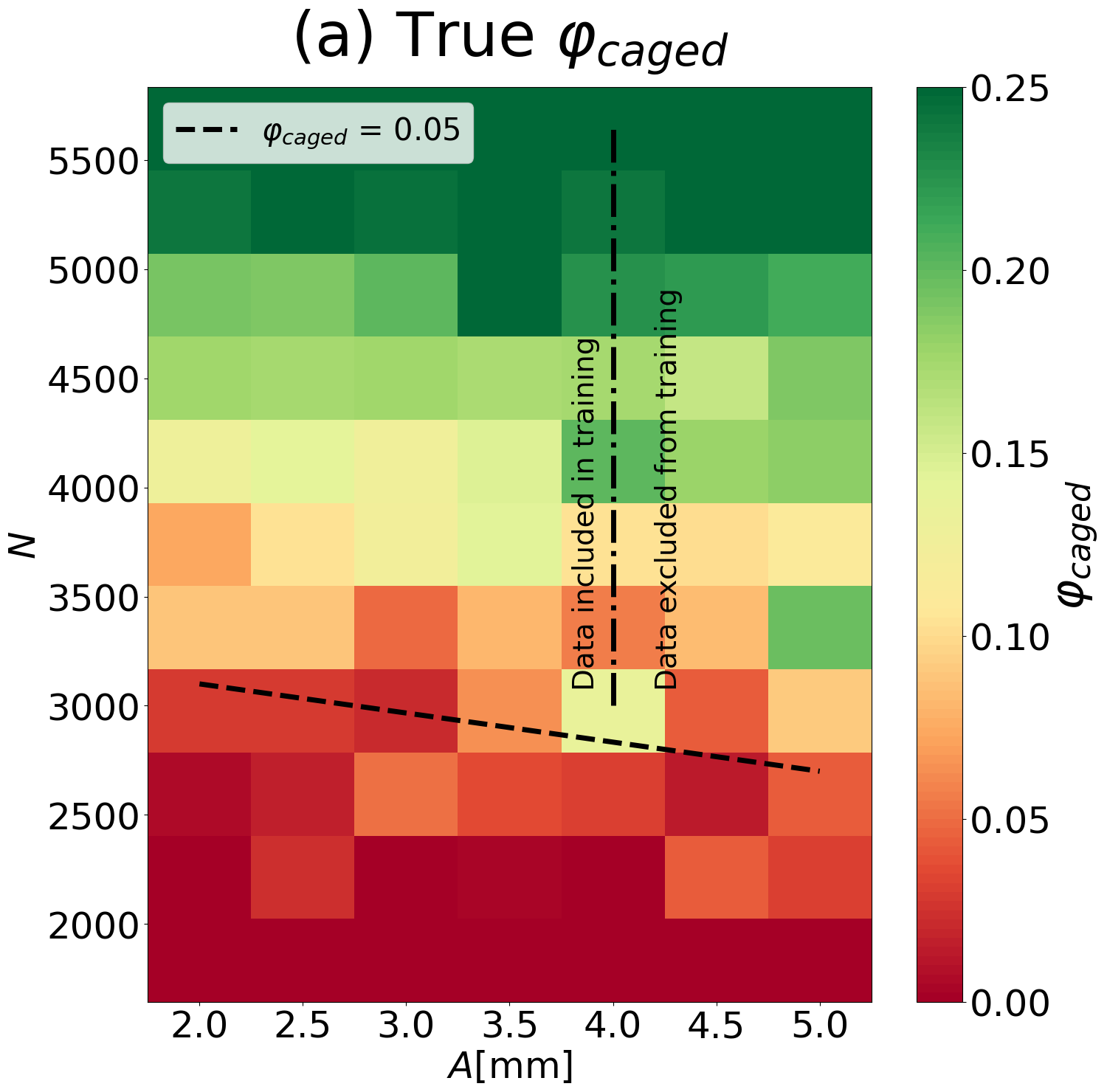} &
        \includegraphics[width=7.0cm]{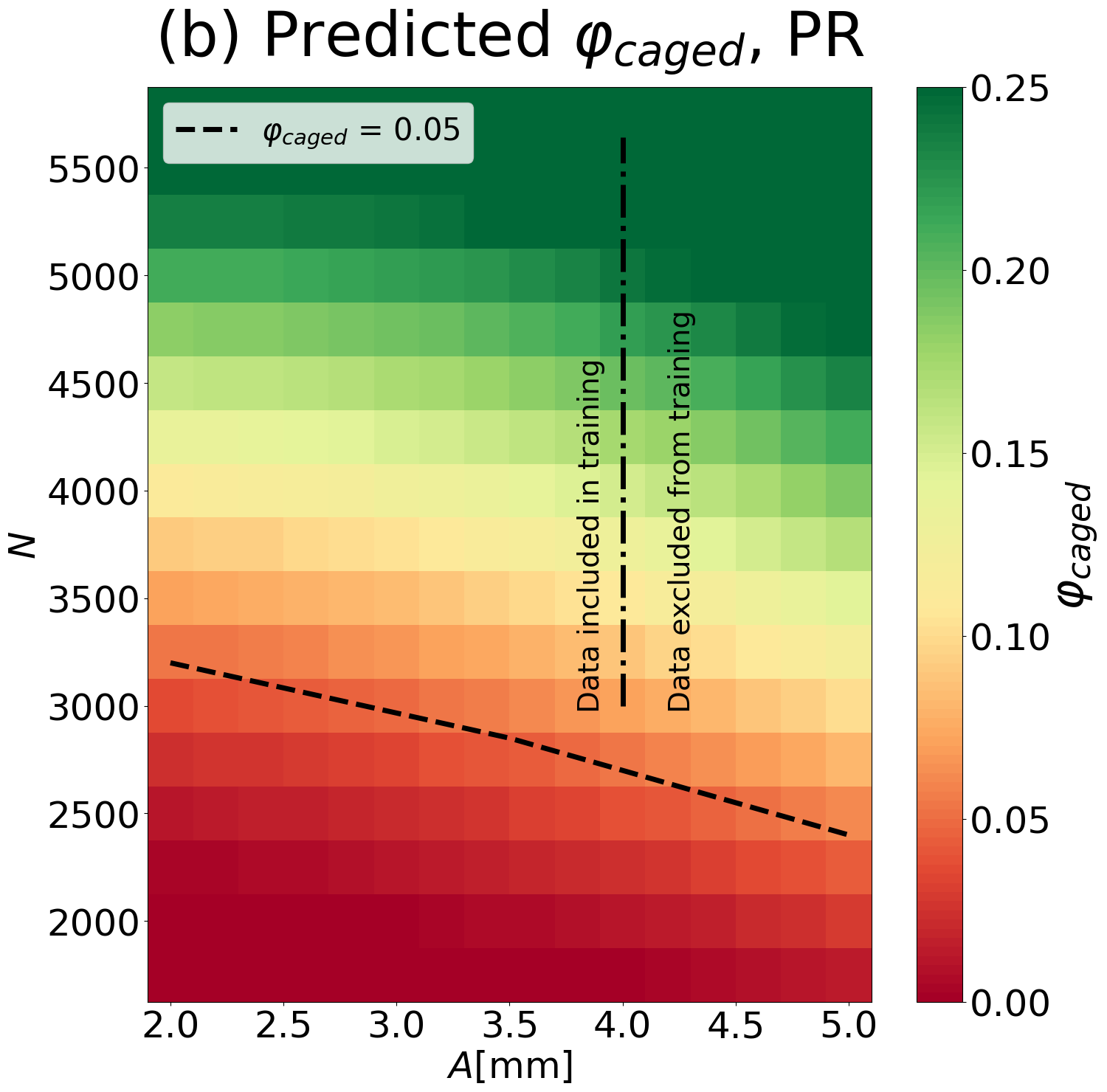} \\
        \includegraphics[width=7.0cm]{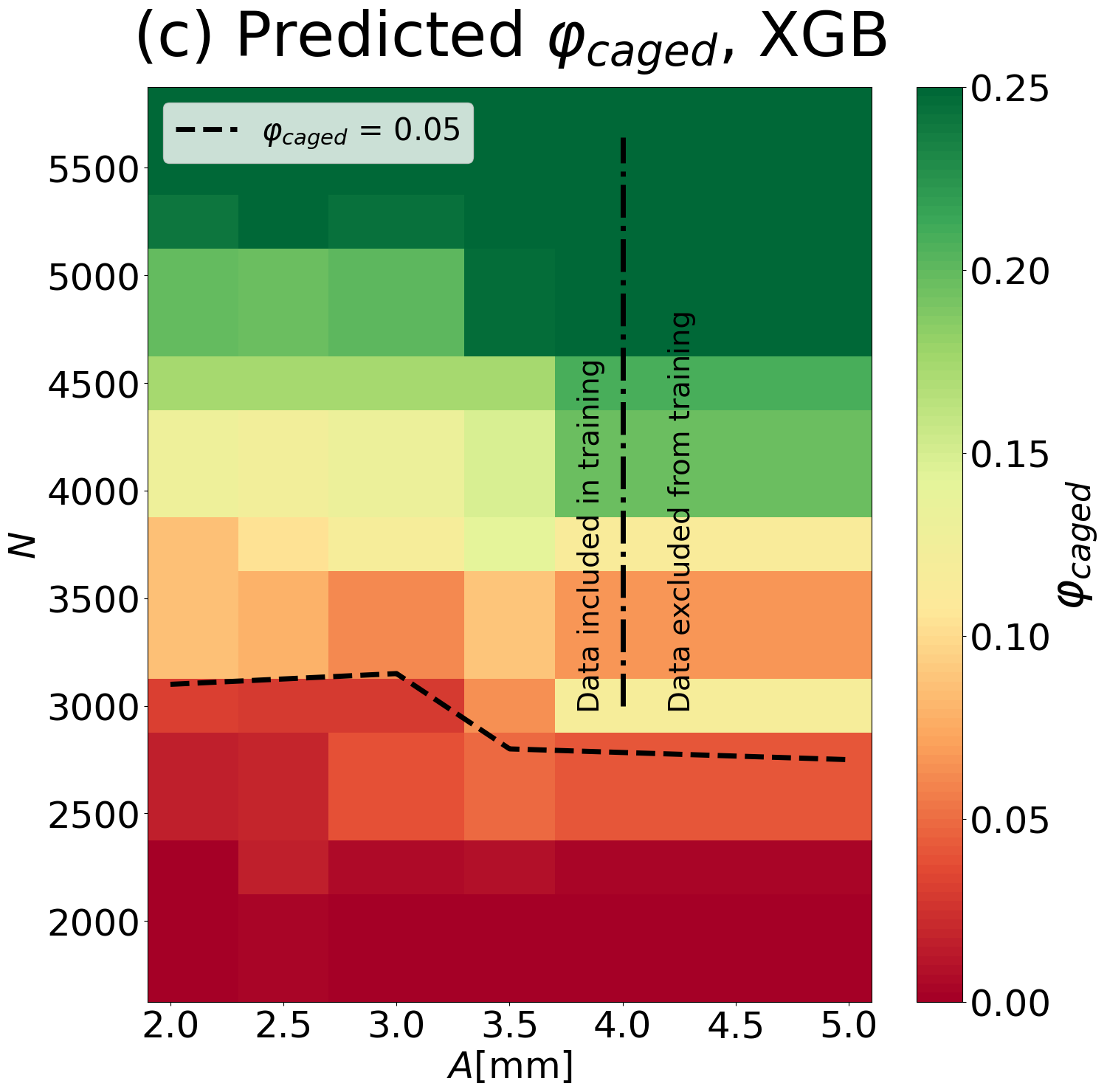} &
        \includegraphics[width=7.0cm]{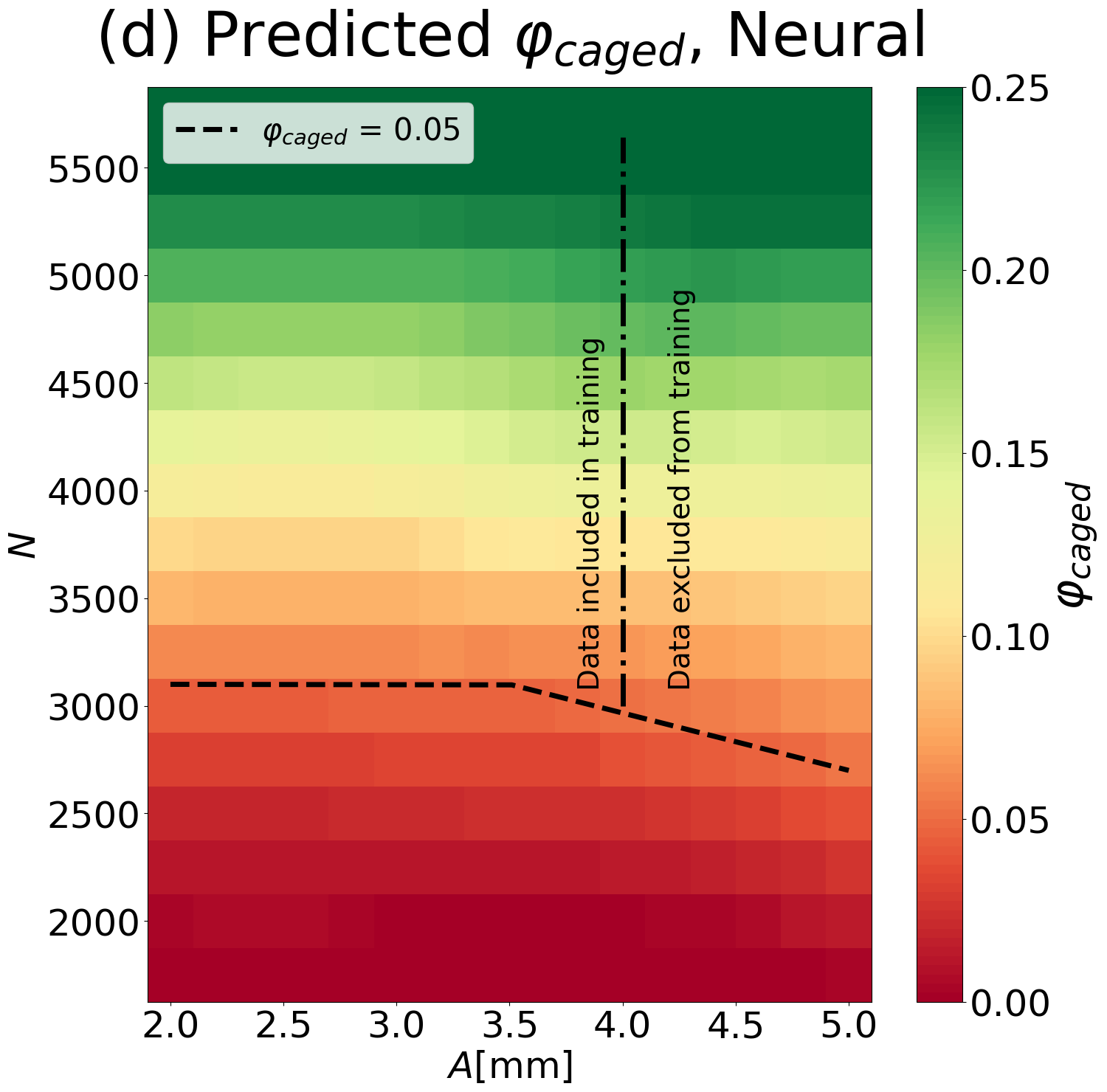}\\
    \end{tabular}
    \caption{True (a) and predicted $\varphi_{caged}$ by Polynomial regression (b), XGBoost regression (c) and Neural Network regression (d) models when $N$ and $A$ are varied in the extrapolation to larger $A$. The gas-cluster transition occurs at the dashed line. The region above the black dash-dot was not used to train the ML algorithms. $\theta=\pi$.
    } %
    \label{fig:contour_plots_true_predicted_extrapolation_phi_caged_amplitude}%
\end{figure}

Figures~\ref{fig:contour_plots_true_predicted_extrapolation_phi_caged_amplitude}(a-d) show the true and predicted $\varphi_{\rm caged}$ from Polynomial regression with degree 2, from XGBoost regression and from Neural Network regression models, respectively, for the extrapolation to larger amplitudes $A$. Even though the predictions from XGBoost regression and Neural Network regression are close to the true data, the predicted transition boundary is different. Compared to these three regression models, predictions from Polynomial regression model are close to the true data with similar transition boundaries between true data and predictions. Hence, Polynomial regression is chosen as the most suitable model in the second case.

In summary, Polynomial regression with degree 2 can be used to predict $\varphi_{\rm caged}$ by extrapolation to higher amplitudes $A$ as well as larger particle numbers $N$. Of cause, this was tested only for a limited extension of the parameter space. An extrapolation to broader regions may require more training data and possible further tuning of the prediction algorithms.

\subsubsection{Classification}

For the classification task, we chose the independent variables $N$, $A$, $\theta$, and the dependent binary variable is a label derived from $\varphi_{\rm caged}$. The label `0' is assigned when $\varphi_{\rm caged}\le 0.05$, marking the gaseous state. The label `1' for $\varphi_{\rm caged}>0.05$ marks the cluster state. Interpolation and extrapolation capabilities are tested separately.   

\begin{figure}[ht]%
    \centering
    \includegraphics[width=8cm]{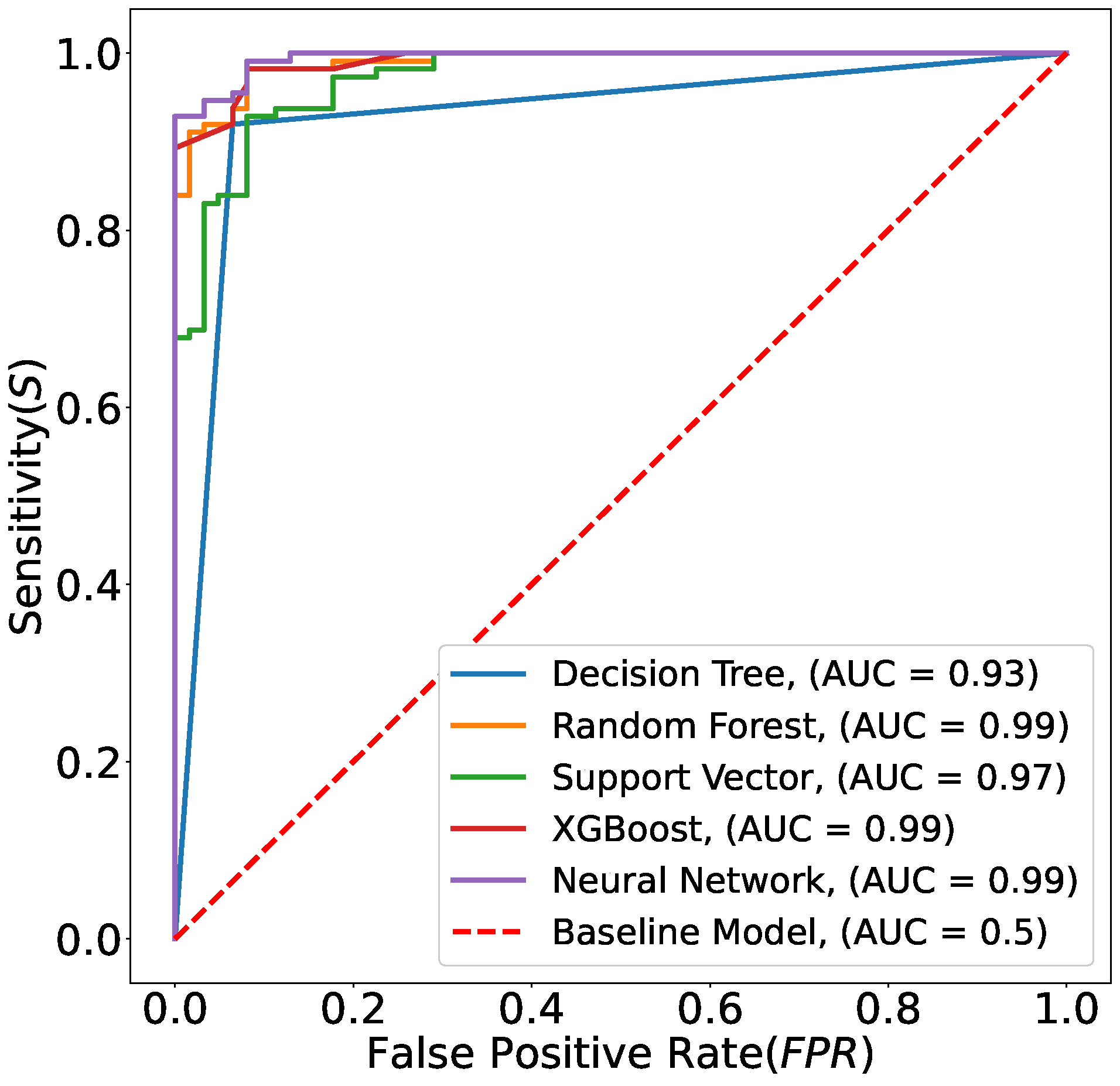}%
    \caption{Comparison of machine learning algorithms by the area under the ROC curve (AUC) for the caging criterion.}%
    \label{fig:roc_auc_curve_interpolation_caging}%
    
\end{figure}

The synthetic dataset was split into training, validation and test subsets in the same way as in the regression tests.
The machine learning algorithms mentioned in \ref{mlalgorithms} were trained and evaluated using the classification metrics. Accuracy($A$), Precision($P$), Recall($R$) and F1-Score($F1$) are compared in Tab.~ \ref{testdataset_var_interpolation_classification_caging}. 
XGBoost and Neural Network Classifier achieve better performance than the other models, yet Random Forest works comparably well. Figure \ref{fig:roc_auc_curve_interpolation_caging} compares the area under the ROC curves for these algorithms. XGBoost and Neural Network Classifier achieve performances close to the perfect one with $AUC=0.99$. The confusion matrices are compared in Tab.~ \ref{confusion_matrix_interpolation_ks_test_caging}. From this matrix, one notices that XGBoost and Neural Network classifiers had the lowest rate of false decisions, but all models performed comparably well.   

\begin{table}[htbp]
\caption{Performance metrics for classification of the system state on the basis of $\varphi_{\rm caged}$ by interpolation}
\label{testdataset_var_interpolation_classification_caging}%
\begin{tabular}{@{}lllll@{}}
\toprule
Model & Accuracy & Precision & Recall & F1-Score\\
\midrule
Decision Tree Classifier & 0.925 & 0.93  & 0.95 & 0.942 \\
Random Forest Classifier & 0.948  & 0.955  & 0.964 & 0.966 \\ 
Support Vector Classifier & 0.913  & 0.936 & 0.928 & 0.932 \\ 
XGBoost Classifier & 0.954  & 0.956  & 0.973 & 0.964  \\ 
Neural Network Classifier & 0.954  & 0.956  & 0.973 & 0.964 \\
\\  
\botrule
\end{tabular}
\end{table}

 \begin{table}[ht]
\caption{Confusion matrix for classifications on the basis of $\varphi_{\rm caged}$}
\label{confusion_matrix_interpolation_ks_test_caging}%
\begin{tabular}{@{}lllll@{}}
\toprule
Model & $TP$ & $FP$ & $TN$ & $FN$\\
\midrule
Decision Tree Classifier & 107 & 8 & 54 & 5 \\
Random Forest Classifier & 108 & 5 & 57 & 4 \\ 
Support Vector Classifier & 104 & 7 & 55 & 8 \\ 
XGBoost Classifier & 109 & 5 & 57 & 3 \\
Neural Network Classifier & 109 & 5 & 57 & 3 \\
\botrule
\end{tabular}
\end{table}

\begin{figure}[!ht]%
    \centering
    \includegraphics[width=8cm]{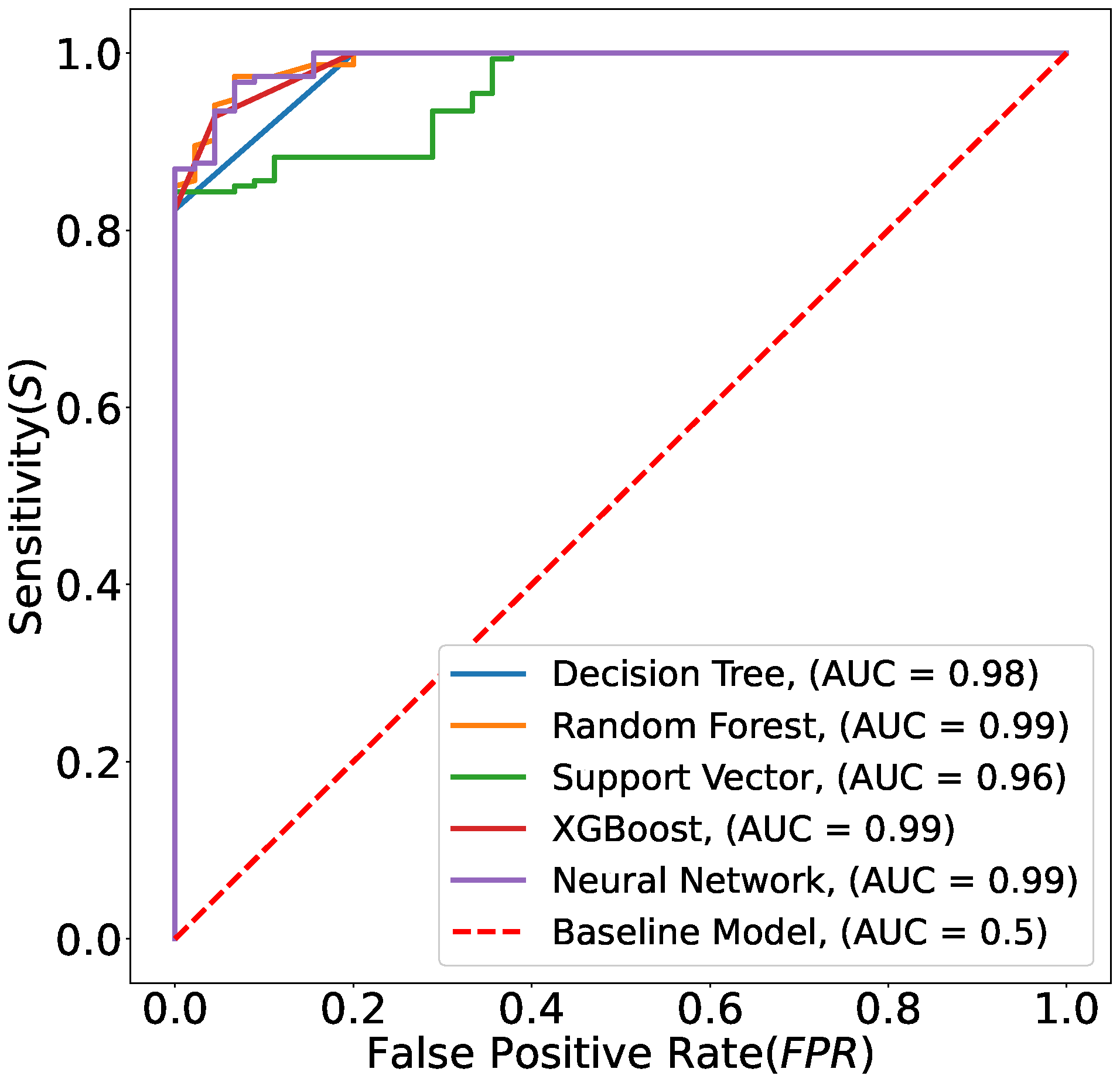}%
    \caption{Comparison of machine learning algorithms by the area under the ROC curve (AUC) for the caging criterion.}%
    \label{fig:roc_auc_curve_extrapolation_caging}%
    
\end{figure}

For the extrapolation test, the data points with amplitudes $A$ between 2 mm and 4 mm were included in the training set, while those between 4.5 mm and 5 mm were included in the test dataset. The performance of these algorithms and their corresponding confusion matrix are tabulated in \ref{testdataset_var_extrapolation_classification_caging} and \ref{confusion_matrix_extrapolation_ks_test_caging} respectively. ROC curves and the area under these curves (AUC) for these algorithms is shown in Fig.~\ref{fig:roc_auc_curve_extrapolation_caging}. By comparing the false predictions and other performance metrics, Neural Network classifier is identified as the best algorithms compared to others for {the extrapolation of the state of the system}.

\begin{table}[!htbp]
\caption{Performance metrics for classifications of the state on the basis of $\varphi_{\rm caged}$ by extrapolation}
\label{testdataset_var_extrapolation_classification_caging}%
\begin{tabular}{@{}lllll@{}}
\toprule
Model & Accuracy & Precision & Recall & F1-Score\\
\midrule
Decision Tree Classifier & 0.95 & 0.94  & 1.0 & 0.97 \\
Random Forest Classifier & 0.86  & 1.0  & 0.82 & 0.90 \\ 
Support Vector Classifier & 0.86  & 1.0 & 0.82 & 0.90 \\ 
XGBoost Classifier & 0.93  & 0.98  & 0.92 & 0.95  \\ 
Neural Network Classifier & 0.954  & 0.956  & 0.99 & 0.97 \\
\\  
\botrule
\end{tabular}
\end{table}

 \begin{table}[!ht]
\caption{Confusion matrix for classifications on the basis of $\varphi_{\rm caged}$}
\label{confusion_matrix_extrapolation_ks_test_caging}%
\begin{tabular}{@{}lllll@{}}
\toprule
Model & $TP$ & $FP$ & $TN$ & $FN$\\
\midrule
Decision Tree Classifier & 153 & 9 & 36 & 0 \\
Random Forest Classifier & 126 & 0 & 45 & 27 \\ 
Support Vector Classifier & 126 & 0 & 45 & 27  \\ 
XGBoost Classifier & 142 & 2 & 43 & 11 \\
Neural Network Classifier & 152 & 7 & 38 & 1 \\
\\  
\botrule
\end{tabular}
\end{table}

In summary, the XGBoost and Neural Network classifiers are comparably well suited as models to predict the state of the system for a given set of system parameters $N, A, \theta$ using the caging criterion by interpolation while the Neural Network classifier is the best model for the extrapolation tasks.

\section{Summary and Conclusions}

We have performed simulations of permanently excited granular ensembles of spheres using the geometry of the VIP-Gran experiment shown in Fig. \ref{fig:VIP_GRAN_setup}. On the basis of these simulations, we utilized two established methods to detect the gas-cluster transitions as a function of three system parameters. 
The data simulated were used to build sets to train and evaluate machine learning algorithms that predict the transition parameters from the homogeneous state to the clustered state. The algorithms were evaluated with various performance metrics. 
We tested the effectiveness of the optimal algorithms by comparing the predicted gas-cluster transitions with simulated test data. The selected algorithms were shown to effectively interpolate and extrapolate to determine the transition parameters and predict the corresponding state without the need for time-consuming and computationally expensive numerical simulations.  
We did not verify here the appropriateness of the two established test methods to define the clustered state
\cite{Opsomer2012,Noirhomme2018}
proposed in earlier studies. This physical aspect remains to be analyzed more thoroughly in future studies. If a better
characterization parameter or set of parameters for clustering can be found, e. g. on the basis of the analysis of the dynamic properties of the ensemble, the ML software can be adapted with little effort to those parameters. 

The advantages of using algorithms based on ML software to predict the behavior of granular gases in microgravity are obvious:
Experimentally, the investigation of such ensembles requires a long lead time and is very cost-intensive. 
During the preparation preceding such experiments, it is desirable to limit the parameter range adequately to obtain the measurement data as effectively as possible. A numerical simulation may offer an opportunity to do this, but it is also very time-consuming. In addition to the time and computational effort required for the numerical simulation, there is also the additional effort to extract the characteristics that are decisive for the tests from the ensemble.

The phase space of the system (number of particles, excitation amplitude and phase) is sufficiently uncomplicated so that it can be predicted 
well by training the software with comparatively few samples, whereby the predictions can also be extrapolated beyond the range used for training.
This system therefore lends itself to making predictions based on artificial intelligence from a number of already known simulations,
with which the entire phase space, or at least a large part of it, can be constructed with sufficient accuracy.
Training is carried out on the basis of a known number of simulated or experimental data sets. In the present study, these were taken on a 
regular grid, but in principle they can also be statistically distributed over the phase space. The method presented here 
should also be useful for many other applications in which a dynamic system with a very large number of degrees of freedom is statistically
classified by a few parameters.


\section*{Acknowledgements}

The authors acknowledge funding by the German Aerospace Center (DLR) under grants 50WM2252 and 50WK2348. We coordially thank our colleagues from Granular Matter group at Otto von Guericke University Magdeburg (OVGU) and Brandenburg University of Applied Sciences,  Ra\'ul Cruz Hidalgo from the Universidad de Navarra for providing the numerical simulation software, Kirsten Harth from TH Brandenburg for comments and feedback. We would also like to thank the Space Grains team for permission to reproduce a figure.

\section*{Funding}

This research was supported by German Aerospace Center (DLR) under projects "Vorraussagen von Inhomogentit\"at und Clusterdynamik mittels k\"unstlichter Intelligenz"(VICKI) [grant number: 50WM2252] and "K\"unstliche Intelligenz zur Objektverfolgung in Vielteilchensystemen"(EVA-II) [grant number: 50WK2348].

\section*{Author contributions}

S.P.S. performed the simulations, worked on the data analysis and prepared the manuscript. D.P. conceived the study, D.P., B.N., and R.S. analyzed and interpreted the results. All authors edited and reviewed the manuscript.


\begin{thebibliography}{42}
\ifx \bisbn   \undefined \def \bisbn  #1{ISBN #1}\fi
\ifx \binits  \undefined \def \binits#1{#1}\fi
\ifx \bauthor  \undefined \def \bauthor#1{#1}\fi
\ifx \batitle  \undefined \def \batitle#1{#1}\fi
\ifx \bjtitle  \undefined \def \bjtitle#1{#1}\fi
\ifx \bvolume  \undefined \def \bvolume#1{\textbf{#1}}\fi
\ifx \byear  \undefined \def \byear#1{#1}\fi
\ifx \bissue  \undefined \def \bissue#1{#1}\fi
\ifx \bfpage  \undefined \def \bfpage#1{#1}\fi
\ifx \blpage  \undefined \def \blpage #1{#1}\fi
\ifx \burl  \undefined \def \burl#1{\textsf{#1}}\fi
\ifx \doiurl  \undefined \def \doiurl#1{\url{https://doi.org/#1}}\fi
\ifx \betal  \undefined \def \betal{\textit{et al.}}\fi
\ifx \binstitute  \undefined \def \binstitute#1{#1}\fi
\ifx \binstitutionaled  \undefined \def \binstitutionaled#1{#1}\fi
\ifx \bctitle  \undefined \def \bctitle#1{#1}\fi
\ifx \beditor  \undefined \def \beditor#1{#1}\fi
\ifx \bpublisher  \undefined \def \bpublisher#1{#1}\fi
\ifx \bbtitle  \undefined \def \bbtitle#1{#1}\fi
\ifx \bedition  \undefined \def \bedition#1{#1}\fi
\ifx \bseriesno  \undefined \def \bseriesno#1{#1}\fi
\ifx \blocation  \undefined \def \blocation#1{#1}\fi
\ifx \bsertitle  \undefined \def \bsertitle#1{#1}\fi
\ifx \bsnm \undefined \def \bsnm#1{#1}\fi
\ifx \bsuffix \undefined \def \bsuffix#1{#1}\fi
\ifx \bparticle \undefined \def \bparticle#1{#1}\fi
\ifx \barticle \undefined \def \barticle#1{#1}\fi
\bibcommenthead
\ifx \bconfdate \undefined \def \bconfdate #1{#1}\fi
\ifx \botherref \undefined \def \botherref #1{#1}\fi
\ifx \url \undefined \def \url#1{\textsf{#1}}\fi
\ifx \bchapter \undefined \def \bchapter#1{#1}\fi
\ifx \bbook \undefined \def \bbook#1{#1}\fi
\ifx \bcomment \undefined \def \bcomment#1{#1}\fi
\ifx \oauthor \undefined \def \oauthor#1{#1}\fi
\ifx \citeauthoryear \undefined \def \citeauthoryear#1{#1}\fi
\ifx \endbibitem  \undefined \def \endbibitem {}\fi
\ifx \bconflocation  \undefined \def \bconflocation#1{#1}\fi
\ifx \arxivurl  \undefined \def \arxivurl#1{\textsf{#1}}\fi
\csname PreBibitemsHook\endcsname

\bibitem[\protect\citeauthoryear{Opsomer et~al.}{2012}]{Opsomer2012}
\begin{barticle}
\bauthor{\bsnm{Opsomer}, \binits{E.}},
\bauthor{\bsnm{Ludewig}, \binits{F.}},
\bauthor{\bsnm{Vandewalle}, \binits{N.}}:
\batitle{Dynamical clustering in driven granular gas}.
\bjtitle{Europhys. Lett.}
\bvolume{99}(\bissue{4}),
\bfpage{40001}
(\byear{2012})
\doiurl{10.1209/0295-5075/99/40001}
\end{barticle}
\endbibitem

\bibitem[\protect\citeauthoryear{Aumaitre et~al.}{2018}]{Aumaitre2018}
\begin{barticle}
\bauthor{\bsnm{Aumaitre}, \binits{S.}},
\bauthor{\bsnm{Behringer}, \binits{R.P.}},
\bauthor{\bsnm{Cazaubiel}, \binits{A.}},
\bauthor{\bsnm{Cl\'ement}, \binits{E.}},
\bauthor{\bsnm{Crassous}, \binits{J.}},
\bauthor{\bsnm{Durian}, \binits{D.J.}},
\bauthor{\bsnm{Falcon}, \binits{E.}},
\bauthor{\bsnm{Fauve}, \binits{S.}},
\bauthor{\bsnm{Fischer}, \binits{D.}},
\bauthor{\bsnm{Garcimart\'in}, \binits{A.}},
\bauthor{\bsnm{Garrabos}, \binits{Y.}},
\bauthor{\bsnm{Hou}, \binits{M.}},
\bauthor{\bsnm{Jia}, \binits{X.}},
\bauthor{\bsnm{Lecoutre}, \binits{C.}},
\bauthor{\bsnm{Luding}, \binits{S.}},
\bauthor{\bsnm{Maza}, \binits{D.}},
\bauthor{\bsnm{Noirhomme}, \binits{M.}},
\bauthor{\bsnm{Opsomer}, \binits{E.}},
\bauthor{\bsnm{Palencia}, \binits{F.}},
\bauthor{\bsnm{P\"oschel}, \binits{T.}},
\bauthor{\bsnm{Schockmel}, \binits{J.}},
\bauthor{\bsnm{Sperl}, \binits{M.}},
\bauthor{\bsnm{Stannarius}, \binits{R.}},
\bauthor{\bsnm{Vandewalle}, \binits{N.}},
\bauthor{\bsnm{Yu}, \binits{P.}}:
\batitle{An instrument for studying granular media in low-gravity environment}.
\bjtitle{Rev. Sci. Instr.}
\bvolume{89},
\bfpage{075103}
(\byear{2018})
\end{barticle}
\endbibitem

\bibitem[\protect\citeauthoryear{Adachi et~al.}{2019}]{Adachi2019}
\begin{barticle}
\bauthor{\bsnm{Adachi}, \binits{M.}},
\bauthor{\bsnm{Yu}, \binits{P.}},
\bauthor{\bsnm{Sperl}, \binits{M.}}:
\batitle{Magnetic excitation of a granular gas as a bulk thermostat}.
\bjtitle{npj Micrograv.}
\bvolume{5},
\bfpage{19}
(\byear{2019})
\end{barticle}
\endbibitem

\bibitem[\protect\citeauthoryear{Harth et~al.}{2015}]{Harth2015}
\begin{barticle}
\bauthor{\bsnm{Harth}, \binits{K.}},
\bauthor{\bsnm{Trittel}, \binits{T.}},
\bauthor{\bsnm{May}, \binits{K.}},
\bauthor{\bsnm{Wegner}, \binits{S.}},
\bauthor{\bsnm{Stannarius}, \binits{R.}}:
\batitle{Three-dimensional (3d) experimental realization and observation of a granular gas in microgravity}.
\bjtitle{Adv. Space Res.}
\bvolume{55},
\bfpage{1901}--\blpage{1912}
(\byear{2015})
\end{barticle}
\endbibitem

\bibitem[\protect\citeauthoryear{Noirhomme et~al.}{2018}]{Noirhomme2018}
\begin{barticle}
\bauthor{\bsnm{Noirhomme}, \binits{M.}},
\bauthor{\bsnm{Cazaubiel}, \binits{A.}},
\bauthor{\bsnm{Darras}, \binits{A.}},
\bauthor{\bsnm{Falcon}, \binits{E.}},
\bauthor{\bsnm{Fischer}, \binits{D.}},
\bauthor{\bsnm{Garrabos}, \binits{Y.}},
\bauthor{\bsnm{Lecoutre-Chabot}, \binits{C.}},
\bauthor{\bsnm{Merminod}, \binits{S.}},
\bauthor{\bsnm{Opsomer}, \binits{E.}},
\bauthor{\bsnm{Palencia}, \binits{F.}},
\bauthor{\bsnm{Schockmel}, \binits{J.}},
\bauthor{\bsnm{Stannarius}, \binits{R.}},
\bauthor{\bsnm{Vandewalle}, \binits{N.}}:
\batitle{Threshold of gas-like to clustering transition in driven granular media in low-gravity environment}.
\bjtitle{Europhys. Lett.}
\bvolume{123}(\bissue{1}),
\bfpage{14003}
(\byear{2018})
\end{barticle}
\endbibitem

\bibitem[\protect\citeauthoryear{Goldhirsch and Zanetti}{1993}]{Goldhirsch1993}
\begin{barticle}
\bauthor{\bsnm{Goldhirsch}, \binits{I.}},
\bauthor{\bsnm{Zanetti}, \binits{G.}}:
\batitle{Clustering instability in dissipative gases}.
\bjtitle{Phys. Rev. Lett.}
\bvolume{70},
\bfpage{1619}--\blpage{1622}
(\byear{1993})
\end{barticle}
\endbibitem

\bibitem[\protect\citeauthoryear{Falcon et~al.}{2006}]{Falcon2006}
\begin{barticle}
\bauthor{\bsnm{Falcon}, \binits{E.}},
\bauthor{\bsnm{Aumaitre}, \binits{S.}},
\bauthor{\bsnm{Evesque}, \binits{P.}},
\bauthor{\bsnm{Palencia}, \binits{F.}},
\bauthor{\bsnm{Lecoutre-Chabot}, \binits{C.}},
\bauthor{\bsnm{Fauve}, \binits{S.}},
\bauthor{\bsnm{Beysens}, \binits{D.}},
\bauthor{\bsnm{Garrabos}, \binits{Y.}}:
\batitle{Collision statistics in a dilute granular gas fluidized by vibrations in low gravity}.
\bjtitle{Eur. Phys. Lett.}
\bvolume{74},
\bfpage{830}--\blpage{836}
(\byear{2006})
\end{barticle}
\endbibitem

\bibitem[\protect\citeauthoryear{Wu et~al.}{2020}]{Wu2020}
\begin{barticle}
\bauthor{\bsnm{Wu}, \binits{Q.-L.}},
\bauthor{\bsnm{Hou}, \binits{M.-Y.}},
\bauthor{\bsnm{Yang}, \binits{L.}},
\bauthor{\bsnm{Wang}, \binits{W.}},
\bauthor{\bsnm{Yang}, \binits{G.-H.}},
\bauthor{\bsnm{Tao}, \binits{K.-W.}},
\bauthor{\bsnm{Chen}, \binits{L.-W.}},
\bauthor{\bsnm{Zhang}, \binits{S.}}:
\batitle{Parametric study of the clustering transition in vibration driven granular gas system}.
\bjtitle{Chin. Phys. B}
\bvolume{29}(\bissue{5}),
\bfpage{054502}
(\byear{2020})
\doiurl{10.1088/1674-1056/ab8217}
\end{barticle}
\endbibitem

\bibitem[\protect\citeauthoryear{Puzyrev et~al.}{2021}]{PuzyrevDmitry2021}
\begin{barticle}
\bauthor{\bsnm{Puzyrev}, \binits{D.}},
\bauthor{\bsnm{Cruz~Hidalgo}, \binits{R.}},
\bauthor{\bsnm{Fischer}, \binits{D.}},
\bauthor{\bsnm{Harth}, \binits{K.}},
\bauthor{\bsnm{Trittel}, \binits{T.}},
\bauthor{\bsnm{Stannarius}, \binits{R.}}:
\batitle{Cluster dynamics in dense granular gases of rod-like particles}.
\bjtitle{EPJ Web of Conf.}
\bvolume{249},
\bfpage{04004}
(\byear{2021})
\doiurl{10.1051/epjconf/202124904004}
\end{barticle}
\endbibitem

\bibitem[\protect\citeauthoryear{Sata et~al.}{2025}]{Sata2025_preprint_1}
\begin{barticle}
\bauthor{\bsnm{Sata}, \binits{S.P.}},
\bauthor{\bsnm{Stannarius}, \binits{R.}},
\bauthor{\bsnm{Puzyrev}, \binits{D.}}:
\batitle{{Criteria for dynamical clustering in permanently excited granular gases: Comparison and estimation with machine learning, PREPRINT (Version 1)}}.
\bjtitle{{available at Research Square}}
(\byear{2025})
\doiurl{10.21203/rs.3.rs-6243611/v1}
\end{barticle}
\endbibitem

\bibitem[\protect\citeauthoryear{Hidalgo et~al.}{2013}]{Hidalgo2013}
\begin{barticle}
\bauthor{\bsnm{Hidalgo}, \binits{R.}},
\bauthor{\bsnm{Kanzaki~Cabrera}, \binits{T.}},
\bauthor{\bsnm{Alonso-Marroquin}, \binits{F.}},
\bauthor{\bsnm{Luding}, \binits{S.}}:
\batitle{On the use of graphics processing units (gpus) for molecular dynamics simulation of spherical particles}.
\bjtitle{AIP Conference Proceedings}
\bvolume{1542},
\bfpage{169}--\blpage{172}
(\byear{2013})
\doiurl{10.1063/1.4811894}
\end{barticle}
\endbibitem

\bibitem[\protect\citeauthoryear{}{}]{SpaceGrains}
\begin{botherref}
Space Grains Team.
\url{https://spacegrains.org/page/2/}
\end{botherref}
\endbibitem

\bibitem[\protect\citeauthoryear{Yu et~al.}{2020}]{Yu2019}
\begin{barticle}
\bauthor{\bsnm{Yu}, \binits{P.}},
\bauthor{\bsnm{St\"ark}, \binits{E.}},
\bauthor{\bsnm{Blochberger}, \binits{G.}},
\bauthor{\bsnm{Kaplik}, \binits{M.}},
\bauthor{\bsnm{Offermann}, \binits{M.}},
\bauthor{\bsnm{Tran}, \binits{D.}},
\bauthor{\bsnm{Adachi}, \binits{M.}},
\bauthor{\bsnm{Sperl}, \binits{M.}}:
\batitle{Magnetically excited granular matter in low gravity}.
\bjtitle{Rev. Sci. Instrum.}
\bvolume{90},
\bfpage{054501}
(\byear{2020})
\end{barticle}
\endbibitem

\bibitem[\protect\citeauthoryear{Falcon et~al.}{2013}]{Falcon2013}
\begin{barticle}
\bauthor{\bsnm{Falcon}, \binits{E.}},
\bauthor{\bsnm{Bacri}, \binits{J.-C.}},
\bauthor{\bsnm{Laroche}, \binits{C.}}:
\batitle{Equation of state of a granular gas homogeneously driven by particle rotations}.
\bjtitle{Eur. Phys. Lett.}
\bvolume{103},
\bfpage{64004}
(\byear{2013})
\end{barticle}
\endbibitem

\bibitem[\protect\citeauthoryear{Falcon et~al.}{2017}]{Falcon2017}
\begin{barticle}
\bauthor{\bsnm{Falcon}, \binits{E.}},
\bauthor{\bsnm{Bacri}, \binits{J.-C.}},
\bauthor{\bsnm{Laroche}, \binits{C.}}:
\batitle{Dissipated power within a turbulent flow forced homogeneously by magnetic particles}.
\bjtitle{Phys. Rev. Fluids}
\bvolume{2},
\bfpage{102601}
(\byear{2017})
\doiurl{10.1103/PhysRevFluids.2.102601}
\end{barticle}
\endbibitem

\bibitem[\protect\citeauthoryear{Leidenfrost}{1756}]{Leidenfrost1756}
\begin{bbook}
\bauthor{\bsnm{Leidenfrost}, \binits{J.G.}}:
\bbtitle{De Aquae Communis Nonnulis Qualitatibus Tractatus}.
\bpublisher{Ovenius},
\blocation{Duisburg}
(\byear{1756})
\end{bbook}
\endbibitem

\bibitem[\protect\citeauthoryear{Eshuis et~al.}{2005}]{Eshuis2005}
\begin{barticle}
\bauthor{\bsnm{Eshuis}, \binits{P.}},
\bauthor{\bsnm{Weele}, \binits{J.P.}},
\bauthor{\bsnm{Meer}, \binits{D.}},
\bauthor{\bsnm{Lohse}, \binits{D.}}:
\batitle{Granular leidenfrost effect: Experiment and theory of floating particle clusters}.
\bjtitle{Phys. Rev. Lett.}
\bvolume{95},
\bfpage{258001}
(\byear{2005})
\end{barticle}
\endbibitem

\bibitem[\protect\citeauthoryear{Menendez et~al.}{2020}]{Menendez2020}
\begin{barticle}
\bauthor{\bsnm{Menendez}, \binits{H.T.}},
\bauthor{\bsnm{Sack}, \binits{A.}},
\bauthor{\bsnm{P\"oschel}, \binits{T.}}:
\batitle{Granular leidenfrost effect in microgravity}.
\bjtitle{Granular Matter}
\bvolume{22},
\bfpage{67}
(\byear{2020})
\end{barticle}
\endbibitem

\bibitem[\protect\citeauthoryear{Puzyrev et~al.}{2021}]{Puzyrev2021}
\begin{barticle}
\bauthor{\bsnm{Puzyrev}, \binits{D.}},
\bauthor{\bsnm{Fischer}, \binits{D.}},
\bauthor{\bsnm{Harth}, \binits{K.}},
\bauthor{\bsnm{Trittel}, \binits{T.}},
\bauthor{\bsnm{Hidalgo}, \binits{R.C.}},
\bauthor{\bsnm{Falcon}, \binits{E.}},
\bauthor{\bsnm{Noirhomme}, \binits{M.}},
\bauthor{\bsnm{Opsomer}, \binits{E.}},
\bauthor{\bsnm{Vandewalle}, \binits{N.}},
\bauthor{\bsnm{Garrabos}, \binits{Y.}},
\bauthor{\bsnm{Lecoutre}, \binits{C.}},
\bauthor{\bsnm{Palencia}, \binits{F.}},
\bauthor{\bsnm{Stannarius}, \binits{R.}}:
\batitle{{Visual analysis of density and velocity profiles in dense 3D granular gases}}.
\bjtitle{Scientific Reports}
\bvolume{11}(\bissue{1}),
\bfpage{10621}
(\byear{2021})
\doiurl{10.1038/s41598-021-89949-z}
\end{barticle}
\endbibitem

\bibitem[\protect\citeauthoryear{Rubio et~al.}{2016}]{Rubio2016}
\begin{botherref}
\oauthor{\bsnm{Rubio}, \binits{S.}},
\oauthor{\bsnm{Maza}, \binits{D.}},
\oauthor{\bsnm{Hidalgo}, \binits{R.}}:
Large-scale numerical simulations of polydisperse particle flow in a silo.
Computational Particle Mechanics
\textbf{4}
(2016)
\doiurl{10.1007/s40571-016-0133-4}
\end{botherref}
\endbibitem

\bibitem[\protect\citeauthoryear{Hidalgo et~al.}{2016}]{Hidalgo2016}
\begin{barticle}
\bauthor{\bsnm{Hidalgo}, \binits{R.C.}},
\bauthor{\bsnm{Serero}, \binits{D.}},
\bauthor{\bsnm{Pöschel}, \binits{T.}}:
\batitle{Homogeneous cooling of mixtures of particle shapes}.
\bjtitle{Phys. Fluids}
\bvolume{28}(\bissue{7}),
\bfpage{073301}
(\byear{2016})
\doiurl{10.1063/1.4954670}
\end{barticle}
\endbibitem

\bibitem[\protect\citeauthoryear{Rubio et~al.}{2015}]{Rubio2015Disentangling}
\begin{barticle}
\bauthor{\bsnm{Rubio}, \binits{S.}},
\bauthor{\bsnm{Janda}, \binits{A.}},
\bauthor{\bsnm{Maza}, \binits{D.}},
\bauthor{\bsnm{Zuriguel}, \binits{I.}},
\bauthor{\bsnm{Hidalgo}, \binits{R.}}:
\batitle{Disentangling the free-fall arch paradox in silo discharge}.
\bjtitle{Physical Review Letters}
\bvolume{114},
\bfpage{238002}
(\byear{2015})
\doiurl{10.1103/PhysRevLett.114.238002}
\end{barticle}
\endbibitem

\bibitem[\protect\citeauthoryear{Wang}{2023}]{Wang2023}
\begin{botherref}
\oauthor{\bsnm{Wang}, \binits{J.}}:
Statistical dynamics of soft low-friction grains.
PhD thesis,
Otto von Guericke University Magdeburg
(2023)
\end{botherref}
\endbibitem

\bibitem[\protect\citeauthoryear{Pongó}{2022}]{pongo-2022}
\begin{botherref}
\oauthor{\bsnm{Pongó}, \binits{T.}}:
Particle flows in silos, significance of particle shape, stiffness and friction.
PhD thesis,
Universidad de Navarra
(2022)
\end{botherref}
\endbibitem

\bibitem[\protect\citeauthoryear{Hertz}{1882}]{hertz1882contact}
\begin{barticle}
\bauthor{\bsnm{Hertz}, \binits{H.}}:
\batitle{On the contact of elastic solids}.
\bjtitle{Journal fur die Reine und Angewandte Mathematik}
\bvolume{92},
\bfpage{156}--\blpage{171}
(\byear{1882})
\end{barticle}
\endbibitem

\bibitem[\protect\citeauthoryear{Mindlin}{1949}]{Mindlin1949}
\begin{barticle}
\bauthor{\bsnm{Mindlin}, \binits{R.D.}}:
\batitle{{Compliance of Elastic Bodies in Contact}}.
\bjtitle{Journal of Applied Mechanics}
\bvolume{16}(\bissue{3}),
\bfpage{259}--\blpage{268}
(\byear{1949})
\end{barticle}
\endbibitem

\bibitem[\protect\citeauthoryear{Puzyrev et~al.}{2024}]{Puzyrev2024}
\begin{barticle}
\bauthor{\bsnm{Puzyrev}, \binits{D.}},
\bauthor{\bsnm{Trittel}, \binits{T.}},
\bauthor{\bsnm{Harth}, \binits{K.}},
\bauthor{\bsnm{Stannarius}, \binits{R.}}:
\batitle{Cooling of a granular gas mixture in microgravity}.
\bjtitle{npj Microgravity}
\bvolume{10}(\bissue{1}),
\bfpage{36}
(\byear{2024})
\doiurl{10.1038/s41526-024-00369-5}
\end{barticle}
\endbibitem

\bibitem[\protect\citeauthoryear{Wu et~al.}{2019}]{Wu2019}
\begin{botherref}
\oauthor{\bsnm{Wu}, \binits{Y.}},
\oauthor{\bsnm{Kirillov}, \binits{A.}},
\oauthor{\bsnm{Massa}, \binits{F.}},
\oauthor{\bsnm{Lo}, \binits{W.-Y.}},
\oauthor{\bsnm{Girshick}, \binits{R.}}:
Detectron2.
\url{https://github.com/facebookresearch/detectron2}
(2019)
\end{botherref}
\endbibitem

\bibitem[\protect\citeauthoryear{}{}]{statistical_significance_0_05}
\begin{botherref}
Statistical significance: p value, 0.05 threshold, and applications to radiomics—reasons for a conservative approach.
\url{https://rdcu.be/dv7LL}
\end{botherref}
\endbibitem

\bibitem[\protect\citeauthoryear{Pedregosa et~al.}{2011}]{scikit-learn}
\begin{barticle}
\bauthor{\bsnm{Pedregosa}, \binits{F.}},
\bauthor{\bsnm{Varoquaux}, \binits{G.}},
\bauthor{\bsnm{Gramfort}, \binits{A.}},
\bauthor{\bsnm{Michel}, \binits{V.}},
\bauthor{\bsnm{Thirion}, \binits{B.}},
\bauthor{\bsnm{Grisel}, \binits{O.}},
\bauthor{\bsnm{Blondel}, \binits{M.}},
\bauthor{\bsnm{Prettenhofer}, \binits{P.}},
\bauthor{\bsnm{Weiss}, \binits{R.}},
\bauthor{\bsnm{Dubourg}, \binits{V.}},
\bauthor{\bsnm{Vanderplas}, \binits{J.}},
\bauthor{\bsnm{Passos}, \binits{A.}},
\bauthor{\bsnm{Cournapeau}, \binits{D.}},
\bauthor{\bsnm{Brucher}, \binits{M.}},
\bauthor{\bsnm{Perrot}, \binits{M.}},
\bauthor{\bsnm{Duchesnay}, \binits{E.}}:
\batitle{Scikit-learn: Machine learning in {P}ython}.
\bjtitle{Journal of Machine Learning Research}
\bvolume{12},
\bfpage{2825}--\blpage{2830}
(\byear{2011})
\end{barticle}
\endbibitem

\bibitem[\protect\citeauthoryear{Chaya}{2020}]{rf}
\begin{botherref}
\oauthor{\bsnm{Chaya}}:
{Random Forest Regression}.
\url{https://levelup.gitconnected.com/random-forest-regression-209c0f354c84}
(2020)
\end{botherref}
\endbibitem

\bibitem[\protect\citeauthoryear{Guo et~al.}{2020}]{xgb}
\begin{barticle}
\bauthor{\bsnm{Guo}, \binits{R.}},
\bauthor{\bsnm{Zhao}, \binits{Z.}},
\bauthor{\bsnm{Wang}, \binits{T.}},
\bauthor{\bsnm{Liu}, \binits{G.}},
\bauthor{\bsnm{Zhao}, \binits{J.}},
\bauthor{\bsnm{Gao}, \binits{D.}}:
\batitle{Degradation state recognition of piston pump based on iceemdan and xgboost}.
\bjtitle{Applied Sciences}
\bvolume{10},
\bfpage{6593}
(\byear{2020})
\doiurl{10.3390/app10186593}
\end{barticle}
\endbibitem

\bibitem[\protect\citeauthoryear{Ustaoglu et~al.}{2008}]{NN_regression}
\begin{barticle}
\bauthor{\bsnm{Ustaoglu}, \binits{B.}},
\bauthor{\bsnm{Cigizoglu}, \binits{H.}},
\bauthor{\bsnm{Karaca}, \binits{M.}}:
\batitle{Forecast of daily mean, maximum and minimum temperature time series by three artificial neural network methods}.
\bjtitle{Meteorological Applications}
\bvolume{15},
\bfpage{431}--\blpage{445}
(\byear{2008})
\doiurl{10.1002/met.83}
\end{barticle}
\endbibitem

\bibitem[\protect\citeauthoryear{Agarap}{2018}]{Agarap2018}
\begin{botherref}
\oauthor{\bsnm{Agarap}, \binits{A.F.}}:
Deep learning using rectified linear units (relu).
CoRR
\textbf{abs/1803.08375}
(2018)
{\href{https://arxiv.org/abs/1803.08375}{{1803.08375}}}
\end{botherref}
\endbibitem

\bibitem[\protect\citeauthoryear{Breiman et~al.}{1984}]{breiman1984classification}
\begin{bbook}
\bauthor{\bsnm{Breiman}, \binits{L.}},
\bauthor{\bsnm{Friedman}, \binits{J.}},
\bauthor{\bsnm{Olshen}, \binits{R.A.}},
\bauthor{\bsnm{Stone}, \binits{C.J.}}:
\bbtitle{Classification and Regression Trees},
\bedition{1st} edn.,
p. \bfpage{368}.
\bpublisher{Chapman and Hall/CRC},
\blocation{New York, USA}
(\byear{1984}).
\doiurl{10.1201/9781315139470}
\end{bbook}
\endbibitem

\bibitem[\protect\citeauthoryear{Breiman}{2001}]{breiman2001random}
\begin{barticle}
\bauthor{\bsnm{Breiman}, \binits{L.}}:
\batitle{Random forests}.
\bjtitle{Machine Learning}
\bvolume{45}(\bissue{1}),
\bfpage{5}--\blpage{32}
(\byear{2001})
\doiurl{10.1023/A:1010933404324}
\end{barticle}
\endbibitem

\bibitem[\protect\citeauthoryear{}{}]{SupportVectorRegression}
\begin{botherref}
LIBSVM: A Library for Support Vector Machines.
\url{https://www.csie.ntu.edu.tw/~cjlin/papers/libsvm.pdf}
\end{botherref}
\endbibitem

\bibitem[\protect\citeauthoryear{NVIDIA}{}]{nvidia_xgboost}
\begin{botherref}
\oauthor{\bsnm{NVIDIA}}:
XGBoost.
\url{https://www.nvidia.com/en-us/glossary/xgboost/}
\end{botherref}
\endbibitem

\bibitem[\protect\citeauthoryear{PyTorch}{}]{bcewithlogits}
\begin{botherref}
\oauthor{\bsnm{PyTorch}}:
BCEWithLogitsLoss.
\url{https://docs.pytorch.org/docs/stable/generated/torch.nn.BCEWithLogitsLoss.html}
\end{botherref}
\endbibitem

\bibitem[\protect\citeauthoryear{NVIDIA}{2025}]{nvidia2025}
\begin{botherref}
\oauthor{\bsnm{NVIDIA}}:
A Comprehensive Overview of Regression Evaluation Metrics
(2025).
\url{https://developer.nvidia.com/blog/a-comprehensive-overview-of-regression-evaluation-metrics}
\end{botherref}
\endbibitem

\bibitem[\protect\citeauthoryear{Google}{}]{google_metrics}
\begin{botherref}
\oauthor{\bsnm{Google}}:
Classification: Accuracy, recall, precision, and related metrics.
\url{https://developers.google.com/machine-learning/crash-course/classification/accuracy-precision-recall}
\end{botherref}
\endbibitem

\bibitem[\protect\citeauthoryear{Opitz}{2024}]{Opitz2024}
\begin{barticle}
\bauthor{\bsnm{Opitz}, \binits{J.}}:
\batitle{A closer look at classification evaluation metrics and a critical reflection of common evaluation practice}.
\bjtitle{Transactions of the Association for Computational Linguistics}
\bvolume{12},
\bfpage{820}--\blpage{836}
(\byear{2024})
\doiurl{10.1162/tacl_a_00675}
{\href{https://arxiv.org/abs/https://direct.mit.edu/tacl/article-pdf/doi/10.1162/tacl\_a\_00675/2465598/tacl\_a\_00675.pdf}{{https://direct.mit.edu/tacl/article-pdf/doi/10.1162/tacl\_a\_00675/2465598/tacl\_a\_00675.pdf}}}
\end{barticle}
\endbibitem

\end{thebibliography}
\end{document}